\documentclass[iop,numberedappendix]{emulateapj}

\usepackage{longtable}
\usepackage{amssymb,amsmath}
\usepackage{array,booktabs}
\usepackage{graphicx}
\usepackage{multirow}
\usepackage{hyperref}
\usepackage{color}
\usepackage{soul}
\usepackage{rotating}

\definecolor{darkgreen}{rgb}{0.0,,0.0}

\newcommand{\be}{\begin{equation}}
\newcommand{\ee}{\end{equation}}
\newcommand{\bea}{\begin{eqnarray}}
\newcommand{\eea}{\end{eqnarray}}

\begin{document}

\title{The Multiplicity of M-dwarfs in Young Moving Groups}

\author{Yutong Shan$^{1}$, Jennifer C. Yee$^{2}$, Brendan P. Bowler$^{3, 10}$, Lucas A. Cieza$^{4}$, Benjamin T. Montet$^{5, 11}$, H{\'e}ctor C{\'a}novas$^{6}$, Michael C. Liu$^{7}$ \\
And\\
Laird M. Close$^{8}$, Phil M. Hinz$^{8}$, Jared R. Males$^{8,11}$, Katie M. Morzinski$^{8,11}$, Amali Vaz$^{8}$, Vanessa P. Bailey$^{9}$, Katherine B. Follette$^{9}$ \\
(MagAO Team)}
\altaffiltext{1}{Department of Astronomy, Harvard University, 60 Garden Street, Cambridge, MA 02138, USA}
\altaffiltext{2}{Smithsonian Astrophysical Observatory, 60 Garden Street, Cambridge, MA 02138, USA}
\altaffiltext{3}{McDonald Observatory and the Department of Astronomy, The University of Texas at Austin, Austin, TX 78712, USA}
\altaffiltext{4}{Universidad Diego Portales,  Facultad de Ingeniería y Ciencias. Av. Ejército 441, Santiago, Chile}
\altaffiltext{5}{Department of Astronomy and Astrophysics, University of Chicago, 5640 S. Ellis Ave., Chicago, IL 60637, USA}
\altaffiltext{6}{Departamento de F\'isica Te\'orica, Universidad Aut\'onoma de Madrid, Cantoblanco, 28049 Madrid, Spain}
\altaffiltext{7}{Institute for Astronomy, University of Hawaii, 2680 Woodlawn Drive, Honolulu, HI 96822, USA}
\altaffiltext{8}{Steward Observatory, University of Arizona, 933 North Cherry Ave, Tucson, AZ 85721, USA}
\altaffiltext{9}{Kavli Institute for Particle Astrophysics and Cosmology, Stanford University, Stanford, CA 94305, USA}
\altaffiltext{10}{NASA Hubble Fellow}
\altaffiltext{11}{NASA Sagan Fellow}

\begin{abstract}

We image 104 newly identified low-mass (mostly M-dwarf) pre-main sequence members of nearby young moving groups with Magellan Adaptive Optics (MagAO) and identify 27 binaries with instantaneous projected separation as small as 40 mas. 15 were previously unknown. The total number of multiple systems in this sample including spectroscopic and visual binaries from the literature is 36, giving a raw multiplicity rate of at least $35^{+5}_{-4}\%$ for this population. In the separation range of roughly 1 - 300 AU in which infrared AO imaging is most sensitive, the raw multiplicity rate is at least $24^{+5}_{-4}\%$ for binaries resolved by the MagAO infrared camera (Clio). The M-star sub-sample of 87 stars yields a raw multiplicity of at least $30^{+5}_{-4}\%$ over all separations, $21^{+5}_{-4}\%$ for secondary companions resolved by Clio from 1 to 300 AU ($23^{+5}_{-4}\%$ for all known binaries in this separation range). A combined analysis with binaries discovered by the Search for Associations Containing Young stars shows that multiplicity fraction as a function of mass and age over the range of 0.2 to 1.2 $M_\sun$ and 10 - 200 Myr appears to be linearly flat in both parameters and across YMGs. This suggests that multiplicity rates are largely set by 100 Myr without appreciable evolution thereafter. After bias corrections are applied, the  multiplicity fraction of low-mass YMG members ($< 0.6 M_\sun$) is in excess of the field.              
\bigskip

\end{abstract}

%%%%%%%%%%%%%%%%%%%%%%%%%%%%%%%%%%%%%%%%%%%%%%%%%%%%%%%%%%%

\section{Introduction}\label{sec:intro}

Young moving groups (YMGs) are groups comprising several dozens of coeval, kinematically-associated stars that share a common birth environment. Typically aged 10 - 150 Myr, they are in the process of dispersing into the field \citep{zs04, tor08}. This population represents an excellent laboratory for directly imaging giant planets \citep[e.g.][]{del13,bow17}.  

Because stars from a given YMG formed from the same collapsing gas cloud, they offer an opportunity to directly test stellar evolution models \citep[e.g. ][]{zs04,yj10,jef16}. In particular, the binary systems among them are important for these tests in 3 ways. First, it is often possible to recover more detailed information on individual binaries, such as dynamical mass measurements, making them a stronger test of stellar evolution \citep[e.g. ][]{azu15,mon15}. Second, when considering an ensemble of YMG members, it is important to have accurate properties, which might be inferred improperly if a close binary is thought to be a single star. Finally, binarity information is crucial to reaching dependable YMG membership conclusions, since unresolved binaries may affect the measured photometry as well as the radial velocity of a system used for its YMG assignment \citep[e.g.][]{mal13,kra14}.   

In addition, the multiplicity of YMGs themselves may provide important insight into the formation and evolution of binary star systems. In general, primordial binarity could play an important role in protostellar disk evolution as well as planet formation \citep{cie09, kra12, kra16}. Furthermore, YMG members are representative of a transitional stage between stars in their birth clusters and those in the field \citep{l-s06}. Their sparse present-day density and youth suggest that they likely represent a pristine population that has not undergone and will not undergo much dynamical processing. Therefore, their multiplicity likely reflects primordial conditions.

The Search for Associations Containing Young stars (SACY) consortium has been actively studying the properties of stars in nearby YMGs \citep{tor08}. Based on their survey of FGK-dwarfs, \citet{ell14,ell15,ell16} find that stellar multiplicity is independent of primary mass. This result is consistent with findings from studies of very young star-forming regions \citep{lei93,koh00}, as well as young (5 Myr) OB-associations \citep{laf14}, but very different from what is observed in the field \citep[e.g.][and references therein]{dk13}.

In this work, we present adaptive optics (AO) imaging for recently discovered, nearby, K and M-type YMG members that complement the multiplicity study of \citet{ell14,ell15,ell16}. We identify visual binaries from 3 - 300 AU and examine their overall multiplicity fractions as a function of mass, age, and YMG, as well as distributions in separation and mass ratio. 

This paper is organized as follows. The sample is presented in Section \ref{sec:sample}. We document our observations and the data reduction procedure in Section \ref{sec:obs-dat}. Section \ref{sec:ana} describes the binaries in our observed targets. The visual binary discoveries are validated through, among others methods, co-moving analysis combined with measurements from literature (Section \ref{subsubsec:valid}). We perform multiplicity statistics measurements on our sample in Section \ref{sec:multi-stats}. Therein we first describe the construction of detection completeness maps (Section \ref{subsec:cc}) and then the evaluation of physical parameters for the systems (Section \ref{subsec:physpars}). From there we calculate overall raw multiplicity fractions (Section \ref{subsec:mf1}) and distributions in separation and mass ratio for our sample (Section \ref{subsec:physep&q}). In Section \ref{subsec:sacy}, we combine our sample with that of SACY for a joint analysis. This extends YMG multiplicity fraction measurements from 0.2 to $1.2 M_\sun$. Possible biases and their effects are discussed at length in Section \ref{subsec:bias}. We conclude with Section \ref{sec:conclusion}.

\bigskip
%==============================================

\section{A Sample of Low-Mass YMG Members}\label{sec:sample}

%----------------------------------------------------------------
Our target list is drawn primarily from \citealt{mal13,mal14a} (64 targets), and \citealt{kra14} (34 targets), and supplemented by targets from \citealt{moo13} (5 targets) and \citealt{rod13} (2 targets). Stars in these catalogues are members of young ($<200$ Myr), local ($\lesssim 100$ pc) YMGs. In selecting targets, we had a strong preference for late-type dwarfs (K5V - M5V, see Figure \ref{fig:spt-distr}) with $\delta < -40^\circ$ and $H < 11$mag. From these catalogues, we eliminated many of the known visual binaries with separately resolvable spectral types. Section \ref{subsubsec:selectionbias} discusses the potential effect this selection strategy has on our statistics. We do not take into account whether or not a target is a known spectroscopic binary (SB). At the time of observation, most of these targets have not been previously imaged with AO.

Table \ref{table:ymgs} displays properties of the seven YMGs with members included in this study. The age for each YMG comes from Table 2 in \citet{bel15} except for Argus which is from Table 1 in \citet{mal13}.

\citet{jan16} [hereafter J16] independently surveyed southern YMGs and published a catalogue of low-mass visual binaries among them. The J16 observations were performed using lucky imaging in the $z'$-band on the 3.5m ESO New Technology telescope, with very comparable angular resolution sensitivity. There are 56 targets in common between J16 and our sample. Twenty-four of our targets were also previously observed with AO by \citet{jan12} [hereafter J12].  
 
% --- Table on YMG properties ---
\begin{table}[htb]
\begin{center}
\caption{Properties of Young Moving Groups Contributing to Our Sample}
\label{table:ymgs}
\begin{tabular}{lcrccc}
\hline\hline
Group & Age & Iso- & Dist. & Number \\
 & & chrone & & \\
 & (Myr) & Age \textsuperscript{a} & Range \textsuperscript{b} & of Stars \\
  &  &  (Myr) & (pc) & Imaged \textsuperscript{c} \\
\hline
Tucana-Horologium & 41-49 & 40 & 36-71 & 51 (2) \\
(TucHor) & &  & \\
Columba (Col) & 38-48 & 40 & 35-81 & 15 (2) \\
$\beta$ Pictoris (bPic) & 21-27 & 25 & 9-73 & 13 (1) \\
Argus (Arg) & 30-50 & 40 & 8-68 & 10 (1) \\
AB Doradus & 130-200 & 120 & 7-77 & 6 (2) \\
(ABDor) & &  & \\
Carina (Car) & 38-56 & 50 & 46-88 & 6 (0)\\
TW Hydrae (TWA) & 7-13 & 10 & 28-92 & 4 (1) \\
\hline\hline
\end{tabular}
\end{center}
Notes: \\
\textsuperscript{a} The ages of the closest isochrones from \citet{bar15} adopted for stellar mass calculations, see Section \ref{subsec:clio-vbs}. \\
\textsuperscript{b} YMG distance ranges are from Table 1 in \citet{mal13}. \\
\textsuperscript{c} Unbracketed value denotes the total number of observed targets attributed to the group. In parentheses are the number of targets which have non-negligible probability (i.e. $> 10\%$) of belonging to another YMG. \\
\end{table}
% --------------------------------------

\bigskip
%==============================================

\section{Observations \& Data}\label{sec:obs-dat}

%----------------------------------------------------------------
\subsection{MagAO Observations}\label{subsec:obs}

Adaptive Optics observations were conducted on the 6.5m Magellan Clay Telescope at the Las Campanas Observatory in Chile using the MagAO instrument \citep{mor14}. Images were taken with two science cameras simultaneously: Clio in the near-infrared and VisAO in the optical. Since late-type dwarfs are brighter in the NIR, we use Clio images for our science analysis. We allocate the visible light as follows: 96\% is directed towards the wavefront sensor whereas VisAO receives 4\% of the remainder. Consequently, VisAO images tend to have poor signal-to-noise and are only used to vet some close binaries.     

We use Clio's narrow camera mode (according to \citealt{mor15}'s measurements: plate scale = $15.846 \pm 0.064$ mas/pixel; instrument north angle = $-1.8^\circ \pm 0.34$, FOV = $16 \times 8''$) and conducted most of our observations through the H-bandpass, except in one run where the Ks-filter was used. These filters are only slightly offset from their 2MASS counterparts \citep[see][Tables 7 \& 8 therein]{males14}, hence throughout this paper we will use Clio photometry interchangeably with 2MASS. For each target, one or two sets of 4 images were taken in an ABBA nodding scheme. Each image is composed of 20 co-added exposures, each 280 ms long, which prevents saturation. A summary of the run dates and filters used is presented in Table \ref{table:rundates}. 

% --- Table on run properties ---
\begin{table}[htb]
\begin{center}
\caption{Summary of the MagAO observing runs}
\label{table:rundates}
\begin{tabular}{rcc}
\hline\hline
Date & Filter (Clio) & \# of images \\ %\HL{seeing conditions?} \\
\hline
17 - 21 Apr 2014 & H & 4\\
10 - 12 Nov 2014 & Ks & 4\\
30 Nov - 2 Dec 2014 & H or Ks & 4 \\
10 - 12 May 2015 & H & 4 \\
26 - 27 Nov 2015 & H & 8\\
\hline\hline
\end{tabular}
\end{center}
%Notes: \\
%\textsuperscript{a} 2 targets from this run were observed through the Ks-filter. \\ 
\end{table}
% --------------------------------------

\begin{figure}[htb]
\center
\vspace{0cm}
\includegraphics[scale=0.5]{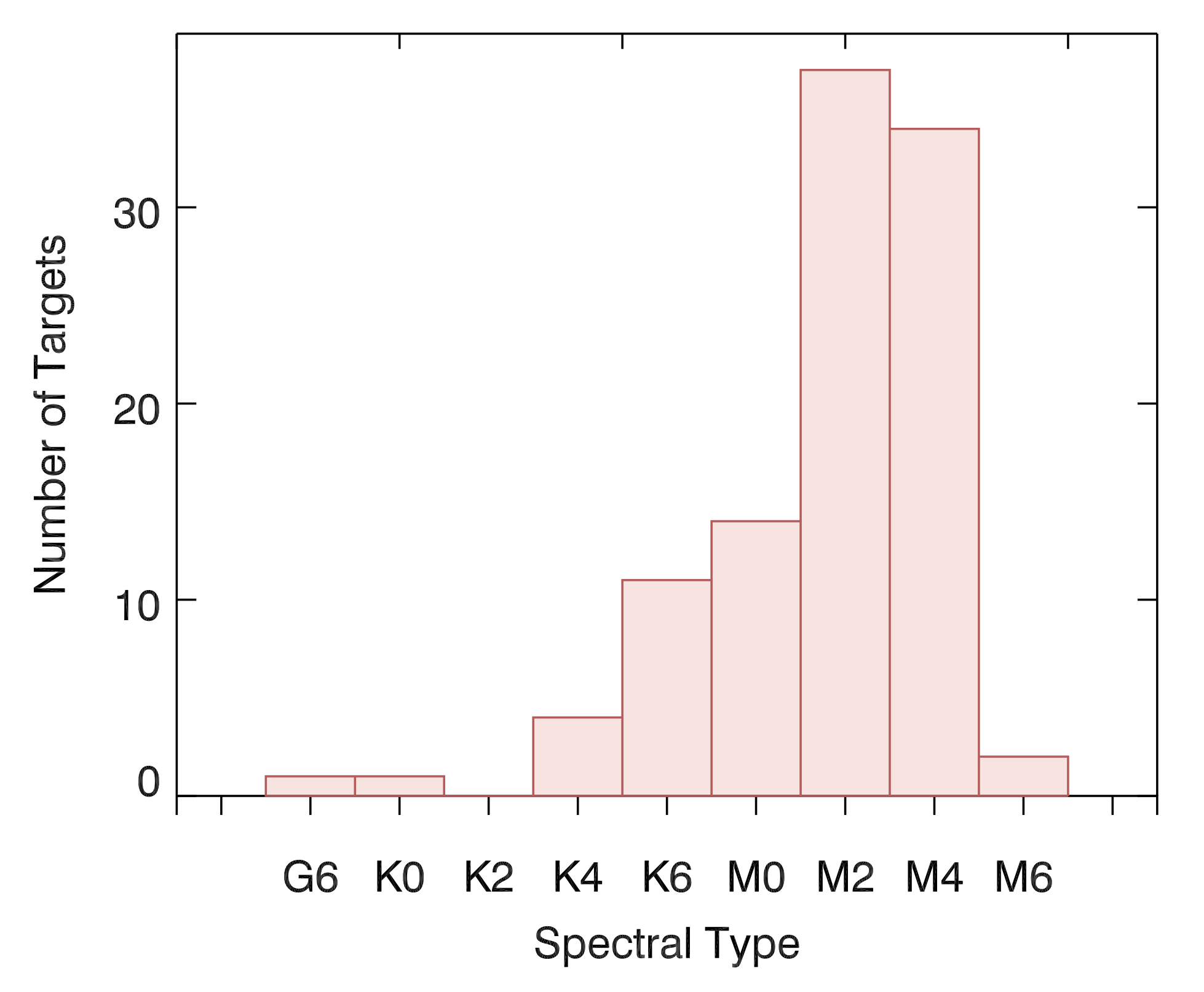}
\caption{The distribution of spectral types for our observed targets. They are chiefly pre-main sequence K/M-dwarfs}
\label{fig:spt-distr}
\end{figure}

105 targets were observed over 5 epochs, 30 of which have been imaged in multiple epochs. Table \ref{table:rundates} summarizes these observations, and Table \ref{table:sample-prop1} lists the observed targets and their relevant properties. A distribution of their spectral types as listed in their source references may be found in Figure \ref{fig:spt-distr}. They are chiefly pre-main sequence K/M-dwarfs. 

\begin{table*}[htb]
\begin{center}
\caption{YMG Low-Mass Target Sample}
\label{table:sample-prop1}
\begin{tabular}{llcccllrrrc}
\hline\hline
Target & Abbrv. & RA & DEC & SpT & Distance\textsuperscript{b}  & Ref.& J & H & K & $M_{\rm{prim}} \textsuperscript{g}$ \\
2MASS ID & Name & & & & (pc) & & & & & ($M_\sun$) \\
\hline\hline
\multicolumn{11}{c}{\bf{Tucana-Horologium (TucHor) }}\\
\hline
                       J00125703-7952073&      J0012-7952&     00:12:57.03&     -79:52:07.3&  M2.9&    48$~\pm$    2&                   3&   9.68&   9.05&   8.75&   0.49 \\
                       J00144767-6003477&      J0014-6003&     00:14:47.67&     -60:03:47.7&  M3.6&    42$~\pm$    2&                   3&   9.71&   9.10&   8.83&   0.39 \\
                       J00152752-6414545&      J0015-6414&     00:15:27.52&     -64:14:54.5&  M1.8&    50$~\pm$    3&                   3&   9.32&   8.69&   8.44&   0.60 \\
                       J00171443-7032021&      J0017-7032&     00:17:14.43&     -70:32:02.1&  M0.5&    63$~\pm$    4&                   1&   9.00&   8.38&   8.15&   0.76 \\
                       J00235732-5531435&      J0023-5531&     00:23:57.32&     -55:31:43.5&  M4.1&    42$~\pm$    2&                   3&  11.11&  10.55&  10.24&   0.16 \\
                       J00273330-6157169&      J0027-6157&     00:27:33.30&     -61:57:16.9&  M4.0&    44$~\pm$    2&                   3&  10.33&   9.73&   9.47&   0.27 \\
                       J00284683-6751446&      J0028-6751&     00:28:46.83&     -67:51:44.6&  M4.5&    46$~\pm$    2&                   3&  11.40&  10.77&  10.50&   0.15 \\
                       J00302572-6236015&      J0030-6236&     00:30:25.72&     -62:36:01.5&  M2.2&    44$~\pm$    2&                   3&   8.44&   7.80&   7.55&   0.54 \\
                       J00332438-5116433&      J0033-5116&     00:33:24.38&     -51:16:43.3&  M3.4&    42$~\pm$    2&                   3&   9.86&   9.27&   9.01&   0.35 \\
                       J00421010-5444431&      J0042-5444&     00:42:10.10&     -54:44:43.1&  M2.9&    46$~\pm$    2&                   3&   9.81&   9.21&   8.93&   0.42 \\
                       J00485254-6526330&      J0048-6526&     00:48:52.54&     -65:26:33.0&  M3.2&    50$~\pm$    3&                   3&  10.41&   9.85&   9.55&   0.31 \\
                       J00493566-6347416&      J0049-6347&     00:49:35.66&     -63:47:41.6&  M1.7&    46$~\pm$    2&                   3&   9.28&   8.66&   8.43&   0.56 \\
                       J01024375-6235344&      J0102-6235&     01:02:43.75&     -62:35:34.4&  M2.9&    46$~\pm$    2&                   3&   9.64&   9.04&   8.80&   \nodata \\
                       J01505688-5844032&      J0150-5844&     01:50:56.88&     -58:44:03.2&  M3.0&    46$~\pm$    2&                   3&   9.54&   8.87&   8.64&   0.50 \\
                       J01521830-5950168&      J0152-5950&     01:52:18.30&     -59:50:16.8&M2-3.0&    40$~\pm$    2&                   2&   8.94&   8.33&   8.14&   0.57 \\
                       J02125819-5851182&      J0212-5851&     02:12:58.19&     -58:51:18.2&  M1.9&    48$~\pm$    2&                   3&   9.32&   8.65&   8.44&   0.58 \\
                       J02205139-5823411&      J0220-5823&     02:20:51.39&     -58:23:41.1&  M3.2&    38$~\pm$    2&                   3&   9.67&   9.09&   8.83&   0.35 \\
                       J02224418-6022476&      J0222-6022&     02:22:44.18&     -60:22:47.6&  M4.0&    32$~\pm$    2&                   2&   8.99&   8.39&   8.10&   0.43 \\
                       J02294569-5541496&      J0229-5541&     02:29:45.69&     -55:41:49.6&  M4.8&    46$~\pm$    2&                   3&  11.10&  10.54&  10.26&   0.11 \\
                       J02414683-5259523&     J0241-5259A&     02:41:46.83&     -52:59:52.3&  K6.0&    43$~\pm$    1\textsuperscript{c}&2&   7.58&   6.93&   6.76&   0.73 \\
                       J02414730-5259306&     J0241-5259B&     02:41:47.30&     -52:59:30.6&  M2.5&    44$~\pm$    3&                   2&   8.48&   7.85&   7.64&   0.56 \\
                       J02423301-5739367&      J0242-5739&     02:42:33.01&     -57:39:36.7&  K5.0&    48$~\pm$    1\textsuperscript{c}&2&   8.56&   7.97&   7.78&   0.73 \\
                       J02474639-5804272&      J0247-5804&     02:47:46.39&     -58:04:27.2&  M1.8&    44$~\pm$    2&                   3&   9.36&   8.67&   8.45&   0.52 \\
                       J02543316-5108313&      J0254-5108&     02:54:33.16&     -51:08:31.3&  M1.1&    44$~\pm$    2&                   3&   8.67&   8.07&   7.78&   0.67 \\
    J02564708-6343027\textsuperscript{a}&      J0256-6343&     02:56:47.08&     -63:43:02.7&  M4.0&    53$~\pm$    3&                   1&   9.86&   9.22&   9.01&   0.49 \\
                       J02572682-6341293&      J0257-6341&     02:57:26.82&     -63:41:29.3&  M3.6&    63$~\pm$    3&                   3&  10.16&   9.57&   9.33&   0.32 \\
                       J03114544-4719501&      J0311-4719&     03:11:45.44&     -47:19:50.1&  M3.2&    44$~\pm$    2&                   3&  10.44&   9.89&   9.57&   0.25 \\
                       J03315564-4359135&      J0331-4359&     03:31:55.64&     -43:59:13.5&  K6.0&    45$~\pm$    1\textsuperscript{c}&2&   8.30&   7.68&   7.47&   0.73 \\
                       J03512287-5154582&      J0351-5154&     03:51:22.87&     -51:54:58.2&  M4.0&    50$~\pm$    3&                   3&  10.61&  10.03&   9.77&   0.28 \\
                       J04053964-4014103&      J0405-4014&     04:05:39.64&     -40:14:10.4&  M4.2&    48$~\pm$   3\textsuperscript{f}&                   5&   9.82&   9.26&   8.98&   0.30 \\
                       J04074372-6825111&      J0407-6825&     04:07:43.72&     -68:25:11.1&  M3.2&    60$~\pm$    3&                   3&  10.41&   9.78&   9.52&   0.42 \\
                       J04133609-4413325&      J0413-4413&     04:13:33.14&     -52:31:58.6&  M2.4&    50$~\pm$    3&                   3&  10.00&   9.35&   9.12&   0.42 \\
                       J04133314-5231586&      J0413-5231&     04:13:36.10&     -44:13:32.5&  M3.9&    60$~\pm$    3&                   3&  10.77&  10.19&   9.91&   0.32 \\
                       J04440099-6624036&      J0444-6624&     04:44:00.99&     -66:24:03.6&  M0.5&    55$~\pm$    4&                   2&   9.47&   8.75&   8.58&   0.62 \\
                       J04475779-5035200&      J0447-5035&     04:47:57.79&     -50:35:20.0&  M4.0&    55$~\pm$    3&                   3&  10.06&   9.43&   9.21&   0.28 \\
                       J04470041-5134405&      J0447-5134&     04:47:00.41&     -51:34:40.5&  M1.9&    60$~\pm$    3&                   3&  10.87&  10.30&  10.02&   0.30 \\
                       J05332558-5117131&      J0533-5117&     05:33:25.58&     -51:17:13.1&  K7.0&    55$~\pm$    1\textsuperscript{c}&2&   8.99&   8.36&   8.16&   0.63 \\
                       J17080882-6936186&      J1708-6936&     17:08:08.82&     -69:36:18.6&  M3.5&    49$~\pm$    3&                   2&   9.06&   8.42&   8.20&   0.54 \\
                       J19225071-6310581&      J1922-6310&     19:22:50.71&     -63:10:58.1&  M3.0&    61$~\pm$    4&                   2&   9.45&   8.82&   8.58&   0.66 \\
                       J20423672-5425263&      J2042-5425&     20:42:36.72&     -54:25:26.3&  M4.0&    48$~\pm$    2&                   3&  10.75&  10.16&   9.86&   0.24 \\
                       J21083826-4244540&      J2108-4244&     21:08:38.26&     -42:44:54.0&  M4.4&    44$~\pm$    2&                   3&  10.14&   9.57&   9.24&   0.20 \\
                       J21100614-5811483&      J2110-5811&     21:10:06.14&     -58:11:48.3&  M4.0&    52$~\pm$    3&                   3&  10.89&  10.33&  10.07&   0.24 \\
                       J21143354-4213528&      J2114-4213&     21:14:33.54&     -42:13:52.8&  M3.9&    52$~\pm$    3&                   3&  11.38&  10.78&  10.53&   0.18 \\
                       J21163528-6005124&      J2116-6005&     21:16:35.28&     -60:05:12.4&  M3.5&    48$~\pm$    2&                   3&  10.19&   9.56&   9.31&   0.35 \\
                       J21354554-4218343&      J2135-4218&     21:35:45.54&     -42:18:34.3&  M5.2&    58$~\pm$    3&                   3&  11.68&  11.15&  10.81&   0.17 \\
                       J21490499-6413039&      J2149-6413&     21:49:04.99&     -64:13:03.9&  M4.5&    44$~\pm$    2&                   2&  10.35&   9.80&   9.47&   0.18 \\
                       J22025453-6440441&      J2202-6440&     22:02:54.53&     -64:40:44.1&  M1.8&    46$~\pm$    2&                   3&   9.06&   8.41&   8.16&   0.62 \\
                       J22440873-5413183&      J2244-5413&     22:44:08.73&     -54:13:18.3&  M4.0&    49$~\pm$    4&                   2&   9.36&   8.71&   8.47&   0.42 \\
                       J22463471-7353504&      J2246-7353&     22:46:34.71&     -73:53:50.4&  M2.3&    52$~\pm$    3&                   3&   9.66&   9.05&   8.81&   0.54 \\
    J22470872-6920447\textsuperscript{a}&      J2247-6920&     22:47:08.72&     -69:20:44.7&  K6.0&    52$~\pm$    1\textsuperscript{c}&1&   8.89&   8.30&   8.09&   0.68 \\
                       J23474694-6517249&      J2347-6517&     23:47:46.94&     -65:17:24.9&  M1.5&    45$~\pm$    2&                   2&   9.10&   8.39&   8.17&   0.61 \\
\hline
\hline
\end{tabular}
\end{center}
Notes: \\
\textsuperscript{a} denotes YMG designation is ambiguous according to \citealt{mal13,mal14a} \\
\textsuperscript{b} The distance estimate for each target is either statistical/kinematical or from trigonometric parallax, taken from their respective references unless \textsuperscript{c}, \textsuperscript{d}, or \textsuperscript{e}\\
\textsuperscript{c} \citealt{gaia_dr1} \\
\textsuperscript{d} \citealt{don16} \\
\textsuperscript{e} \citealt{wei13} \\
\textsuperscript{f} no distance error is reported in source reference. Error estimate is based on the average error for other targets in this YMG \\
\textsuperscript{g} mass calculations are discussed in Sections \ref{subsec:clio-vbs} and \ref{subsec:physpars} 
\\
References: 1. \citealt{mal13}; 2. \citealt{mal14a}; 3. \citealt{kra14}; 4. \citealt{moo13}; 5. \citealt{rod13}
\end{table*}

\begin{table*}[htb]
\setcounter{table}{2}
\begin{center}
\caption{YMG Low-Mass Target Sample}
\label{table:sample-prop2}
\begin{tabular}{llcccclrrrc}
\hline\hline
Target & Abbrv. & RA & DEC & SpT & Distance\textsuperscript{b}  & Ref.& J & H & K & $M_{\rm{prim}} \textsuperscript{g}$ \\
2MASS ID & Name & & & & (pc) & & & & & ($M_\sun$) \\
\hline\hline
\multicolumn{11}{c}{\bf{Columba (Col) }}\\
\hline
                       J01424689-5126469&      J0142-5126&     01:42:46.89&     -51:26:46.9&  M6.5&    \hphantom{a}66$~\pm$ 6\textsuperscript{f}&                   5&  11.08&  10.58&  10.10&   0.21 \\
    J02365171-5203036\textsuperscript{a}&      J0236-5203&     02:36:51.71&     -52:03:03.6&  M2.0&    39$~\pm$    2&                   1&   8.42&   7.76&   7.50&   0.68 \\
                       J03241504-5901125&      J0324-5901&     03:24:15.04&     -59:01:12.5&  K7.0&    90$~\pm$    5&                   2&   9.55&   8.92&   8.72&   0.78 \\
                       J03320347-5139550&      J0332-5139&     03:32:03.47&     -51:39:55.0&  M2.0&    88$~\pm$    5&                   1&  10.23&   9.58&   9.35&   0.62 \\
                       J03494535-6730350&      J0349-6730&     03:49:45.35&     -67:30:35.0&  K7.0&    81$~\pm$    4&                   2&   9.85&   9.23&   9.03&   0.70 \\
                       J04091413-4008019&      J0409-4008&     04:09:14.13&     -40:08:01.9&  M3.5&    63$~\pm$    5&                   2&  10.65&  10.00&   9.77&   0.38 \\
                       J04240094-5512223&      J0424-5512&     04:24:00.94&     -55:12:22.3&  M2.5&    68$~\pm$    5&                   2&   9.80&   9.16&   8.95&   0.64 \\
                       J04515303-4647309&      J0451-4647&     04:51:53.03&     -46:47:30.9&  M0.0&    76$~\pm$    6&                   2&   9.80&   9.14&   8.89&   0.59 \\
                       J05111098-4903597&      J0511-4903&     05:11:10.98&     -49:03:59.7&  M3.5&    62$~\pm$    6&                   2&  10.64&  10.01&   9.77&   0.38 \\
                       J05164586-5410168&      J0516-5410&     05:16:45.86&     -54:10:16.8&  M3.0&    69$~\pm$    6&                   2&  10.43&   9.78&   9.55&   0.50 \\
                       J05392505-4245211&      J0539-4245&     05:39:25.05&     -42:45:21.1&  M2.0&    41$~\pm$    4&                   1&   9.45&   8.80&   8.60&   0.45 \\
                       J07065772-5353463&      J0706-5353&     07:06:57.72&     -53:53:46.3&  M0.0&    53$~\pm$    5&                   2&   8.54&   7.90&   7.67&   0.78 \\
    J07170438-6311123\textsuperscript{a}&      J0717-6311&     07:17:04.38&     -63:11:12.3&  M2.0&    58$~\pm$    4&                   1&   9.73&   9.09&   8.86&   0.39 \\
                       J08152160-4918303&      J0815-4918&     08:15:21.60&     -49:18:30.3&  G7.0&   120$~\pm$    4\textsuperscript{c}&4&   9.61&   9.19&   9.11&   0.89 \\
                       J09331427-4848331&      J0933-4848&     09:33:14.27&     -48:48:33.1&  K7.0&    \hphantom{a}46$~\pm$    1\textsuperscript{c}&2&   8.94&   8.33&   8.10&   0.63 \\
\hline
\multicolumn{11}{c}{\bf{$\beta$ Pictoris (bPic) }}\\
\hline
                       J00172353-6645124&      J0017-6645&        00:17:24&     -66:45:12.4&  M2.5&    39$~\pm$    2&                   2&   8.56&   7.93&   7.70&   0.60 \\
                       J05332802-4257205&      J0533-4257&     05:33:28.02&     -42:57:20.5&  M4.5&    16$~\pm$    4&                   1&   8.00&   7.40&   7.12&   0.17 \\
                       J08475676-7854532&      J0847-7854&     08:47:56.76&     -78:54:53.2&  M3.0&    66$~\pm$    7&                   1&   9.32&   8.68&   8.41&   0.72 \\
                       J11493184-7851011&      J1149-7851&     11:49:31.84&     -78:51:01.1&  M1.0&    68$~\pm$    6&                   2&   9.45&   8.72&   8.49&   0.72 \\
                       J13545390-7121476&      J1354-7121&     13:54:53.90&     -71:21:47.6&  M2.5&    21$~\pm$    1&                   2&   8.55&   7.92&   7.67&   0.25 \\
    J14252913-4113323\textsuperscript{a}&      J1425-4113&     14:25:29.13&     -41:13:32.3&  M2.5&    66$~\pm$    4&                   2&   8.55&   7.91&   7.61&   0.73 \\
                       J16572029-5343316&      J1657-5343&     16:57:20.29&     -53:43:31.6&  M3.0&    51$~\pm$    3&                   2&   8.69&   8.07&   7.79&   0.53 \\
                       J17292067-5014529&      J1729-5014&     17:29:20.67&     -50:14:52.9&  M3.0&    64$~\pm$    5&                   2&   8.87&   8.19&   7.99&   0.64 \\
                       J18142207-3246100&      J1814-3246&     18:14:22.07&     -32:46:10.0&  M1.5&    90$~\pm$    8&                   2&   9.44&   8.77&   8.54&   0.87 \\
                       J18420694-5554254&      J1842-5554&     18:42:06.94&     -55:54:25.4&  M3.5&    54$~\pm$    4&                   2&   9.49&   8.82&   8.58&   0.54 \\
                       J18465255-6210366&      J1846-6210&     18:46:52.55&     -62:10:36.6&  M1.0&    54$~\pm$    3&                   2&   8.75&   8.05&   7.85&   0.76 \\
                       J19233820-4606316&      J1923-4606&     19:23:38.20&     -46:06:31.6&  M0.0&    70$~\pm$    4&                   2&   9.11&   8.44&   8.27&   0.81 \\
                       J21212873-6655063&      J2121-6655&     21:21:28.73&     -66:55:06.3&  K7.0&    \hphantom{a}32$~\pm$    1\textsuperscript{c}&2&   7.88&   7.26&   7.01&   0.55 \\
\hline
\multicolumn{11}{c}{\bf{Argus (Arg) }}\\
\hline
                       J04464970-6034109&      J0446-6034&     04:46:49.70&     -60:34:10.9&  M1.5&    37$~\pm$    2&                   2&   8.55&   7.95&   7.72&   0.62 \\
                       J05090356-4209199&      J0509-4209&     05:09:03.56&     -42:09:19.9&  M3.5&    51$~\pm$    4&                   2&   9.58&   8.98&   8.76&   0.54 \\
                       J06380031-4056011&      J0638-4056&     06:38:00.31&     -40:56:01.1&  M3.5&    35$~\pm$    2&                   1&  10.35&   9.81&   9.53&   0.19 \\
                       J10252563-4918389&      J1025-4918&     10:25:25.63&     -49:18:38.9&  M4.0&    27$~\pm$    1&                   2&   9.12&   8.51&   8.26&   0.30 \\
                       J11234697-5257393&      J1123-5257&     11:23:46.97&     -52:57:39.3&  M0.0&    \hphantom{a}49$~\pm$    1\textsuperscript{c}&4&   8.79&   8.20&   7.95&   0.70 \\
                       J12233860-4606203&      J1223-4606&     12:23:38.60&     -46:06:20.3&  M4.0&    23$~\pm$    1&                   1&   9.53&   8.94&   8.70&   0.18 \\
                       J14284804-7430205&      J1428-7430&     14:28:48.04&     -74:30:20.5&  M1.0&    76$~\pm$    3&                   1&   9.26&   8.57&   8.35&   0.80 \\
                       J15163224-5855237&      J1516-5855&     15:16:32.24&     -58:55:23.7&  K7.0&    77$~\pm$    3&                   1&   9.10&   8.55&   8.29&   0.67 \\
    J20072376-5147272\textsuperscript{a}&      J2007-5147&     20:07:23.76&     -51:47:27.2&  K6.0&    34$~\pm$    0&1\textsuperscript{c}&   8.16&   7.57&   7.39&   0.66 \\
                       J23532520-7056410&      J2353-7056&     23:53:25.20&     -70:56:41.0&  M3.5&    17$~\pm$    1&                   2&   8.68&   8.10&   7.78&   0.21 \\
\hline
\multicolumn{11}{c}{\bf{Carina (Car) }}\\
\hline
                       J07540718-6320149&      J0754-6320&     07:54:07.18&     -63:20:14.9&  M3.0&    80$~\pm$    5&                   2&  10.33&   9.69&   9.45&   0.45 \\
                       J08094269-5652199&      J0809-5652&     08:09:42.69&     -56:52:19.9&  K0.0&   104$~\pm$    3\textsuperscript{c}&4&   9.38&   8.98&   8.83&   0.90 \\
                       J08185942-7239561&      J0818-7239&     08:18:59.42&     -72:39:56.1&  M0.0&    60$~\pm$    2&                   1&   9.78&   9.15&   8.94&   0.42 \\
                       J09032434-6348330&      J0903-6348&     09:03:24.34&     -63:48:33.0&  M0.5&    66$~\pm$    3&                   2&   9.57&   8.86&   8.69&   0.67 \\
                       J09111581-5014149&      J0911-5014&     09:11:15.81&     -50:14:14.9&  K5.0&   117$~\pm$    5\textsuperscript{c}&4&  10.23&   9.65&   9.50&   0.77 \\
                       J09312541-5314366&      J0931-5314&     09:31:25.41&     -53:14:36.6&  K5.0&   120$~\pm$   19&                   4&  10.19&   9.66&   9.50&   0.76 \\
\hline
\multicolumn{11}{c}{\bf{AB Doradus (ABDor) }}\\
\hline
                       J01484087-4830519&      J0148-4830&     01:48:40.87&     -48:30:51.9&  M1.5&    36$~\pm$    2&                   2&   9.19&   8.55&   8.36&   0.53 \\
                       J05240991-4223054&      J0524-4223&     05:24:09.91&     -42:23:05.4&  M0.5&    52$~\pm$    9&                   2&  10.58&   9.92&   9.72&   0.37 \\
    J05381615-6923321\textsuperscript{a}&      J0538-6923&     05:38:16.15&     -69:23:32.1&  M0.5&    21$~\pm$    2&                   1&   8.96&   8.29&   8.11&   0.37 \\
                       J05531299-4505119&      J0553-4505&     05:53:12.99&     -45:05:11.9&  M0.5&    34$~\pm$    4&                   2&   8.60&   7.94&   7.73&   0.61 \\
    J08465879-7246588\textsuperscript{a}&      J0846-7246&     08:46:58.79&     -72:46:58.8&  K7.0&    45$~\pm$    1&1\textsuperscript{c}&   8.49&   7.81&   7.60&   0.61 \\
                       J15244849-4929473&      J1524-4929&     15:24:48.49&     -49:29:47.3&  M2.0&    23$~\pm$    1&                   2&   8.16&   7.53&   7.30&   0.54 \\
\hline
\multicolumn{11}{c}{\bf{TW Hydrae (TWA) }}\\
\hline
    J11455177-5520456\textsuperscript{a}&      J1145-5520&     11:45:51.77&     -55:20:45.6&  K5.0&    \hphantom{a}43$~\pm$    1\textsuperscript{c}&1&   8.02&   7.41&   7.27&   0.63 \\
                       J12313807-4558593&      J1231-4558&     12:31:38.07&     -45:58:59.3&  M3.0&    78$~\pm$    3&                   2&   9.33&   8.69&   8.41&   0.39 \\
                       J12345629-4538075&      J1234-4538&     12:34:56.29&     -45:38:07.5&  M1.5&    \hphantom{a}78$~\pm$    3\textsuperscript{e}&2&   8.99&   8.33&   8.09&   0.49 \\
                       J12350424-4136385&      J1235-4136&     12:35:04.24&     -41:36:38.5&  M2.0&    \hphantom{a}59$~\pm$    2\textsuperscript{d}&2&   9.12&   8.48&   8.19&   0.50 \\
\hline
\hline
\end{tabular}
\end{center}
Notes: \\
\textsuperscript{a} denotes YMG designation is ambiguous according to \citealt{mal13,mal14a} \\
\textsuperscript{b} The distance estimate for each target is either statistical/kinematical or from trigonometric parallax, taken from their respective references unless \textsuperscript{c}, \textsuperscript{d}, or \textsuperscript{e}\\
\textsuperscript{c} \citealt{gaia_dr1} \\
\textsuperscript{d} \citealt{don16} \\
\textsuperscript{e} \citealt{wei13} \\
\textsuperscript{f} no distance error is reported in source reference. Error estimate is based on the average error for other targets in this YMG \\
\textsuperscript{g} mass calculations are discussed in Sections \ref{subsec:clio-vbs} and \ref{subsec:physpars} 
\\
References: 1. \citealt{mal13}; 2. \citealt{mal14a}; 3. \citealt{kra14}; 4. \citealt{moo13}; 5. \citealt{rod13}
\end{table*}

%\iffalse
% --- Figure showing target SpT distribution ---
% --------------------------------------
%\fi

Fourteen of our YMG targets have Tycho designations and are part of the recent {\emph{Gaia}} Data Release 1 \citep{gaia_dr1}. For those targets, we adopt the Gaia distances, which are consistent with the statistical ones derived from their sources in literature. In addition to the targets in Table \ref{table:sample-prop1}, we observed J06511418-4037510, J12170465-5743558, J10252092-4241539, and J13213722-4421518, which were later eliminated from our sample because of new distance information. The first two have Gaia parallax measurements consistent with being background giants and are thus excluded from analysis. This means 10-20\% of the sample may have incorrect YMG designations. The last two have been rejected by \citet{don16} from membership in TWA also on the basis of parallax. None of these four objects harbours obvious visual companions down to $0.08''$. Future {\emph{Gaia}} releases will clarify the YMG assignments for the remaining targets.

%----------------------------------------------------------------

\subsection{Data Reduction}\label{subsec:data reduce}

\emph{Sky Subtraction --} We perform sky subtraction by directly subtracting the nod pairs that are closest temporally (i.e. `A' and `B' images). This procedure also eliminates the dark and bias current.  

\emph{Bad Pixels --} We exclude bad pixels and obvious cosmic rays from any analysis and source-fitting we perform when extracting measurements from our images. 

\emph{Flat Fielding --} H-band flat fields are constructed from MagAO twilight flats (and darks) taken on the same night as the observations whenever available, and from an adjacent night otherwise. For Ks images, we use the master Ks-flat field from April 2014 supplied by the Magellan MagAO team, available through the online Clio user manual.  

MagAO twilight flats are known to be unreliable (see, e.g. \citealt{mor15}). Thus, we compared photometric and astrometric measurements derived from several images of binaries that had been flat-fielded to the unflattened ones. In terms of astrometry, the two sets of images produce virtually identical values with indistinguishable precision. For the flux ratio measurements, flattening leads to a slight reduction in the errors. Therefore, we opt to use flattened images for our entire analysis.  

%\emph{Non-Linearity? --} As the exposures are sufficiently short, non-linearity effects are negligible.    

\bigskip
%==============================================

~
\section{YMG Binaries}\label{sec:ana}

%----------------------------------------------------------------
\subsection{Clio Binary Detections}\label{subsec:clio-vbs}

\begin{figure}[htb]
\center 
\includegraphics[scale=0.55]{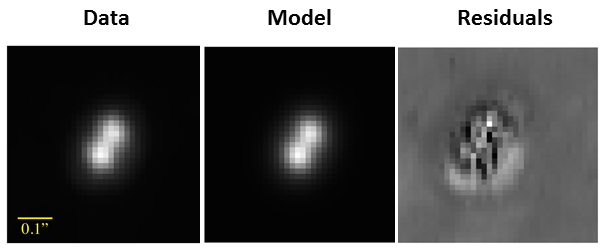}
\caption{An example of a PSF-fit to a binary (J0717-6311), separation = $0.08''$, $\Delta H = 0.09$. Left: pre-processed data image. Middle: best-fit triple bivariate Gaussian model. Right: fit residuals. The stretch on the residuals plot is 1/10th that of the data and model plots.}
\label{fig:amg0717-6311-thumbnail}
\end{figure}

We visually examined every Clio as well as VisAO image in SAOImage DS9 \citep{jm03} at various contrasts to identify visual doubles among our targets. To the Clio images we fit triple bivariate Gaussian PSFs to all suspected sources using {\tt{mpfitfun.pro}}, an IDL routine that uses a Levenberg-Marquardt technique to efficiently explore the $\chi^2$ space of a multi-parameter function \citep{mar09}. This is similar to the algorithm described by \citet{bh15}. The fits are initialized by parameter estimates from visual inspection of the images. For every target whose Clio or VisAO images in a particular epoch exhibit multiple clear, visually identifiable components, we perform a multi-source fit to its Clio images. The PSFs of all components are constrained to share the same PSF shape. The centroid positions and the overall height scalings are allowed to vary independently for each component. Figure \ref{fig:amg0717-6311-thumbnail} shows an example fit to a tight binary. 

For all images, we measure the minimum radius of the PSF cross-section at half-height. We compute the axis ratio of this cross-section as the minimum radius divided by the maximum radius. We also output the East of North orientation of the major axis. We give these parameters in Table \ref{table:fwhm} for all targets measured in their best epochs (determined by the depth of their contrast curves in the annulus between 0\farcs1 and 1\farcs0). For binaries we additionally measure their separation, position angle (PA), and flux ratio from the fitted Gaussians. Binary separation is the distance between the two fitted PSF centroids, in pixel units. This can be converted into angular units using the plate scale (15.846 mas/pixel, \citealt{mor15}), and then into physical units using the distance to the source (as reported in Table \ref{table:sample-prop1}). When comparing the plate scale measured in 2012 \citep{mor15} to that measured from data taken in November 2015, the systematic uncertainty in the plate scale is approximately 0.1 mas/pixel. We use this value as the fiducial error in our plate scale. Position angle (PA) is the angle between the vector connecting the primary to the secondary centroid and north, measured counterclockwise (i.e. east of north). To obtain the true PA, the image angles are derotated using the de-rotation angle (ROTOFF) associated with that observation as well as the true north angle (NORTH\_CLIO = $-1.80 \pm 0.34$; \citealt{mor15}) of the instrument. The flux ratio is determined as the overall secondary-to-primary PSF scaling factor, which is converted into $\Delta$mag.  

The final reported values are averaged over the independent measurements in the set of (4 or 8) consecutive images taken for each target at each epoch. We note that, in the case of PA, we take the average in complex space. We take the measurement errors to be the standard error of the mean for each epoch. The final errors combine, in quadrature, the measurement errors and uncertainties from the instrument calibrations. Physical quantities also include uncertainties due to the distances from the literature. 

For all apparently single sources where the average PSF axis ratio at a given epoch is less than 0.72 (corresponding to a fractional elongation of 40\%), we perform a 2-source fit. Based on our wider binary detections, we define secure close binary detection to be one in which the standard deviations in the measured binaries quantities meet the following criteria: 

\begin{itemize}
\item{Position Angle: $\sigma_{\rm{PA}} < 15^\circ$,} 
\item{Flux ratio: $\sigma_{F_s/F_p} < 0.1$, and}
\item{Separation: $\sigma_{\rm{sep}}/{\rm{Sep}} < 20\%$.}
\end{itemize}

In addition, we require that the 1-source elongated PSF has an orientation angle that is consistent with the 2-source PA within 15 degrees, and that the 2-source PSF is more round than the 1-source counterpart (i.e. axis ratio increases towards 1). This procedure allowed us to obtain robust astrometric measurements for 3 very tight ($<70$ mas) binaries, which were also resolved in VisAO (see Figure \ref{fig:close-vbs}). Furthermore, if a binary is observed in one or more epochs, we perform a 2-source fit to that object in all epochs.

\begin{table*}[htb]
\begin{center}
\caption{Target FWHMs}
\label{table:fwhm}
\begin{tabular}{ccclccrc}
\hline\hline
Target & \# of & Best & Band & Minor & Axis & Orientation & Multiple?\textsuperscript{a} \\
Name & Epochs & Epoch & & Axis (") & Ratio & Angle ($^\circ$) & \\
\hline\hline
      J0012-7952&   1&    14-11-10& Ks&   0.038$~\pm$   0.001&    0.88$~\pm$    0.02&      34$~\pm$       6&  N \\
      J0014-6003&   1&    14-11-10& Ks&   0.039$~\pm$   0.002&    0.91$~\pm$    0.03&       4$~\pm$       6&  N \\
      J0015-6414&   1&    14-11-10& Ks&   0.039$~\pm$   0.008&    0.83$~\pm$    0.06&       4$~\pm$      12&  N \\
      J0017-6645&   1&    14-11-12& Ks&   0.034$~\pm$   0.000&    0.88$~\pm$    0.01&      35$~\pm$       7&  N \\
      J0017-7032&   1&    14-11-10& Ks&   0.041$~\pm$   0.001&    0.90$~\pm$    0.03&     175$~\pm$       7&  Y \\
      J0023-5531&   1&    14-11-10& Ks&   0.043$~\pm$   0.001&    0.84$~\pm$    0.01&      83$~\pm$       7&  N \\
      J0027-6157&   1&    14-11-10& Ks&   0.042$~\pm$   0.000&    0.87$~\pm$    0.02&      92$~\pm$       6&  N \\
      J0028-6751&   1&    14-11-10& Ks&   0.055$~\pm$   0.003&    0.86$~\pm$    0.03&     111$~\pm$       8&  N \\
      J0030-6236&   1&    15-11-26&  H&   0.045$~\pm$   0.007&    0.88$~\pm$    0.02&      32$~\pm$      14&  Y \\
      J0033-5116&   1&    14-11-12& Ks&   0.037$~\pm$   0.000&    0.86$~\pm$    0.04&      49$~\pm$       8&  N \\

\hline
\hline
\end{tabular}
\end{center}
Notes: \\
\textsuperscript{a} N denotes `Single' and Y denotes `Multiple', to the best of our knowledge.\\
Full ASCII table available online. \\
\end{table*}

\begin{table*}[htb]
\begin{center}
\caption{Astrometric Measurements for VB Candidates from This Work}
\label{table:vb-epochs}
\begin{tabular}{lrrllcrcc}
\hline\hline
Target Name & $\mu_\alpha {\rm{cos}} \delta$\textsuperscript{a} & $\mu_\delta$\textsuperscript{a} &Epoch & Band & Ang. Sep. (") & PA ($^\circ$) & $\Delta$mag & Comove?\textsuperscript{c} \\
  & (mas/yr) & (mas/yr) & & & & & & \\
\hline\hline
      J0030-6236&    95.2$~\pm$     0.9&   -48.0$~\pm$     0.9&    15-11-26&  H&   0.121$~\pm$   0.001&   268.6$~\pm$    0.36&    0.14$~\pm$    0.04& CO \\
\hline
      J0102-6235&    88.9$~\pm$     1.2&   -39.3$~\pm$     1.2&    14-12-01&  H&   0.566$~\pm$   0.044&   332.7$~\pm$    2.05&    0.27$~\pm$    0.08&  B \\
&&&    15-11-26&  H&   0.678$~\pm$   0.005&   325.7$~\pm$    0.35&    0.36$~\pm$    0.04& \\
\hline
      J0142-5126&    66.8$~\pm$     4.2&   -12.7$~\pm$     4.2&    15-11-26&  H&   2.014$~\pm$   0.014&   288.6$~\pm$    0.37&    0.89$~\pm$    0.23&  \nodata \\
\hline
      J0229-5541&    91.3$~\pm$     3.4&   -16.0$~\pm$     3.4&    14-11-30&  H&   0.159$~\pm$   0.001&   101.4$~\pm$    0.35&    0.21$~\pm$    0.04&  \nodata \\
\hline
     J0241-5259B&    91.6$~\pm$     1.1&    -3.6$~\pm$     1.0&    14-12-01&  H&   0.075$~\pm$   0.001&   174.5$~\pm$    0.36&    0.05$~\pm$    0.01&  \nodata \\
\hline
      J0257-6341&    64.1$~\pm$     1.9&    11.6$~\pm$     4.1&    14-12-01&  H&   0.423$~\pm$   0.003&   325.4$~\pm$    0.35&    0.07$~\pm$    0.01&  \nodata \\
\hline
      J0324-5901&    37.5$~\pm$     1.1&     9.6$~\pm$     1.1&    14-11-30&  H&   0.466$~\pm$   0.003&   280.2$~\pm$    0.34&    2.33$~\pm$    0.01& CO \\
\hline
      J0331-4359&    85.0$~\pm$     1.4&    -8.2$~\pm$     1.9&    14-11-30&  H&   0.393$~\pm$   0.003&    92.3$~\pm$    0.34&    2.77$~\pm$    0.01& CO \\
\hline
      J0332-5139&    38.2$~\pm$     1.3&    11.6$~\pm$     1.3&    14-11-30&  H&   3.199$~\pm$   0.024&   115.5$~\pm$    0.34&    1.50$~\pm$    0.03&  I \\
\hline
      J0405-4014&    71.6$~\pm$     2.0&    -0.8$~\pm$     2.1&    15-11-27&  H&   0.609$~\pm$   0.004&   102.6$~\pm$    0.34&    0.58$~\pm$    0.04&  \nodata \\
\hline
      J0447-5035&    46.7$~\pm$     2.6&    19.8$~\pm$     2.6&    14-12-01&  H&   0.544$~\pm$   0.004&   121.3$~\pm$    0.34&    0.03$~\pm$    0.03& CO \\
\hline
      J0524-4223&     3.2$~\pm$     1.8&   -13.5$~\pm$     1.6&    14-12-01&  H&   0.248$~\pm$   0.002&    61.0$~\pm$    0.37&    0.88$~\pm$    0.01& CO \\
\hline
      J0533-4257&   -18.6$~\pm$     3.1&    43.1$~\pm$     3.5&    14-12-01&  H&   0.066$~\pm$   0.000&   222.2$~\pm$    1.09&    0.55$~\pm$    0.02&  \nodata \\
&&&    15-11-26&  H&   \nodata &    \nodata &    \nodata & \\
\hline
      J0717-6311&   -12.0$~\pm$     1.4&    45.8$~\pm$     1.4&    14-04-21&  H&   0.077$~\pm$   0.001&   203.2$~\pm$    0.41&    0.07$~\pm$    0.01&  \nodata \\
&&&    15-11-26&  H&   \nodata &     \nodata &    \nodata & \\
\hline
      J0754-6320&    -9.4$~\pm$     2.5&    28.7$~\pm$     2.5&    14-04-21&  H&   0.857$~\pm$   0.005&   156.9$~\pm$    0.34&    0.10$~\pm$    0.01&  C \\
&&&    15-11-26&  H&   0.859$~\pm$   0.005&   156.8$~\pm$    0.34&    0.11$~\pm$    0.03& \\
\hline
      J0809-5652&    -8.1$~\pm$     1.6&    19.6$~\pm$     1.1&    14-12-01&  H&   0.782$~\pm$   0.005&    29.9$~\pm$    0.34&    2.73$~\pm$    0.01&  C \\
&&&    15-11-26&  H&   0.785$~\pm$   0.005&    29.7$~\pm$    0.34&    2.71$~\pm$    0.02& \\
\hline
      J0903-6348&   -33.3$~\pm$     1.4&    33.6$~\pm$     1.4&    14-04-21&  H&   1.132$~\pm$   0.007&    64.1$~\pm$    0.34&    5.02$~\pm$    0.03&  B \\
&&&    14-11-30&  H&   1.124$~\pm$   0.008&    64.4$~\pm$    0.34&    5.04$~\pm$    0.07& \\
&&&    15-05-10&  H&   1.149$~\pm$   0.010&    66.0$~\pm$    0.37&    5.01$~\pm$    0.04& \\
&&&    15-11-26&  H&   1.129$~\pm$   0.007&    65.8$~\pm$    0.35&    5.18$~\pm$    0.02& \\
\hline
      J0931-5314&   -20.8$~\pm$     1.5&     3.4$~\pm$     3.4&    15-05-12&  H&   0.264$~\pm$   0.002&     8.2$~\pm$    0.34&    3.48$~\pm$    0.10& CO \\
&&&    15-11-26&  H&   0.236$~\pm$   0.011&    13.0$~\pm$    2.39&    3.36$~\pm$    0.28& \\
\hline
      J1234-4538&   -47.5$~\pm$     1.3&   -20.2$~\pm$     0.8&    15-05-12&  H&   0.591$~\pm$   0.004&   309.2$~\pm$    0.34&    0.03$~\pm$    0.03& CO \\
\hline
      J1516-5855&   -42.8$~\pm$     3.4\textsuperscript{b} &   -44.2$~\pm$     3.2\textsuperscript{b}&    15-05-10&  H&   2.337$~\pm$   0.015&   207.1$~\pm$    0.34&    0.12$~\pm$    0.06&  C \\
\hline
      J1657-5343&   -13.0$~\pm$     6.3&   -85.1$~\pm$     2.2&    15-05-10&  H&   0.051$~\pm$   0.000&   219.7$~\pm$    0.47&    0.20$~\pm$    0.01&  \nodata \\
\hline
      J1708-6936&   -54.6$~\pm$     1.7&   -81.1$~\pm$     1.7&    15-05-11&  H&   0.435$~\pm$   0.003&     9.1$~\pm$    0.34&    0.72$~\pm$    0.02&  C \\
\hline
      J1729-5014&    -5.8$~\pm$     1.5&   -62.7$~\pm$     5.1&    15-05-11&  H&   0.699$~\pm$   0.005&    16.7$~\pm$    0.34&    0.21$~\pm$    0.05& CO \\
\hline
      J2108-4244&    33.3$~\pm$     1.6&   -99.6$~\pm$     1.5&    14-11-11& Ks&   0.139$~\pm$   0.001&   302.3$~\pm$    0.35&    0.11$~\pm$    0.04& CO \\
&&&    15-11-27&  H&   0.129$~\pm$   0.001&   287.1$~\pm$    0.35&    0.13$~\pm$    0.04& \\
\hline
      J2121-6655&    97.2$~\pm$     1.1&  -104.1$~\pm$     1.6&    15-05-11&  H&   0.063$~\pm$   0.001&   301.6$~\pm$    4.48&    0.90$~\pm$    0.15&  \nodata \\
&&&    15-11-26&  H&  \nodata &     \nodata &    \nodata & \\
\hline
      J2244-5413&    70.7$~\pm$     1.3&   -60.0$~\pm$     1.3&    14-12-01&  H&   0.535$~\pm$   0.004&   298.8$~\pm$    0.35&    0.32$~\pm$    0.01& CO \\
&&&    15-11-26&  H&   0.520$~\pm$   0.003&   298.1$~\pm$    0.34&    0.28$~\pm$    0.02& \\
\hline
      J2247-6920&    70.9$~\pm$     1.6&   -58.9$~\pm$     1.8&    14-12-01&  H&   0.146$~\pm$   0.015&   161.2$~\pm$    4.45&    3.12$~\pm$    0.09& CO \\
&&&    15-05-11&  H&   0.246$~\pm$   0.002&   175.2$~\pm$    0.35&    3.50$~\pm$    0.11& \\
&&&    15-11-26&  H&   0.228$~\pm$   0.003&   166.0$~\pm$    0.59&    3.42$~\pm$    0.19& \\
\hline
\hline
\end{tabular}
\end{center}
Notes:\\
\textsuperscript{a} Proper motions from the {\emph{UCAC4}} catalog \citep{zac12} unless otherwise specified. \\
\textsuperscript{b} Proper motion from the {\emph{NOMAD}} catalog \citep{zac05}, which was used in the analysis by \citealt{mal13}. \\
\textsuperscript{c} C: co-moving; CO: Co-moving with orbital motion; B: Background; I: Indeterminate. 
\\
\end{table*}
\setlength{\tabcolsep}{6pt}

% --- Figure showing blended binary fit ---
% -------------------------------------------

\begin{figure*}[htb] 
\begin{minipage}{0.45\textwidth}
\centering
\hspace{-1cm}
\includegraphics[scale=0.5]{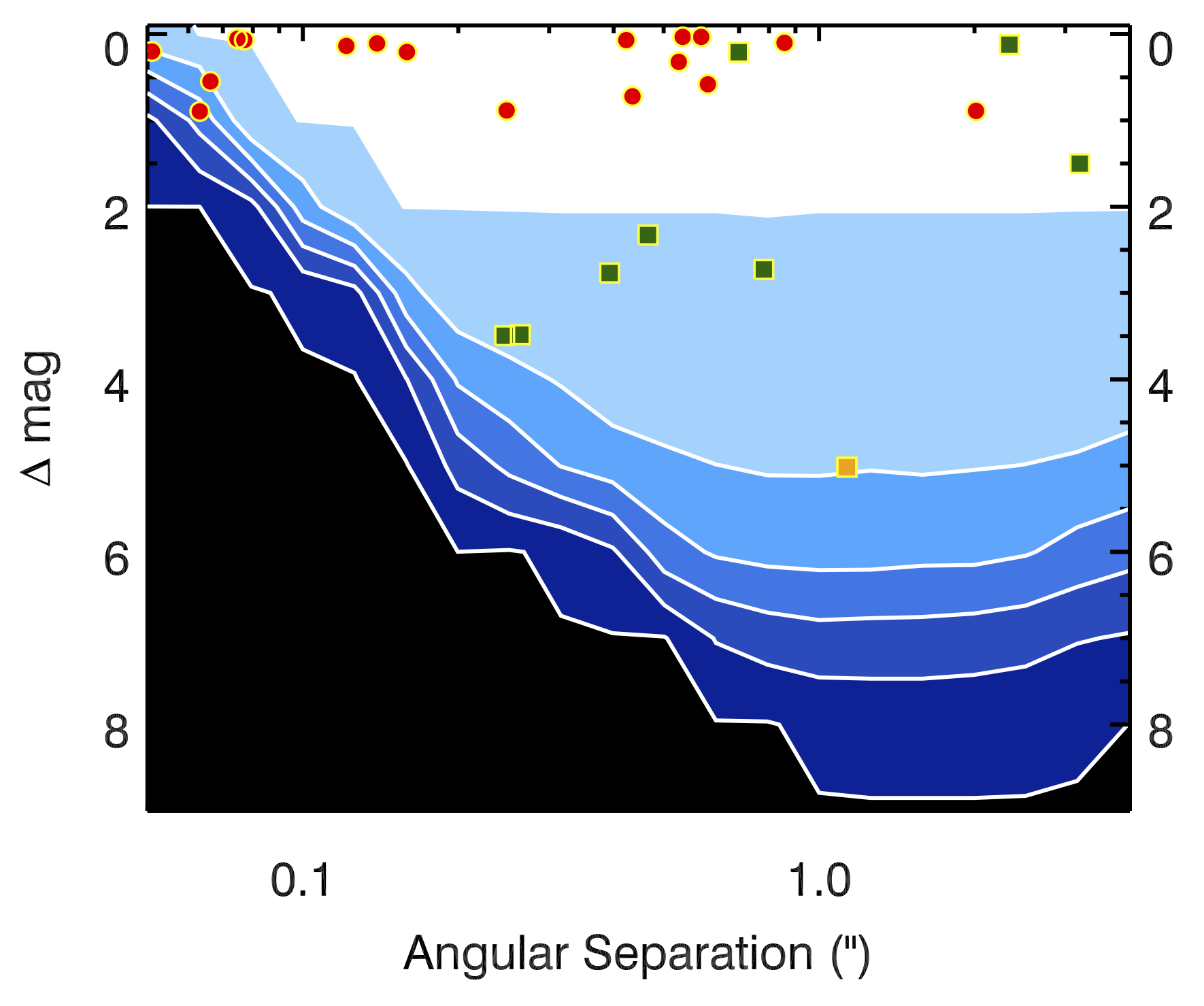}
\end{minipage}%
\begin{minipage}{0.1\textwidth}
\centering
\vspace{-1cm}
\hspace{-1cm}
\includegraphics[scale=0.43]{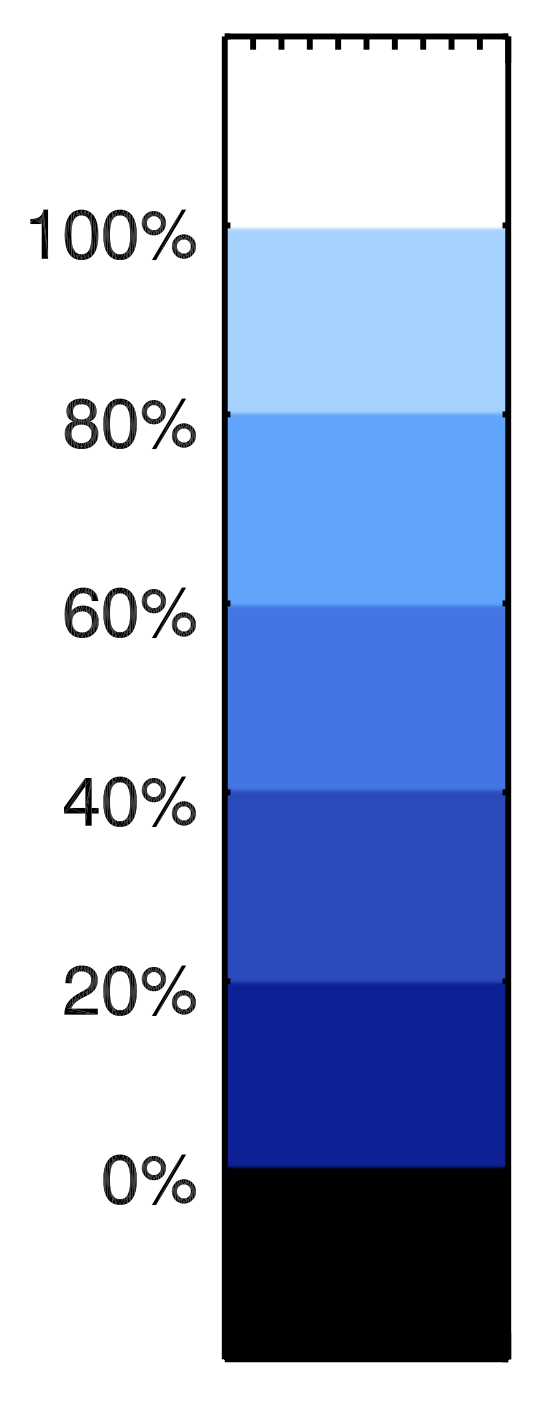}
\end{minipage}%
\begin{minipage}{0.45\textwidth}
\centering
\hspace{-1cm}
\includegraphics[scale=0.5]{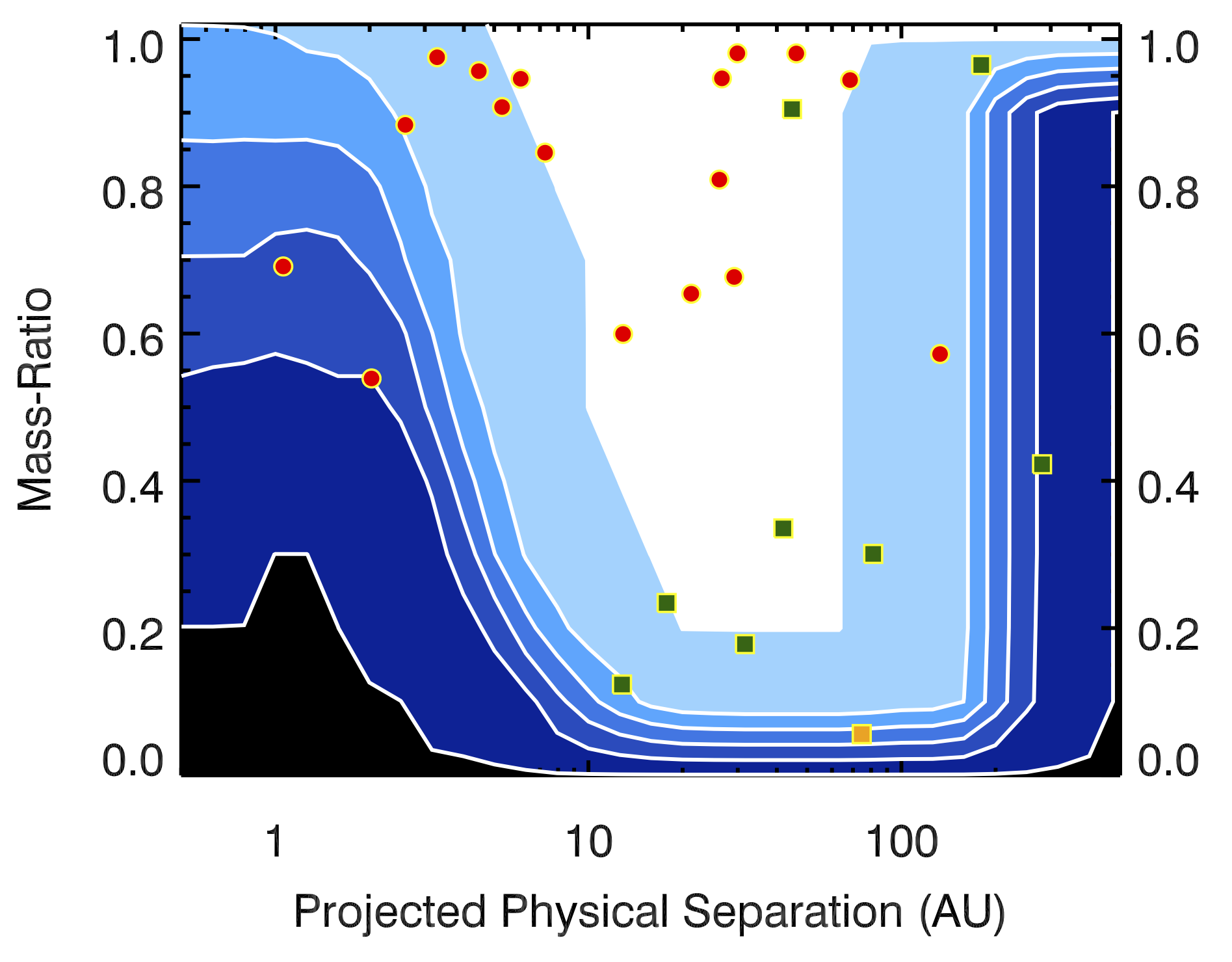}
\end{minipage}
\caption{Left Panel: Clio's completeness map in direct observables for the target ensemble. Filled red circles depict the actual detections with primary masses below $0.6 M_{\sun}$, and filled green squares are systems with $M_{\rm prim} > 0.6 M_{\sun}$. The orange square belongs to J0903-6348, which is consistent with a background object (see Section \ref{subsubsec:valid}). Centre Panel: Colour bar showing sensitivity gradient for overall sample.  Right Panel: Completeness map in physical parameters for the target ensemble. %Red-orange lines depict sensitivity contours for the low-mass ($< 0.5M_{\sun}$) sub-sample. 
%Black line depicts the 100\% sensitivity contour for a low-mass ($< 0.5M_{\sun}$) sub-sample.
Sections \ref{subsec:cc} and \ref{subsec:physpars} describe the method for generating these maps.}
\label{fig:smap}
\end{figure*}

\begin{figure*}[htb]
\center 
\includegraphics[scale=0.7]{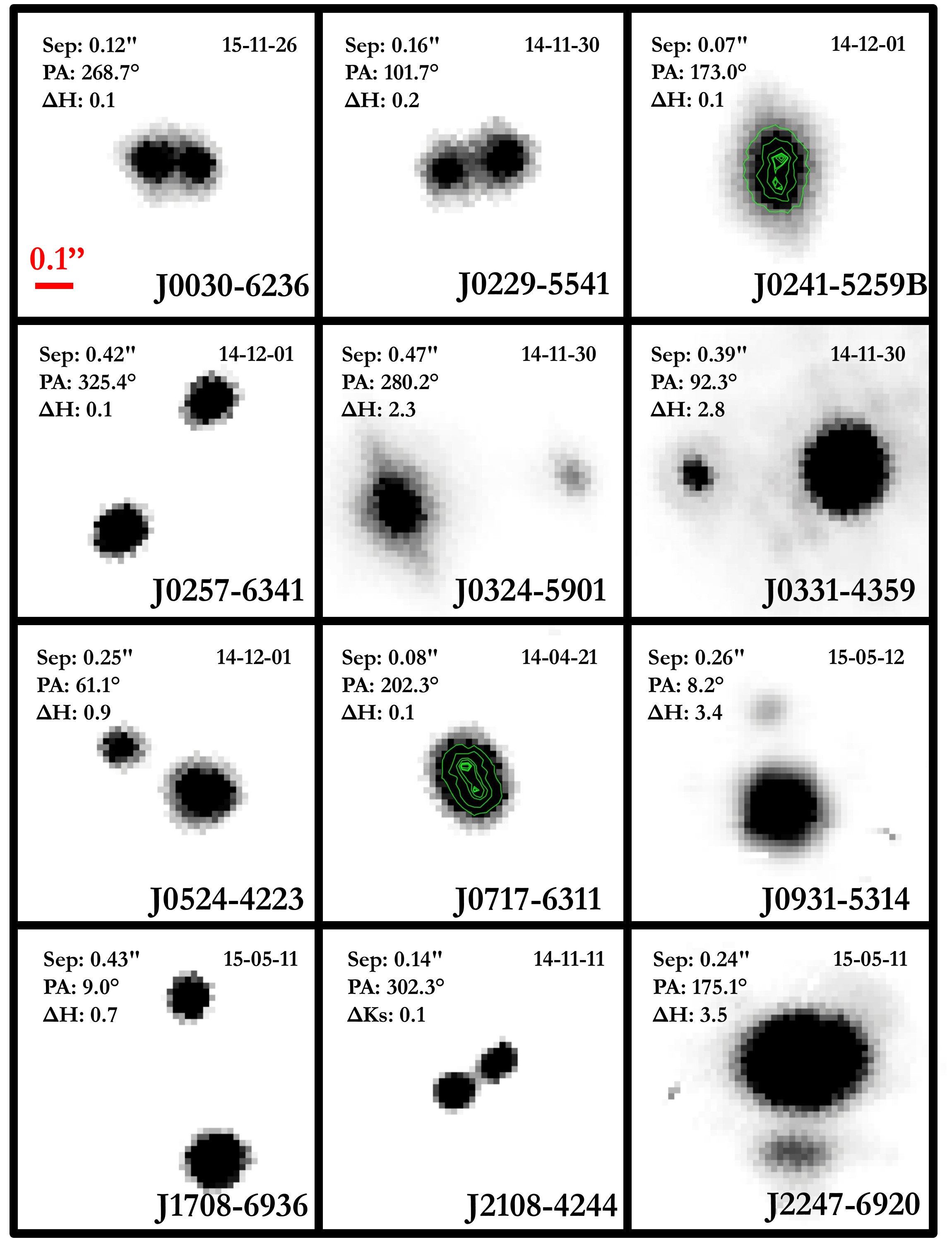}
\caption{Clio images of binaries separated by less than 0\farcs5. Each image is $0\farcs8 \times 0\farcs8$. The images have been oriented so that North points up and East points left. The colour scale is arbitrary for display purposes. Contour plots are shown whenever the binary nature is not obvious. }
\label{fig:thumbnails-small}
\end{figure*}

\begin{figure*}[htb]
\center 
\includegraphics[scale=0.7]{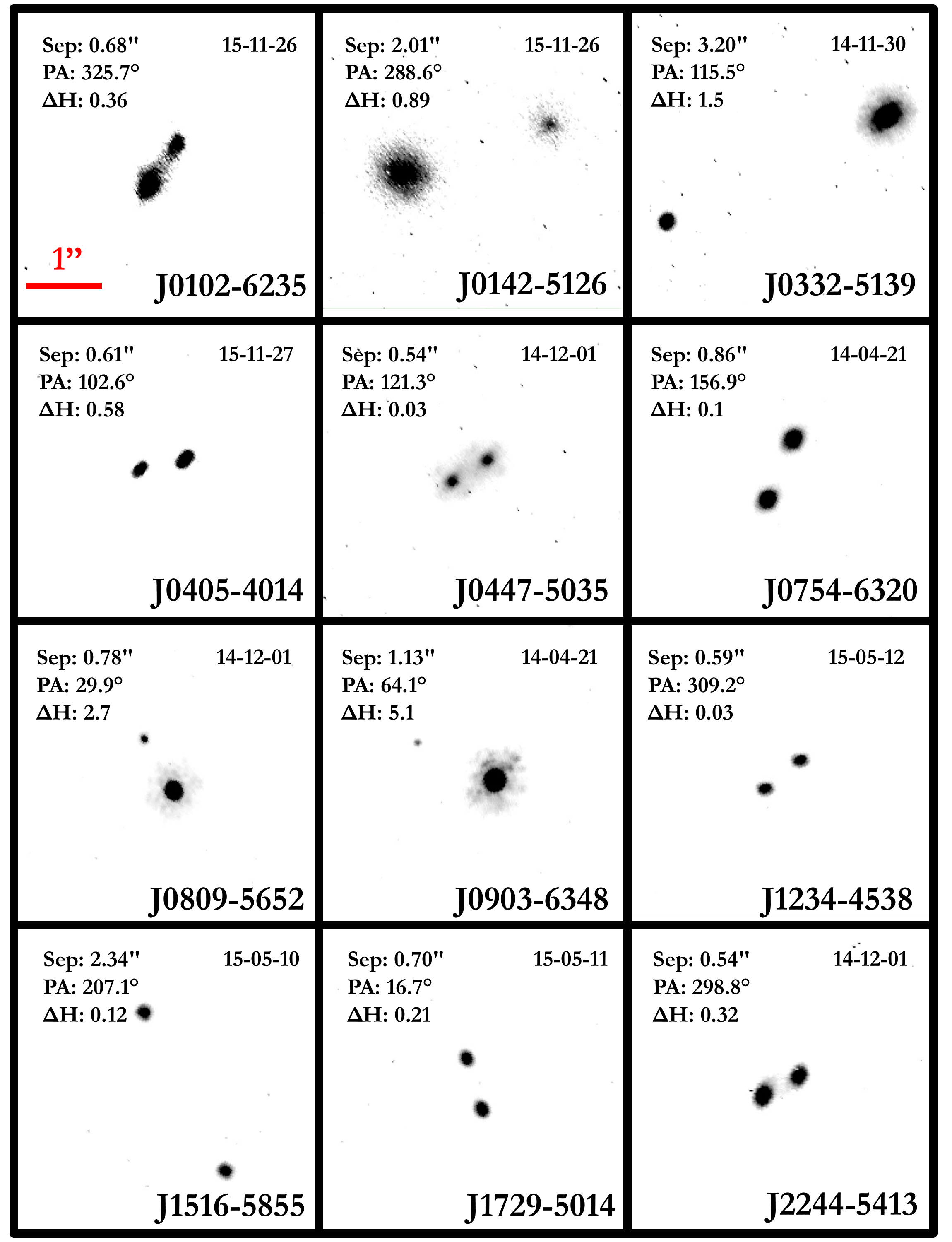}
\caption{Clio images of binaries with angular separations between $0\farcs5$ and $4\farcs0$. Each image is $4\farcs0 \times 4\farcs0$. The images have been oriented so that North points up and East points left. The colour scale is arbitrary for display purposes. }
\label{fig:thumbnails-large}
\end{figure*}

\begin{figure*}[htb]
\center 
\includegraphics[scale=0.65]{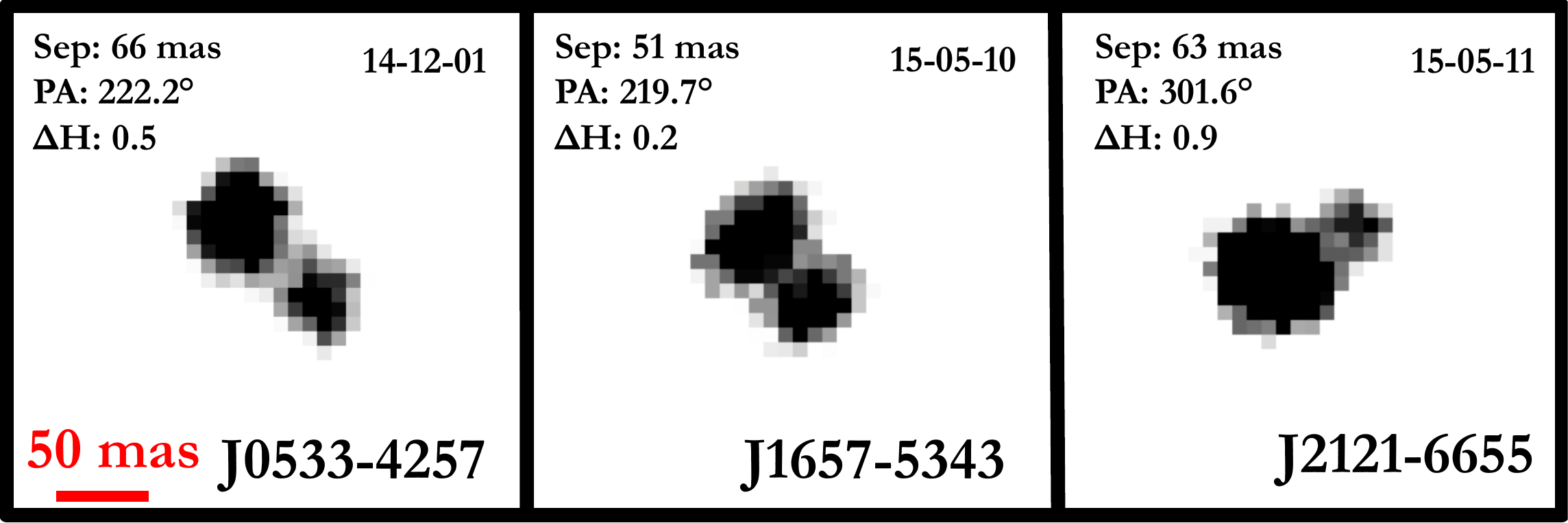}
\caption{VisAO images of binaries with angular separations below 70 mas. Though the components are not cleanly separated in their Clio images, fits to these Clio images with 2-source models yield robust measurements of their binary parameters. Each image is $280 \times 280$ mas. The images have been oriented so that North points up and East points left. The colour scale is arbitrary for display purposes. }
\label{fig:close-vbs}
\vspace{+1cm}
\end{figure*}

In summary, our Clio images revealed 27 close ($<4"$) visual doubles that were confidently fit by 2-source solutions among 105 targets, of which 14 are new discoveries.  The left panel of Figure \ref{fig:smap} shows the distribution of these doubles as a function of $\Delta$mag and projected angular separation. Snapshots of these visual pairs detected are found in Figures \ref{fig:thumbnails-small}, \ref{fig:thumbnails-large}, and \ref{fig:close-vbs}. For targets imaged in previous works, our multiplicity designation are mostly in agreement except in individual cases, listed under Section \ref{subsec:indiv-targs}. 

The right panel of Figure \ref{fig:smap} shows the detected binaries in physical units. To estimate the primary masses we use the 2MASS magnitudes for each target in conjunction with the latest stellar isochrones for low-mass PMS stars from \citet{bar15}, which assumes solar metallicity. Table \ref{table:ymgs} lists the isochrone ages adopted for each YMG. For single stars, this conversion is straightforward. For the doubles, we must account for the fact that the 2MASS magnitudes include both unresolved visual components. We split the corresponding 2MASS flux according to the flux ratios we measure in the Clio H or K-band for the best epoch and use the evolutionary models from \citet{bar15} for both components. Cases in which a tertiary and/or a known SB exists are described individually in Section \ref{subsec:indiv-targs}. 

% --- Figure showing Sensitivity Maps ---

% --- Figure showing Tight VB Images ---
% --------------------------------------

% --- Figure showing Wider VB Images ---
% --------------------------------------
   
% --- Figure showing Very Close VB Images ---
% --------------------------------------  

%----------------------------------------------------------------
\subsubsection{Validation}\label{subsubsec:valid} 

There are three ways in which we validate the Clio visual binaries as bound companions. First, in combination with published information from literature (\citealt{cha10}, J12, \citealt{ell15}, and J16), we have multi-epoch astrometric measurements for 18 visual doubles that can be used to confirm whether or not they are co-moving and therefore bound. Table \ref{table:vb-epochs} presents the astrometric measurements from this work. The figures in Appendix \ref{app:astrometry} show the measured positions of the binaries as compared to the expectation for a background star. This test shows that 15 of the 18 targets are co-moving. For J0332-5139 this test is inconclusive, and we reject the remaining 2 visual companions, J0102-6235 and J0903-6348, as likely background objects (also see Section \ref{subsec:indiv-targs}).

\iffalse
% --- Table on Multiepoch Targets ---
\begin{table}[htb]
\begin{center}
\caption{Summary of VBs with Multiepoch Imaging}
\label{table:multiepoch}
\begin{tabular}{ccc|cc}
\hline\hline
Target & Co-Moving? & Orbital & Epochs & Notes  \\
Abbrv. Name &  &  &  Motion & \\
\hline\hline
J0102-6235 & inconcl. & - & 14-12-01 & -\\
  &  &  & 15-11-26 & 1 \\ 
\hline
J0717-6311 & inconcl. & - & 14-04-21 & -\\
  &  &  & 15-11-26 & 2 \\
\hline
J0754-6320 & yes & no & 14-04-21 & -\\
  &  &  & 15-11-26 & - \\
\hline
J0809-5652 &  yes & no & 14-12-01 & -\\
  &  &  & 15-11-26 & -\\
\hline
J0903-6348 & inconcl. & - & 14-04-21 & 3 \\
  &  &  & 14-11-30 & - \\
  &  &  & 15-05-10 & - \\
  &  &  & 15-11-26 & - \\
\hline
J0931-5314 & inconcl. & - & 15-05-12 &- \\
  &  &  & 15-11-26 & 1 \\
\hline
J2108-4244 & yes & yes & 14-11-11 & - \\
  &  &  & 15-11-27 & -\\
\hline
J2244-5413 & yes & yes & 14-12-01 & -\\
  &  &  & 15-11-26 & -\\
\hline
J2247-6920 & yes & yes & 14-12-01 & 2 \\
  &  &  & 15-05-11 & -\\
  &  &  & 15-11-26 & -\\
\hline\hline
\end{tabular}
\end{center}
Notes: \\
1. astrometric error in this epoch too large \\
2. this epoch does not resolve binary \\
3. 3/4 measurements are consistent with background parallactic movement (see Figure \ref{J0931-5314-ast} in Appendix \ref{app:astrometry}), most likely unbound \\
\end{table}
% -----------------------------------
\fi

For all of the binaries, we also checked finder charts from DSS, 2MASS \citep{skr06}, and WISE \citep{wri10} using NASA's IRSA database\footnote{http://irsa.ipac.caltech.edu}, but find no evidence that any of the binaries are caused by chance alignment. 
%DSS: https://archive.stsci.edu/dss/acknowledging.html

Finally, given the spatial distribution of stars in the galaxy, we can compute the chance alignment probability of a given pair of sources with the measured angular separation and magnitudes. We use the TRILEGAL stellar population synthesis code to calculate the density of stars in the 0.5 deg$^2$ field around each VB candidate. We compute $P_{\rm{bound}}$ for each binary following \citet[][see Section 3.1.2 therein]{duc01}. We find that all binaries have essentially 100\% bound probability. This is due to their small angular separations, relatively large flux ratios, and distributions in the sky being generally out of the galactic plane. A summary of the chance alignment probabilities (that is, $P_{\rm{unbound}} = 100\% - P_{\rm{bound}}$) is recorded in Table \ref{table:vbs}.

\bigskip
%----------------------------------------------------------------
\subsection{Other Binaries}\label{subsec:other-bs}

\begin{figure}[htb]
\center 
\includegraphics[scale=0.55]{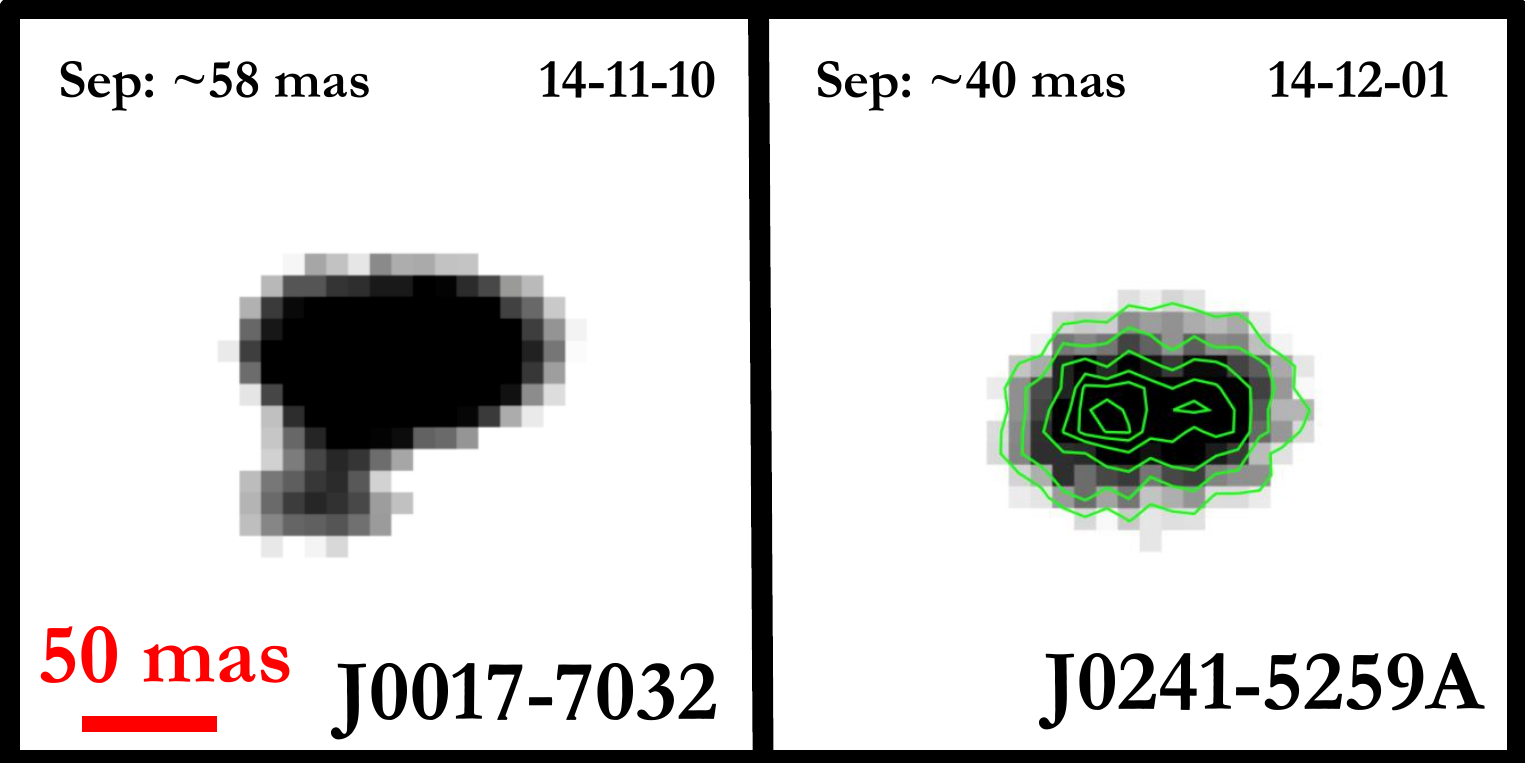}
\caption{Visual binaries resolved through VisAO, band $z'$. The images have been de-roated such that North points up and East points left. The colour scale is arbitrary for display purposes.}
\label{fig:thumbnails-visao}
\end{figure}

Two additional new, close, visual companions were apparent through the VisAO camera but were not resolved in their Clio images. Table \ref{table:vis-vbs} gives their approximate properties and their VisAO images are shown in Figure \ref{fig:thumbnails-visao}. Since only 4\% of the light was sent to the VisAO camera, accurate photometric measurements on these discoveries have not been attempted. Notably, both objects have been flagged as probable binaries by the BANYAN YMG membership analysis from \citet{mal13}.

\begin{table*}[htb]
%\begin{sidewaystable}
%\vspace{5cm}
\begin{center}
\caption{System Properties of VisAO Visual Binaries}
\label{table:vis-vbs}
\begin{tabular}{lcccccccc}
\hline\hline
Target & SpT & Dist. & Epoch & Equal- & $M_{\star}\textsuperscript{a}$ & $M_p\textsuperscript{b}$ & Ang. Sep. & Proj. Phys. \\
Name &  & (pc) & & Mass?\textsuperscript{c} & ($M_\sun$) & ($M_\sun$) & (mas) & Sep. (AU) \\
\hline\hline
      J0017-7032&      M0.5&    63 $\pm$  4&  14-11-10& no &  0.8&  0.8 & 58 & 3.7 \\
     J0241-5259A&      K6.0&    43 $\pm$  1&  14-12-01& yes &  0.9&  0.7& 40 & 1.7 \\
\hline
\end{tabular}
\end{center}
Notes: \\
\textsuperscript{a} Mass derived from single-passband photometry, age, distance, and pre-MS isochrones \citep{bar15}, assuming the star is single. \\
\textsuperscript{b} Same as `a' but now taking into account whether the star appears to be an equal-mass binary. \\
\textsuperscript{c} see Section \ref{subsec:physpars} for explanation. \\
\\
\end{table*}
%\end{sidewaystable}

% --- Figure showing VisAO VB Images ---
% --------------------------------------

Several of the targets are also previously known spectroscopic binaries. These are listed in Table \ref{table:SBs}. 

% --- Table on SBs ---
\begin{table}[htb]
\begin{center}
\caption{List of Spectroscopic Binaries in Our Sample}
\label{table:SBs}
\begin{tabular}{clclcc}
\hline\hline
Target & SpT & $M_{\rm prim}$ & $M_{\rm sec}$ & SB &  Ref. \\ 
Name &  & ($M_\sun$) & ($M_\sun$) & Type  &  \\
\hline
%J0102-6235 & M2.9 & 0.26 & 0.22 & SB3 & 1 \\
%  &  &  &  0.1 & " & "\\
J0222-6022 & M4.0 & 0.47 & - & SB1  & 5 \\
J0451-4647 & M0.0 & 0.59 & 0.48 & SB2 & 1 \\
J0533-5117 & K7.0 & 0.70 & 0.53 & SB2 & 1 \\
J0818-7239 & M0.0 & 0.42 & 0.42 & SB2 & 2 \\
J0846-7246 & K7.0 & 0.61 & 0.61 & SB2 & 2 \\
J1231-4558 & M3.0 & 0.39 & 0.39 & SB2 & 3,4 \\
J1425-4113 & M2.5 & 0.73 & 0.73 & SB2 & 2 \\
J1524-4929 & M2.0 & 0.54 & - & SB1 & 2 \\
J2247-6920 & K6.0 & 0.68 & - & SB1 & 2 \\
\hline\hline
\end{tabular}
\end{center}
Notes: \\
For detailed explanation of the calculation of $M_{\rm{prim}}$ and $M_{\rm{sec}}$, see Sections \ref{subsec:indiv-targs} and \ref{subsec:physpars}.
\\
\\
References: \\
1. \citealt{kra14}; 2. \citealt{mal14a}; 3. \citealt{jay06}; 4. \citealt{ell14}; 5. This work: we infer this target is an SB1 based on the fact that, following independent RV measurements \citet{kra14} and \citet{mal14a} come to different conclusions regarding this object's membership in TucHor. 
\end{table}

% --------------------------------------

\bigskip
%----------------------------------------------------------------

\subsection{Individual Targets of Note}\label{subsec:indiv-targs}

\begin{figure}[htb]
\center 
\includegraphics[scale=0.4]{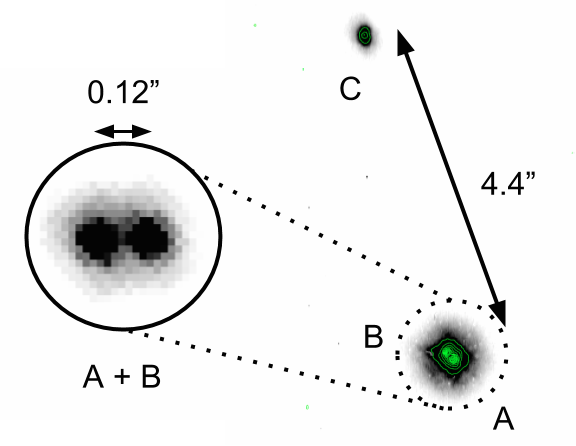}
\caption{The triple system J0030-6236, with its tertiary companion at 4.4''. The primary in itself a tight 0.12'' visual binary.}
\label{J0030-6236}
\end{figure}

\emph{J0030-6236 -- } (see Figure \ref{J0030-6236}) This object consists of a tight binary (separation $\sim 0.12'' \sim 5.3$ AU, $\Delta$H $\sim 0.14$) and a fainter wide visual tertiary (separation $\sim 4.45 '' \sim 200$ AU, $\Delta$H $\sim 2.0$ relative to the inner binary). The inferred masses of the components are 0.54, 0.49, and 0.23 $M_\sun$. Note that the tertiary is outside the $4''$ limit we set for our statistical analysis in Section \ref{sec:multi-stats}, but by virtue of its inner binary, this system is counted as a multiple.

\emph{J0102-6235 -- } \citet{kra14} describe this object to be a member of TucHor and an SB3. We observe this target to be a 0\farcs5 visual double. Our co-moving analysis suggests that the fainter component is a background object (see Table \ref{table:vb-epochs} and Figure \ref{fig:ast1} in Appendix \ref{app:astrometry}). There is insufficient information to determine which of the visual components is the tight binary. Future spectroscopic observations could resolve this issue. In the meantime, we exclude this target from our statistical analysis. 

\emph{J0222-6022 -- } \citet{kra14} find that their single-epoch RV measurement is not consistent with being a member of TucHor, unless it is an SB1. In contrast, \citet{mal14a} deem it a very secure TucHor member based on an independent RV measurement. From these discrepant classifications, we posit that this system is in fact a single-lined spectroscopic binary and treat it as such in the analysis. We do not detect a visual companion in this system, but most SBs are below our detection limits.  

\emph{J0241-5259AB -- } We observe both targets to be tight binaries. This updates J12's multiplicity designation for the `B' component, which they observed to be single.  The nominal A and B components are $\sim 21''$ apart. At their common distance of 43 pc they may be part of the same system separated by 1000 AU. Moreover, since each component is a tight binary in itself (this work), the system is potentially a hierarchical quadruple.   

\emph{J0511-4903 --} J16 tentatively suggest the existence of a very faint, $\Delta z' = 6.1$ companion at $2\farcs4$ whose mass would be $0.015 M_\sun$. According to the Baraffe isochrones corresponding to the age of Columba, for a low-mass companion of this $\Delta z'$, the expected $\Delta H$ is $\sim4.5$. We do not observe such a companion in our short exposures, but we would not expect to. At this separation, these short Clio exposures are sensitive to $\Delta H \sim 4.3$ at $15\sigma$ and $\Delta H \sim 4.8$ at $10\sigma$.      

\emph{J0903-6348 -- } Co-moving analysis shows that the $\sim1\farcs1$ `companion' detected in Clio is a background object (see Table \ref{table:vb-epochs}). J16 also reports a detection of this companion, but their $z'$-band measurement indicates the spectral energy distribution (SED) rises towards the optical bandpasses, consistent with the hypothesis that it is a background object. In addition, there exists a wide (7\farcs9) companion \citep{mal14a, jan16}. Regardless, we include this system as a single-star system in our statistical analysis. 

\emph{J1234-4538 -- } This visual binary was first imaged by J12 in the $i'$ and $z'$ bands. Our additional epoch gives us a $\sim5$-year baseline in total, providing astrometric confirmation (see Table \ref{table:vb-epochs} and Figure \ref{fig:ast2}). Based on NIR fluxes ($\Delta H=0.03$, Section \ref{subsec:physpars}), we determine primary and secondary masses to be roughly equal-mass at 0.49 and 0.48 $M_\sun$, respectively. These are discrepant from the masses estimated by J12 (0.355 and 0.290 $M_\sun$), which are based partly on optical flux ratios $\Delta i' = 0.48$ and $\Delta z' = 0.51$. While the nature of this discrepancy is unclear, it may be because 1) J12 derived their masses from spectral types using relations calibrated for older field stars, and that the spectral type they use (M2.5) is different than that measured by \citet{shk11} (M1.8) and 2) there exists systematic differences between models at this age, mass, and wavelength range. We opt to use our own inferred masses for consistency with the rest of our sample.  

\emph{J1657-5343 -- } We resolve this target into a $\sim51$ mas binary (see Figure \ref{fig:close-vbs}). J16 claims a possible detection of two sub-stellar companions at $3.2''$ and $3.5''$ with $\Delta z' = 5.6$ and $6.59$, respectively, but flags them as probable background objects. Their inferred masses are 0.03 and 0.02 $M_\sun$. There is no evidence of these objects in Clio images, whose sensitivity at these separations is $\Delta H = 7.5$, supporting the hypothesis that these are blue background contaminants.   

\emph{J2149-6413 -- } This object appears single in both the Clio and VisAO images taken on the night of 2015-05-11, for which the VisAO corrections are relatively poor. The minor axis of the half-width-at-half-maximum surface has width 64 mas and the axis ratio is 0.89, meaning the single-source PSF is rather round. Attempted 2-source fits did not find robust solutions. \citet{cha10} find this is a nearly equal-mass ($\Delta$K = 0.2) binary with separation 74 mas in 2005. As in Sections \ref{subsec:clio-vbs} and \ref{subsec:physpars}, we use the $\Delta$K measured from \citet{cha10} to calculate the primary mass to be 0.18 and the secondary mass 0.16 $M_\sun$. At a distance of 44 pc the projected separation corresponds to 3.3 AU and the period of this target assuming equal-mass components ($M_{tot} = 0.18 + 0.16 = 0.34 M_\sun$) would be $\sim10$ years. We include this target in our calculations of multiplicity rates based on all known binaries, but not in the visual binary rate and the separation range and mass ratio distributions which are based on our Clio detections only (see Section \ref{sec:multi-stats}).

\emph{J2247-6920 -- } We detect a faint visual companion at $0.25''\sim12$ AU with Clio. J16 observed the same target but designated it as a single star. This object is also an SB1 and the RV amplitude of $>27$ km/s suggests a tight binary \citep{mal14a}, making the observed companion a tertiary. This object was originally noted by \citet{mal13} to be a possible field interloper. However, it is unclear whether the SB1 is corrupting the RV measurement. With a new parallax from Gaia, J16 re-analyzes this system using BANYAN and assigns it to the field. However, the Gaia distance appears to be consistent with the statistical distance prediction from \citet{mal13} assuming it is a TucHor member. We opt to keep J2247 in our sample.

%----------------------------------------------------------------

\subsection{Comparison to BANYAN}

The Bayesian Analysis for Nearby Young AssociatioNs (BANYAN, \citealt{mal13,gag14}) is an algorithm that assigns to stars a statistical probability of belonging to a given nearby young moving group. It considers the hypothesis that a given object may be an unresolved equal-luminosity binary. If the binary solution gives a higher membership probability, then the object is flagged to reflect its possible binary nature. \citet{mal14a} suggests that follow-up observations could be used to verify the BANYAN methodology for identifying binaries.

Sixty-four of the targets in our sample were drawn from \citet{mal13,mal14a} and therefore have a BANYAN assessment of binarity. We observe 17 visual binaries with Clio and 2 additional binaries with VisAO. Of these, 5 and 2, respectively, were flagged as possible binaries by BANYAN. BANYAN flags a further 10 systems as possible binaries. We do not observe any of these to be VBs. However, one system (J2149-6413, see Section \ref{subsec:indiv-targs}) was identified by \citet{cha10} to be a binary and three more are confirmed SBs (J1231-4558: SB2; J1425-4113: SB2; J1524-4929: SB1; refer to Table \ref{table:SBs}). Of the remaining BANYAN objects in our sample which are not flagged as binaries, 4 are detected to be SB2's (J0533-5117 and J0451-4647 \citep{kra14}, and J0818-7239 and J0846-7246 \citep{mal14a}). J2247-6920 is not only an unequal-mass visual binary (this work) but also an SB1 \citep[][see Section \ref{subsec:indiv-targs}]{mal14a}. In addition, we infer J0222-6022 to be an SB1 (see Section \ref{subsec:indiv-targs}).

While we expect the BANYAN algorithm to miss very high mass-ratio pairs, it is unclear why it failed to identify nearly equal-mass binaries (e.g.J1516-5855, J0717-6311). It would be worthwhile to reassess the YMG membership probabilities for the new binaries presented in this work. 
% =============================================

%==============================================
\bigskip
\section{Multiplicity Statistics}\label{sec:multi-stats}

We compute multiplicity statistics within our sample. For each statistic, we will perform the analysis on two binary sub-samples below 4\farcs0: 1) all known binaries (including SBs) and 2) visual binaries robustly fit by a 2-source model in the Clio images.

%----------------------------------------------------------------
\subsection{Contrast Curves \& Observational Completeness Maps}\label{subsec:cc}

In order to study population statistics, we must first compute the completeness corrections by computing contrast curves for our targets. 
%A contrast curve quantifies our ability to detect a binary companion as a function of flux ratio and angular separation for a particular target. The contrast curves may be used in a forward model to correct a raw observed population statistic, such as distributions of separations or mass ratios.  

We calculate azimuthally-averaged contrast curves using the formalism in \citet{bra06} (see Eqn 1 therein): 
\begin{equation}
\Sigma(R) = -2.5 \log\left[\frac{n\sigma_{\rm pix}(R)\sqrt{N_{\rm PSF}}}{F_{\rm PSF}}\right] \textrm{mag.}
\label{eqn:cc-bra06}
\end{equation}
Here $R$ denotes distance from the primary star's centroid and $\Sigma(R)$ denotes the limiting contrast for the secondary to be detected in terms of magnitudes. $n$ is the threshold SNR determined to serve as a good threshold signal-to-noise corresponding to the detection method. $\sigma_{\rm pix}(R)$ is the RMS in an annulus of $R-\delta R$ to $R+\delta R$ around the primary. $N_{\rm PSF}$ represents the number of pixels within the FWHM of the primary PSF and $F_{\rm PSF}$ is the integrated flux enclosed in the pixels. In the narrow mode, the Clio detector is $8''\times16''$, so we are generally sensitive to companions occurring below $8/2 = 4''$ in all directions. For simplicity, we calculate the contrast curves to $4''$ and use that as our cutoff for companion detections.

For each processed image we use the best-fit triple bivariate Gaussian PSF solution for the primary to numerically compute $F_{\rm PSF}$ and $N_{\rm PSF}$. If there is a companion, we subtract the PSF model of the companion before computing the contrast curve. 

The chosen contrast threshold should result in reliable identification of true stellar companions. The most prevalent false sources are in the form of ghosts and speckles. 

\begin{figure}
\vspace{0.5cm}
\includegraphics[scale=0.52]{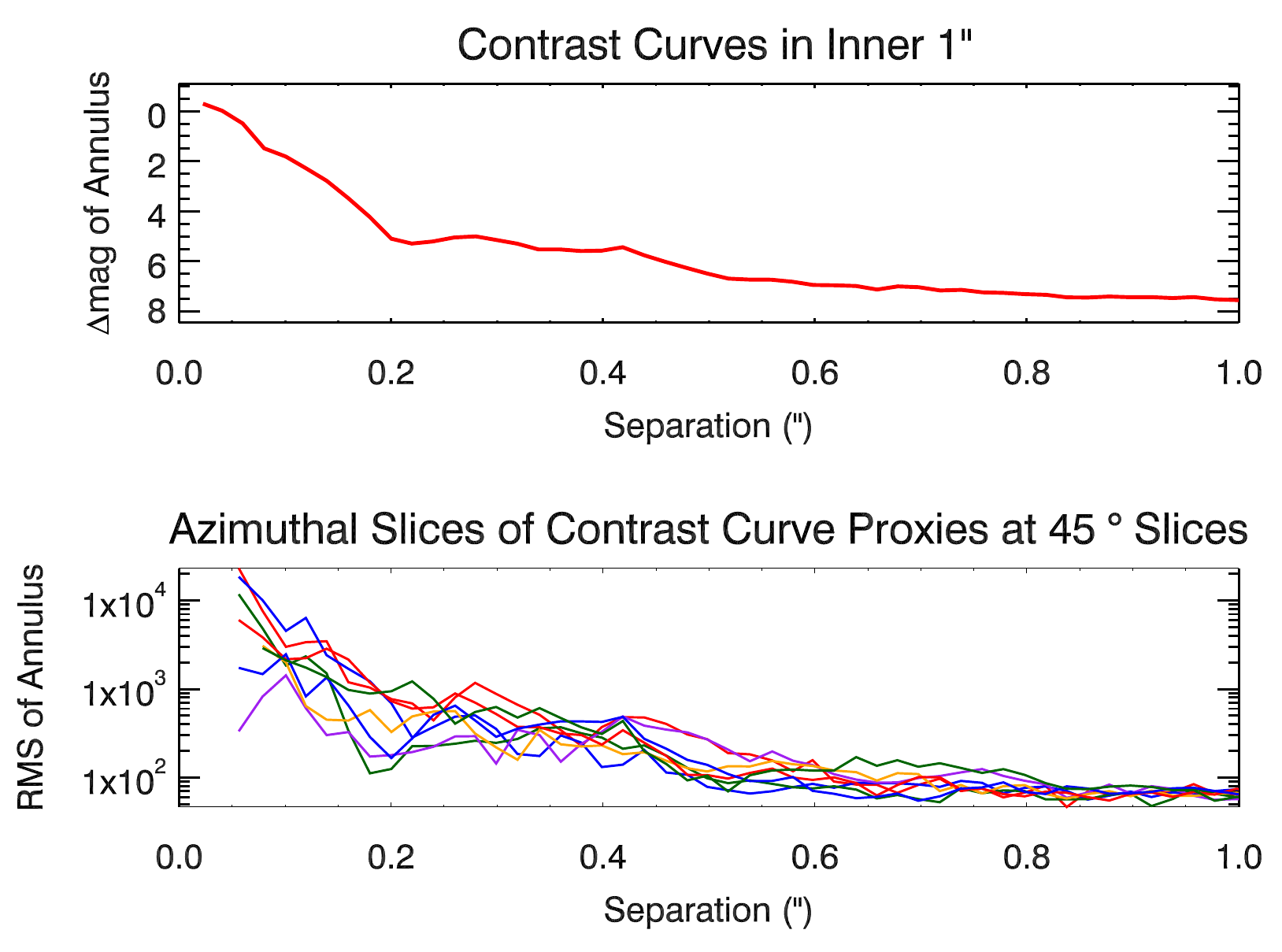}
\caption{15-$\sigma$ contrast curves in the inner $1\farcs$ for J0254-5108, a single star, is typical of that for our sample. Top Panel: 15-$\sigma$ contrast curve. Bottom Panel: Annulus RMS curve in 45-degree slices. }
\label{fig:j0254-5108-cc}
\end{figure}

To determine the appropriate threshold, for every target we constructed residual images by subtracting the empirical, azimuthally-averaged flux profile around the primary centroid. We searched for candidate sources that fell above the local annular RMS in the residual image at $n$-$\sigma$ as follows: First, we isolated each pixel that meets this criteria. We treated each pixel as if it could be the centroid of a source scaled from the fitted PSF of the primary, hence taking the pixel flux to be the peak flux of such a `source'. We demand that, within a square with side length equal to the PSF half-maximum diameter around this pixel,  70\% of all pixels are at least 50\% as bright as the target pixel. For pixels that meet the conditions thus far, we cross-match them to within 3 pixels of the same position with respect to the primary in the other images for a given target and epoch. The last step is very effective in screening out bright ghosts associated with the instrument whose location depends on the dithering position. Inspection reveals that speckles sometimes exceed $\sim 10\sigma$, but are virtually nonexistent at $\sim 15\sigma$. Therfore, we choose $n = 15$ for the contrast curves. Figure \ref{fig:j0254-5108-cc} shows a representative contrast curve constructed for a typical image. The lower panel shows that the dependence on azimuthal angle is weak. 

% ---------- CC example ------------
% -----------------------------------

\citet{maw14} points out that speckles increase the noise level above the naive Gaussian assumption for 2-3 $\lambda/D$. In contrast, Eqn. (\ref{eqn:cc-bra06}) assumes Gaussian noise and so our computation of the contrast curves does not explicitly take this effect into account. At the same time, we do not expect this to have a substantial effect on our results because we choose our detection threshold based on the level of the speckle noise.

For each target, we averaged all the contrast curves for each epoch. We chose the best among the epochs for our detectability proxy. Here `best' is defined as the curve with the most sensitivity between $0\farcs1$ and $1\farcs0$ from the primary. %\HLb{A compilation of these contrasts is tabulated in Table \ref{table:cc} under Appendix \ref{app:cc}.}

From these contrast curves we construct an observational completeness map for our sample. At any given $\Delta$mag and log(separation), we use the contrast curves to calculate the fraction of stars in the sample for which a companion with these properties would be detectable. The completeness map in these observed parameters is displayed in the left panel of Figure \ref{fig:smap}.

\bigskip
% -------------------------------------------------
\subsection{Calculation of Physical Parameters}\label{subsec:physpars}

Physical projected separations are calculated using the reported distances to each target (Table \ref{table:sample-prop1}). The primary mass is derived from the 2MASS photometry in conjunction with the latest stellar isochrones for PMS stars \citep{bar15}. If the star is a detected VB, then we divide the 2MASS magnitude between the components according to the measured flux ratio (see also Section \ref{subsec:clio-vbs}). For the 2 binaries detected in VisAO and not Clio, there is not enough throughput for a reliable extrapolation from the observed visual flux ratio to the NIR flux ratio, thus we used the following approximation: if we judge the flux ratio to be large, we assigned all of the 2MASS flux to the primary; otherwise, we split the light equally between the two components, equivalent to obtaining the mass of each component from the system magnitude revised upwards by 0.75 (see Table \ref{table:vis-vbs}). If the target is a known SB (see Section \ref{subsec:other-bs}), we take that information into account when calculating the mass. For SB1s, we assume the primary is the dominant component in flux and mass and neglect any contribution from the secondary. For SB2s, unless an explicit measurement of flux and/or mass ratio is available, we assume an equal-mass binary. For J2149-6413 and the triple system J0030-6236, see Section \ref{subsec:indiv-targs}. 

For most targets, the dominant source of mass error comes from uncertainties in the distance, resulting in an error of $\sim 0.02 M_\sun$ on average. For comparison, typical uncertainties on 2MASS photometry and measured $\Delta$ mag ($\sim0.03$) correspond to an error of $\sim 0.005 M_\sun$. The youngest targets ($< 30$ Myr) may be subjected to greater mass error due to uncertainty in age, since luminosity evolves most steeply at lower ages. For TWA ($10\pm3$ Myr) and $\beta$ Pictoris ($24\pm3$ Myr), the typical mass error from age indeterminacy can be up to $\sim0.08 M_\sun$ and $\sim0.03 M_\sun$, respectively. Inherent stellar model uncertainties can introduce greater errors and are briefly discussed in Section \ref{subsubsec:model-uncertainties}. We also note that Baraffe models are computed for solar abundances. Since young stars tend to be more metal-rich on average, any metallicity-dependence of the isochrones could introduce a systematic error in the inferred mass.   

We then transform the observational completeness map into one in mass ratio and projected physical separation. To calculate the binary separations we again use the reported distances to each target (Table \ref{table:sample-prop1}). We convert each target's best contrast curve into one in the space of mass ratio vs projected physical separation. In the case of known unresolved binaries (J2149-6413, VisAO binaries, and SBs), we calculate the mass ratio relative to the sum of the masses of the binary components (see Tables \ref{table:vis-vbs} and \ref{table:SBs}). Our outer detection limit of 4" set by the detector's field of view translates into a variable limit in projected physical separation. The resulting map is shown in the right panel of Figure \ref{fig:smap}. Our completeness to mass ratios greater than 0.2 between 10 and 100 AU is $\sim90\%$. The 100\% detectability contour for the sub-sample of primaries with masses less than $0.5 M_\Sun$ is overlaid in black. In general, less massive primaries are fainter and allow for greater sensitivity to closer-in and lower-mass companions.     

Based on a lack of $q < 1$ binary discoveries beyond 20 AU in their sample of young M-dwarfs, \citet{dae07} suggest that, for M-dwarf primaries, there may be a desert of low-mass ratio companion occurrence beyond 25 AU. In this work we have found 10 companions to M-type stars beyond 20 AU with mass ratios between $\sim 0.5$ and 0.99, hence it is unclear whether such a desert really exists. 

\bigskip
%----------------------------------------------------------------
\subsection{Integrated Multiplicity Fractions}\label{subsec:mf1}

Out of the total sample of 104 imaged systems (after discarding J0102-6235, see Section \ref{subsec:indiv-targs}), 25 are found to have a NIR visual companion whose projected separation lies between $\sim 0\farcs05$ and $4\farcs0$ (Table \ref{table:vbs}). The widest companion in this angular separation range lies at $\sim300$ AU. At least 1 of these systems also harbour SB signatures (J2247-6920). VisAO resolved 2 more binaries (Table \ref{table:vis-vbs}). From the literature, a further 9 systems have known RV or close visual companions (Tables \ref{table:SBs}, J2149-6413). At least 2 of these systems may be triples (J0030-6236 and J2247-6920), and 1 might be a $\sim1000$ AU hierarchical quadruple. Counting the quadruple as two separate systems, the resulting measurement of the raw multiplicity fraction (MF $\equiv N_{\rm multiple}/N_{\rm{all}}$) for our sample is $36/104 = 35^{+5}_{-4}\%$ for all separations below 300 AU, and $25/104 = 24^{+5}_{-4}\%$ for Clio VBs only. 

When referring to M-dwarfs for pre-main sequence stars we are faced with a dilemma: should we designate objects as `M-dwarfs' on account of spectral type or stellar mass? On the main sequence, the two are tightly correlated, hence can be used interchangeably (e.g. a M0 star has mass $\sim 0.6 M_\sun$ \citealt{bc96}). However, the correspondence is sensitive to age for pre-main sequence stars. While its mass stays relatively constant, a young star's observable spectroscopic parameters change with time during the PMS phase as it contracts towards the main sequence. Stellar mass is a more fundamental and physically meaningful parameter that is also relatively invariant with age. However, characterizing a system by spectral type is attractive because it is a direct observable and less sensitive to any binarity information to which we are ignorant. Hereafter, for any measurements made for the `M-dwarf sub-sample', we will specify whether they refer to systems of the M-spectral type or with $M_{\rm{prim}} \leq  0.6 M_\sun$, whichever is more appropriate.

The sub-sample of M-dwarfs (by spectral type) exhibits a raw multiplicity rate of $26/87 = 30\%^{+5}_{-4}$ below 300 AU and $18/87 = 21\%^{+5}_{-4}$ for visual companions identified in Clio imaging. For the sample of systems for which $M_{\rm{prim}} \leq 0.6 M_\sun$, the rates are $23/64 = 36\pm6\%$ and $17/64 = 27\%^{+6}_{-5}$, respectively.

To account for our detection incompleteness to binaries of particular mass ratios and physical separations, we compute the correction factor for each Clio VB detection as the inverse-detectability evaluated at its location in the \emph{physical} completeness map. After applying these corrections, the Clio VB rate for the total sample is $34\pm5\%$. For the M-type sub-sample the VB multiplicity with this correction becomes $27\pm5\%$. For the $M_{\rm{prim}} \leq  0.6 M_\sun$ sample, the revised VB rate after accounting for detectability is $36\pm6\%$. When all binaries are counted, the fractions are $42\pm5\%$ for the entire Clio sample, $34\pm\%$ for M-stars, and $44\pm6\%$ for $M_\star \leq  0.6 M_\sun$. Note that in this calculation we have explicitly excluded those VBs in regions with detectability $>0\%$ that we resolved by VisAO (J0017-7032 and J0241-5259A) or noted in previous works (J2149-6413) but were missed by Clio, as they should have been accounted for in the detectability corrections.

\bigskip
% --------------------------------------
\subsection{Distribution in Projected Physical Separation and Mass Ratio}\label{subsec:physep&q}

% --- Figure showing Separation Distribution & Lognormal Fit and q-Distribution & Power-law Fit ---
\begin{figure*}[htb] 
\begin{minipage}{0.5\textwidth}
\centering
%\vspace{1cm}
\includegraphics[scale=0.4]{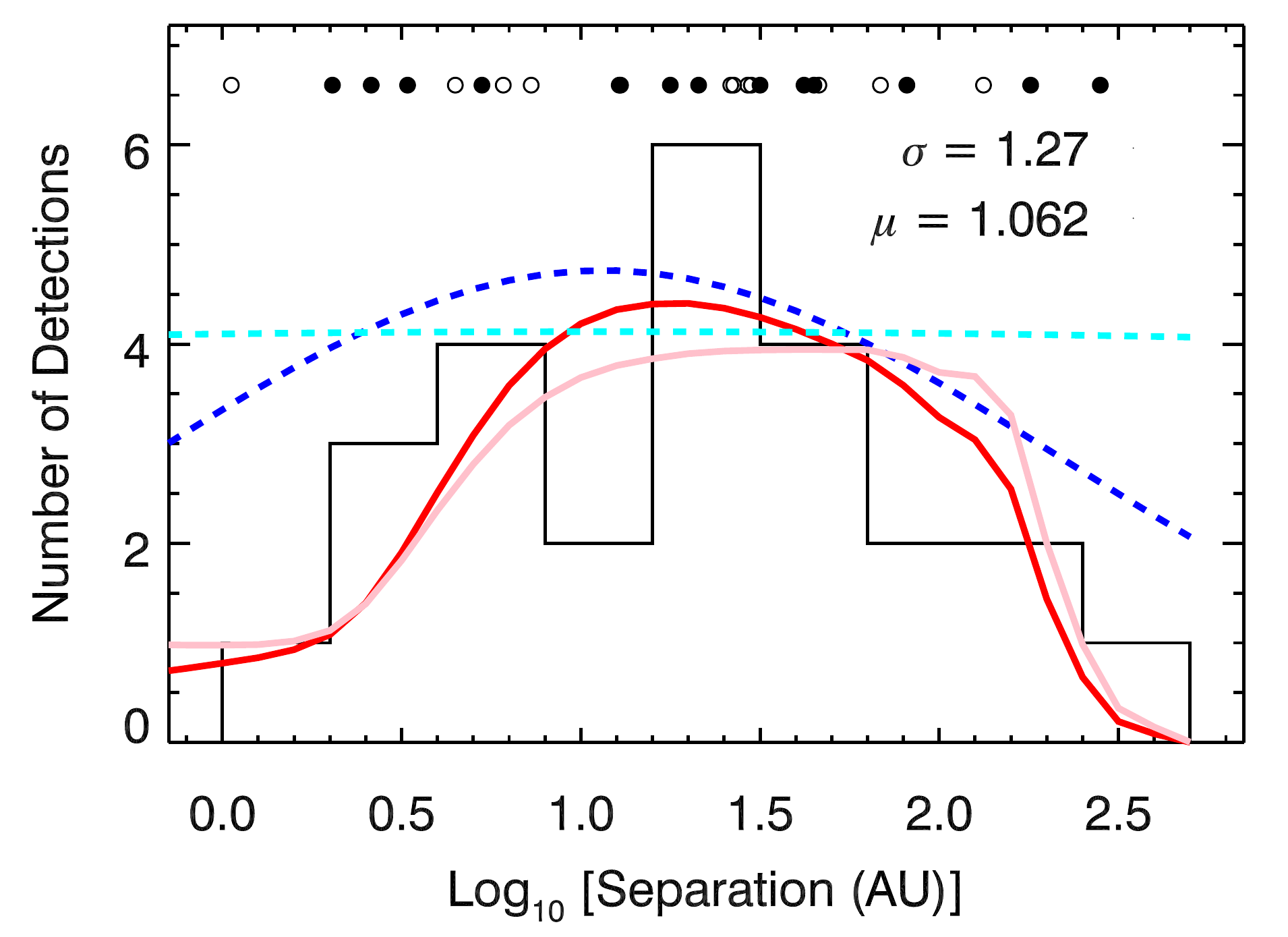}
\end{minipage}%
\begin{minipage}{0.5\textwidth}
\centering
%\vspace{-0.2cm}
\includegraphics[scale=0.4]{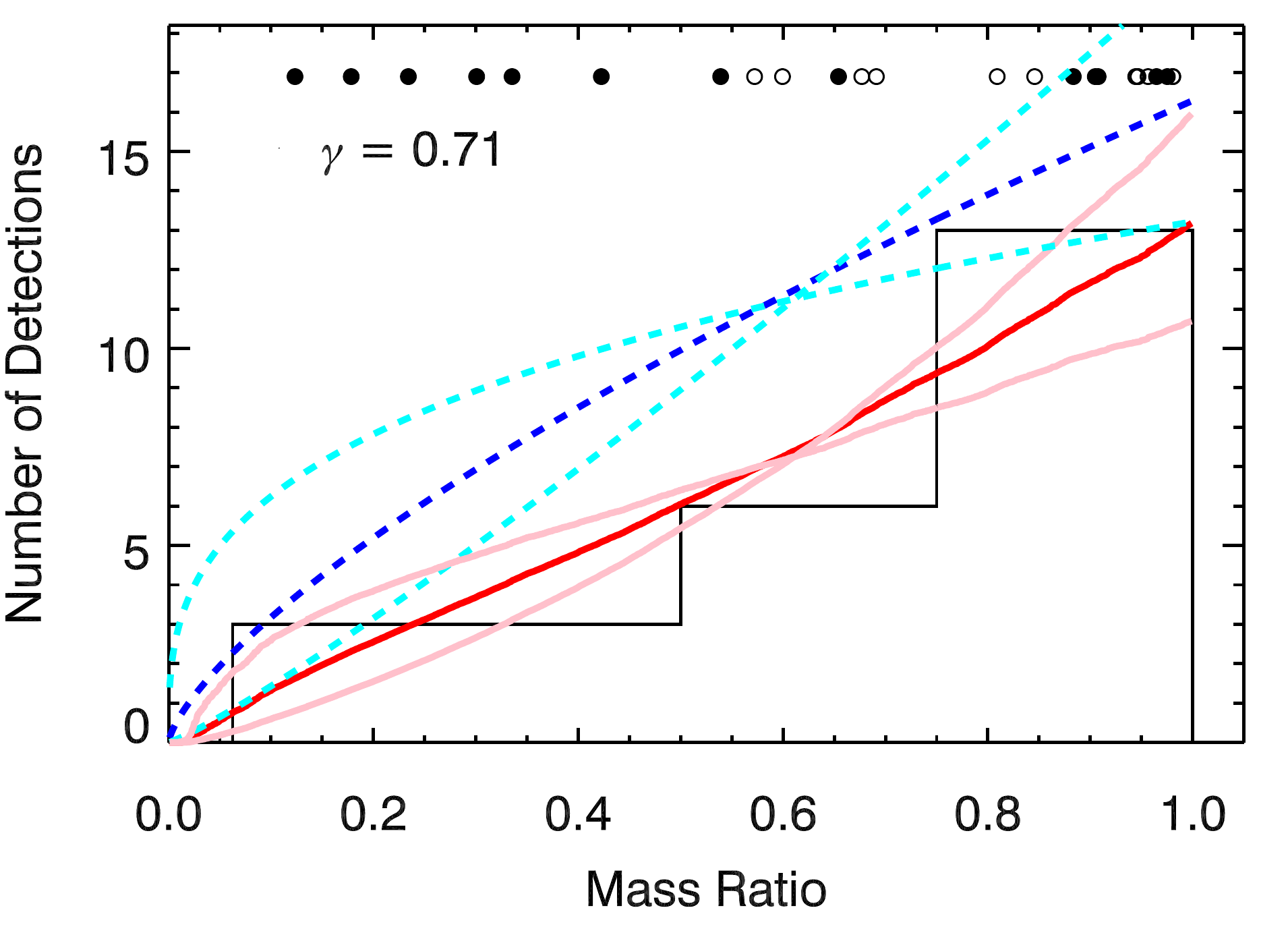}
\end{minipage}
%\vspace{1cm}
\caption{Left Panel: The distribution of projected physical separations for the Clio VBs. Black points at the top of the plot denote positions of individual detections, closed and open points denote those for which $M_{\rm prim} > 0.6 $ and $ < 0.6 M_\sun$, respectively. The Blue broken line is the underlying log-normal distribution fit to the detections, whereas the red solid line represents that of the incompleteness discounted (i.e. observed) distribution. Shown in Cyan is a log-normal distribution with width $\sigma = 10)$, i.e. virtually flat. The pink curve is the predicted observation. The two scenarios are virtually indistinguishable. Right Panel: The distribution of mass ratios for the Clio VBs. A power law of form $q^\gamma$ is fit to the detections with the best fit functions shown in blue dashed lines and its completeness corrected version in solid red. In cyan are dashed lines representing the solutions whose likelihood deviates from the best fit by $1\sigma$. The pink lines are corrected for detectability effects.  
}
\label{fig:sepqdistr}
\end{figure*}
% --------------------------------------

The upper left panel of Figure \ref{fig:sepqdistr} shows the distribution of all the Clio visual binary discoveries between $0\farcs05 - 4\farcs0$ in log projected physical separation. We attempted to fit a log-normal function with mean $\mu$ and intrinsic width $\sigma$ to the detections, modified by the empirical completeness map from the right panel of Figure \ref{fig:smap}. Using a maximum likelihood prescription we find the distribution to be both consistent with being log-normal and flat. Thus, we cannot make a meaningful comparison to the field, where FGKM binaries exhibit separations adequately described by a log-normal \citep[e.g.][]{rag10,jan12, dk13,wd15}.

The right panel of Figure \ref{fig:sepqdistr} shows the mass ratio ($q$) distributions of the Clio visual binaries. We fit a power-law ($q^{\gamma}$), modified by the physical completeness map in Figure \ref{fig:smap}, to each individual detection in a maximum likelihood framework. The best-fit $\gamma$ is $0.7 \pm 0.4$, hence inconsistent with a flat distribution. It is apparent that there is a strong preference for equal-mass pairings in this sample. To compare, in the field, FGK binaries mass distributions can generally be characterized by $\gamma \sim 0.3 - 0.4$ \citep{dk13}. Of course, the mass for a stellar companion has a lower bound at the hydrogen-burning limit ($0.08 M_\sun$). When considering stellar companions, the lower-bound on the relevant mass ratios increases for decreasing primary mass.

%----------------------------------------------------------------
\subsection{Combined Analysis with SACY}\label{subsec:sacy}

The Search for Associations Containing Young stars (SACY) consortium has been actively studying the properties of stars in nearby YMGs \citep{tor08}. \citet{ell14, ell15} present results from their multiplicity survey of SBs and visual binaries. SACY covers many of the same YMGs as in our work over a comparable separation range but their objects are generally more massive --- predominantly greater than $ M_\sun$. Here we combine our samples (see \citealt{ell15} Tables 1 and 5) and repeat our calculation of multiplicity fractions. This joint analysis commands greater statistical power and over an extended range in parameter space.    

Of the 109 targets in the SACY sample with secure YMG memberships, we discard BD-20 1111 from the joint sample due to ambiguity over its YMG membership status (see Notes under Table 5 in \citealt{ell15}). Of the remaining 108 objects, there are secure detections (i.e. bound probability exceeds 95\%) for 32 visual multiples, 5 of which are triples (3 visual triples, 2 visual binaries + SB), and an additional 5 SBs. In the joint sample analysis, we use measurements from our campaign for the 9 targets that were also observed by SACY (2 VBs, 1 SB) . Our mass, astrometry, and multiplicity measurements are consistent with SACY when accounting for the modest distance discrepancies (up to 5pc) except in the following cases: 

\begin{itemize}
\item{{\emph{V* DZ Cha (J1149-7851):}} SACY assigns this star to $\eta$-Cha at distance 102.7 pc (see also \citealt{luh07,mur13}), deriving a mass of $0.91 M_\sun$, while \citet{mal14a} lists it under $\beta$-Pic at 68 pc, which is used to derive a stellar mass of $0.72 M_\sun$ in this work. This lower stellar mass is more consistent with the spectral type (M0-1V), therefore we use this mass and the \citet{mal14a} YMG designation. Nevertheless, we acknowledge that youth signatures found by \citet{mur13} argue in favour the $\eta$-Cha interpretation. A {\emph{Gaia}} parallax will settle this ambiguity.} 

\item{{\emph{CD-53 544 (J0241-5259A):}} this target is observed to be a 40 mas binary using VisAO imaging (see Table \ref{table:vis-vbs}), which is below the imaging resolution of SACY.} 

\item{{\emph{TYC 8098-414-1 (J0533-5117):}} \citet{kra14} identify this object as an SB2.}  

\item{{\emph{GSC 08350-01924 (J1729-5014)}} \& {\emph{CD-44 1173 (J0331-4359):}} Small differences exists between our astrometry measurements for these objects which can be explained by expected orbital motion over the time span of $\sim$2 - 10 years separating our images from the SACY work. Our primary and secondary mass calculations are consistent with each other. }

\end{itemize}

In total, for the joint sample of 203 targets there are at least 52 visual multiples with projected separations above 75 mas, 6 visual binaries below 75 mas, 13 SBs which have no noted VB counterparts, and 7 triples, all of which also manifested as visual multiples. This gives a combined raw multiplicity fraction of $35\pm3\%$ and a triple fraction of $3\pm1\%$ for MKG-dwarfs in YMGs. The companion fraction (CF $\equiv (N_{\rm binary} + 2N_{\rm triple} + ... )/N_{\rm all}$ is then $38\pm3\%$. The visual binarity rate is $29\pm3\%$.  

% --- Figure showing raw MF as function of Mp ---
\begin{figure}[htb]
\center 
\hspace{-0.5cm}
\includegraphics[scale=0.55]{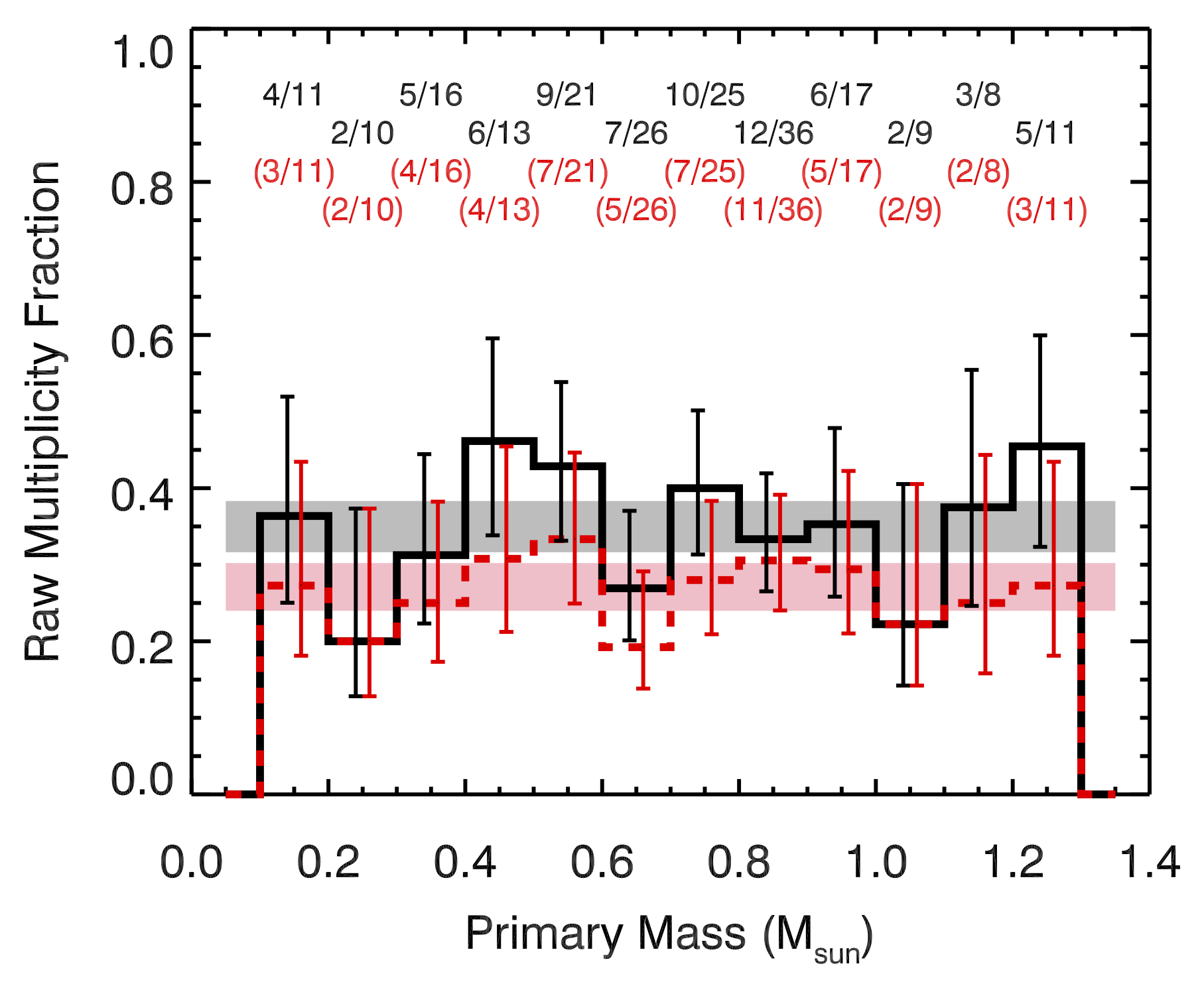}
\caption{The raw multiplicity fraction of our joint YMG sample with SACY. The black solid histogram denotes multiplicity calculated from all known binaries (including VisAO and SBs) over the entire sample whereas the red dashed histogram signifies the visual binary detections with physical projected separations between 1 and 300 AU. $1\sigma$ binomial confidence intervals are shown. Fractions near the top of the plot correspond to number of total known multiple systems out of number of systems observed, and in parentheses the number of VBs (excluding J2149-6413 and VisAO binaries unresolved by Clio). The gray and pink shaded regions depict the average overall and visual multiplicities and their uncertainties across all masses. }
\label{fig:MF-Mp-SACY}
\end{figure}
% --------------------------------------

Figure \ref{fig:MF-Mp-SACY} shows the raw multiplicity fraction of all known multiple systems (black) and that of detected visual multiples (red) for the joint sample of 203 systems spanning a primary mass range of 0.1 - 1.2 $M_\sun$. In mass bins for which each survey contributes comparable numbers of targets (0.6 - 0.8 $M_\sun$), the binarity rates we measure agree. Below  0.6 $M_\sun$ the statistics are dominated by our campaign whereas above 0.8 $M_\sun$ they are chiefly determined by SACY. With the added statistical power in each mass bin and an extended range of masses, we still see no indication of deviation from flatness in the raw multiplicity fraction observed. This finding appears to extend the same conclusion from \citet{ell15} for FGK-dwarfs in YMGs down to $0.2 M_\sun$ (but see below).

\begin{figure*}[htb] 
\begin{minipage}{0.5\textwidth}
\centering
\hspace{-1cm}
\includegraphics[scale=0.55]{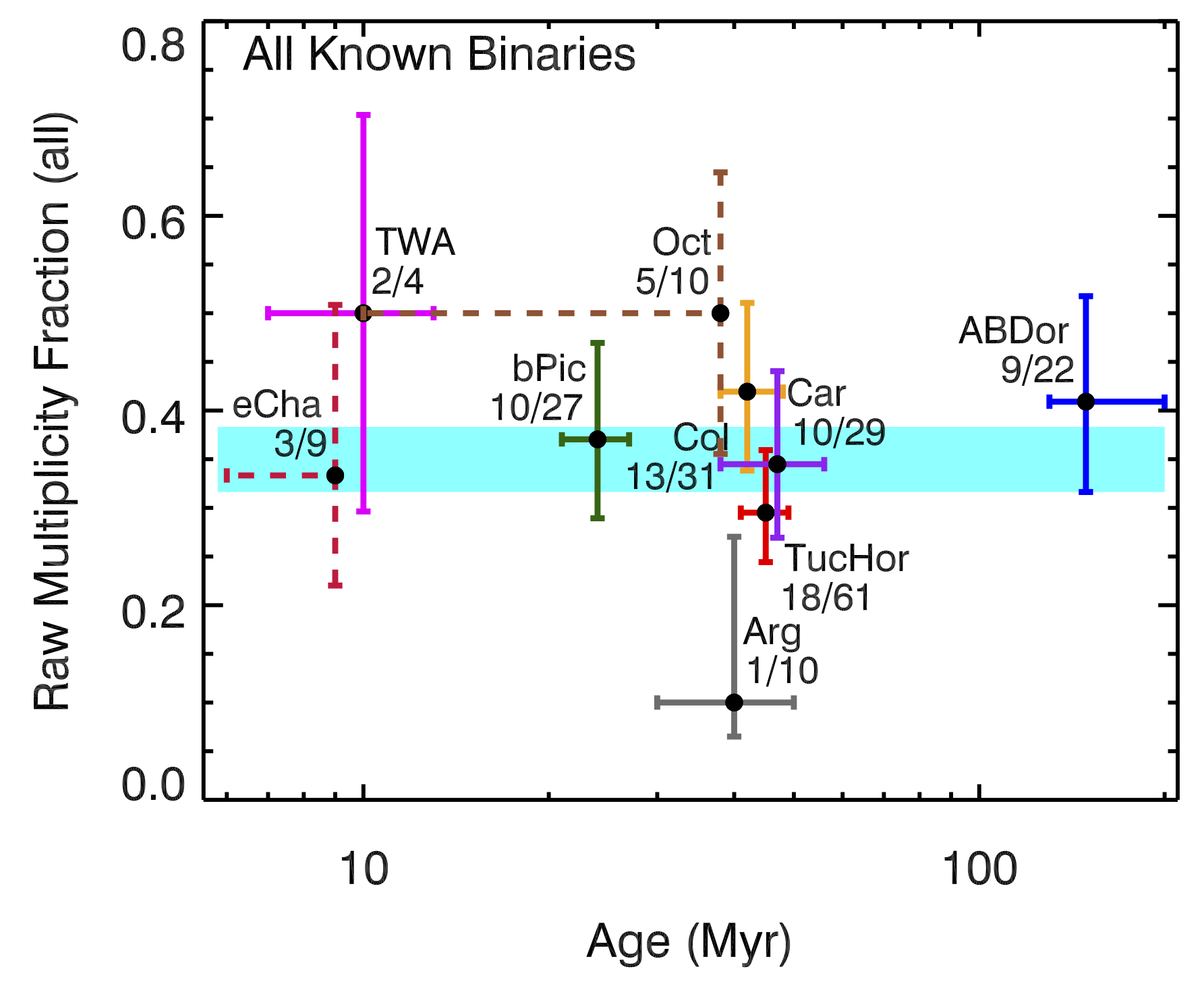}
\end{minipage}%
\begin{minipage}{0.5\textwidth}
\centering
\hspace{-1cm}
\includegraphics[scale=0.55]{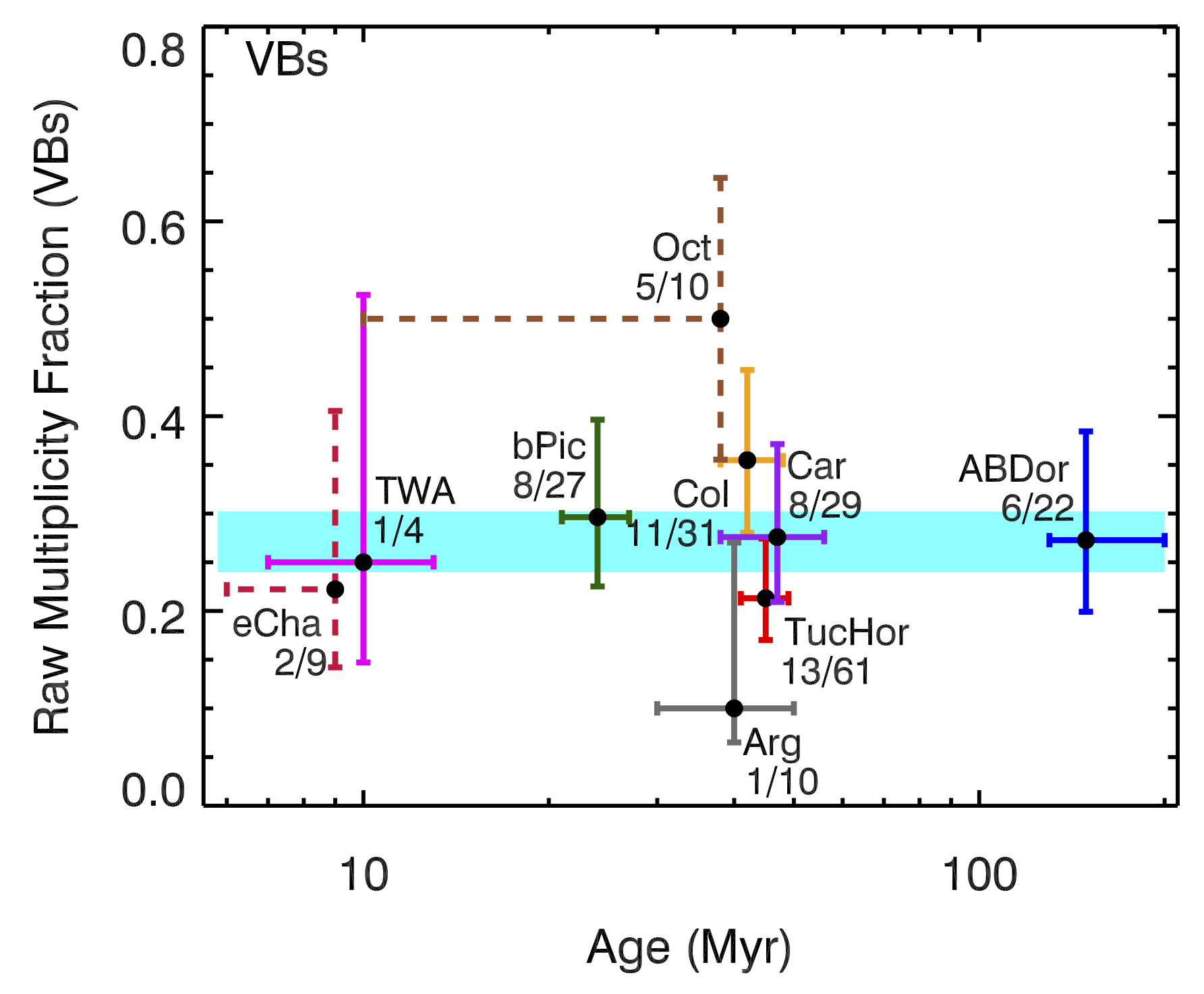}
\end{minipage}
\caption{Left Panel: raw multiplicity fraction counting all known multiple systems by YMG and its age for our joint YMG sample with SACY. Vertical errors are $1\sigma$ binomial confidence intervals. Horizontal errors reflect the uncertainty in group age (see Table \ref{table:ymgs}). Range shaded in cyan indicates the overall multiplicity rate ($35\pm3\%$). Right Panel: same as the left, but only for known VBs between 1 and 300 AU, excluding J2149-6413 and VisAO binaries unresolved by Clio. The overall visual multiplicity rate for VBs in this projected separation range is $28\pm3\%$. On both plots we use dashed error bars to indicate YMGs for which the measurements come entirely from SACY.  }
\label{fig:MF-Age-SACY}
\end{figure*}

Figure \ref{fig:MF-Age-SACY} shows the fraction of multiple systems among the targets in our sample grouped by YMG membership and plotted against their ages for the joint sample. Once again, there does not appear to be a dependence on age or environment in the range spanned by this set of YMGs.

\bigskip
%----------------------------------------------------------------
\subsection{Sources of Bias}\label{subsec:bias}

As with all population studies, a number of observational biases could have influenced our results and interpretations. Under Section \ref{subsec:mf1} we have used completeness maps to correct the bias due to non-uniform sensitivity to the mass ratios and separation ranges of companions for each target. This is done for our assessments of the distributions in separation and mass ratio as well as the overall visual binarity. In this subsection we discuss other relevant biases. Their net effects are summarized in Section \ref{subsubsec:netbiases}.

%----------------------------------------------------------------

\subsubsection{Branch Bias}\label{subsubsec:branch} 

\citet{bra76} cautioned that a key observational bias affecting binary statistics from magnitude-limited surveys is a systematic difference in the volume out to which binaries are visible as compared to single stars. The net outcome is a binary rate boosted by a factor dependent on the intrinsic binarity rate as well as the underlying distribution of companion flux ratios \citep[see, e.g.][Section 6.1, eqn. (4-5)]{bur03}. The relative enhancement is greater for populations with lower binarity and those consisting of systems biased towards equal-mass pairings. Similarly, it tends to inflate a population's mass ratio distribution at the high end. Furthermore, in general we expect the bias to affect M-dwarfs most severely because they are intrinsically fainter. 

However, it is difficult to evaluate a quantitative Branch boost factor for our study. This is in part due to our relative ignorance over the intrinsic mass ratio distribution of our target population, but also because our sample is not strictly magnitude-limited -- sample selection is primarily motivated by membership in YMGs. Compared to a magnitude-limited study in uniform space, the Branch bias should be markedly weaker for populations which have a finite spatial extent tangentially and radially, as for YMGs. This is because the increased volume for binaries from larger distances goes as $r$ rather than $r^3$. Furthermore, when there exists an upper bound in distance, stars which are brighter than the threshold survey magnitude even at the far end of the population will not suffer from this bias at all. This implies that earlier-type stars are systematically less affected than later-type ones. Below we attempt to estimate the possible effect of Branch bias in our multiplicity measurement.     

There are two ways in which Branch bias could manifest in our study. First, we need to examine whether such bias may already be present in the target source catalogues. None of the catalogues claim to be complete and it is difficult to assess their intrinsic selection biases. To address the level of their completeness to low-mass stars we ask the following: whether the catalogues document a substantial population of members down to M4 in each YMG (they do), and whether the later types are dominated by known multiples over singles (they are not, except in distant Carina). We conclude that catalogue bias is probably sub-dominant to the next source of Branch bias.

The other source of Branch bias is associated with our own target selection strategy, which has prioritized brighter objects and resulted in the majority of observed stars being $K < 10$. 

To understand this bias for our specific selection of YMGs and target spectral types, we constructed toy models of our YMGs at their corresponding ages and distances. We distributed young, low-mass star systems, a controlled fraction of which are equal-mass binaries, in a Gaussian around the average YMG distance. The Gaussian widths are defined such that $2\sigma$ incorporates the range of distances spanned by the known group members. After computing their apparent K-magnitudes using Baraffe models \citep{bar15}, we calculated the ratio of injected to observed binary fraction ($F\equiv f_{\rm obs}/f_{\rm true}$) for a given limiting {\it{K}}-magnitude. Based on the magnitudes of the observed targets, we set the limiting magnitude to be $K < 10$mag, and injected $f_{\rm true} = 0.3$. 

We arrive at the following estimates: the mass-weighted mean $F$ for each YMG is less than $1.3$ over masses $0.2 - 0.6 M_\sun$, and the overall weighted mean is $\sim1.16$. In all groups, there is no bias at the upper end of the mass bracket ($> 0.5 M_\sun$). Of course, the Branch effect is more pronounced for lower masses. Carina is most extreme, where in our toy model $F \sim 3$ for $M_\star = 0.2 M_\sun$, followed by TucHor and Columba at $F \sim 2.3$. The rest have $F < 1.6$ for $M_\star = 0.2 M_\sun$. In reality, only 6\% of our survey targets are drawn from Carina. Since all 6 targets are relatively massive ($>0.4 M_\sun$), Carina does not actually contribute to the low-mass Branch bias. Nevertheless, in the lowest-mass range ($0.1 - 0.3 M_\sun$) the mean $F$ is still relatively high at  $\sim 1.8$ due to the high representation of TucHor members in our sample.    

It is worth remembering that the estimates above are conservative in the sense that the toy models include equal-mass binaries only. Branch bias is reduced for a realistic population with a range of mass ratios. Furthermore, Branch bias is not only a function of a star's intrinsic brightness (therefore mass and age) but also highly dependent on the geometry of a population's spatial distribution. 

In sum, we expect Branch bias to inflate the average overall M-dwarf multiplicity measured here by no more than a multiplicative factor of 1.25 (i.e. requires multiplying the observed rate by 0.8 to correct). However, a more than 0.6 downward adjustment may be needed in the lowest mass bins.

% ------------------------------------------------------------------
\subsubsection{Selection Bias}\label{subsubsec:selectionbias}
One obvious bias affecting observed multiplicity is the sample selection bias. As described in Section \ref{sec:sample}, the initial sample was culled for \citet{mal13} binaries for which components have been separately resolved and have individual spectral types identified. The net effect of such a selection bias is to depress the observed binarity. 

We can estimate the extent to which our measurement is affected by this selection bias, as follows. We assemble the full pool of M-type candidate YMG members with declinations below $-40^\circ$ from the 5 source catalogues (see Section \ref{sec:sample}) without regard to binarity. All but one (J1814-3246) of our M-type targets have $\delta < -40^\circ$. There are 192 objects in this pool, 86 of which have visual binarity constraints provided by Clio imaging (excluding J0102-6235, see Section \ref{subsec:indiv-targs}). 19 other candidates were part of J12's AO campaign, screened for visual binarity as close as 0\farcs075, who found 13 more multiples. Another 34 objects were observed by J16, who identified a further 9 visual multiples in addition to 2 VBs and 2 SBs already noted by \citet{mal14a}. In addition, we have identified two more binaries (J03210395-6816475 and J20143542-5430588) in the full sample pool based on guider camera imaging. 

For the 51 remaining objects for which there are no binarity constraints available, including J0102-6235, we have assigned to them an `average' $<300$ AU multiplicity equalling the overall rate derived from our own M-star sub-sample ($34 \pm 5\%$). After accounting for Branch bias ($34\% \times 0.8 = 27\%$, see Section \ref{subsubsec:branch}), we get an additional $51 \times 0.27 = 13.8$ binaries. In total, the multiplicity rate for the YMG M-type sample pool becomes $[13+9+2+2+13.8+26]/192 = 34\%$. This is $\sim4\%$ higher than the raw value of $30\%$.     

We conclude that sample selection bias has an appreciable effect on our raw multiplicity fraction and, independent of other biases, it means our measured value of raw multiplicity is underestimated by another additive factor of $\sim 4\%$.  

\bigskip
%----------------------------------------------------------------

\subsubsection{Separation Distribution Bias}\label{subsubsec:sepbias}

Both SACY and our own multiplicity survey are sensitive to a limited range of projected separations, meaning we are potentially probing only a subset of all binaries. The fraction we are missing depends on the underlying separation distribution. 

In the field, the separation distribution is a function of primary mass. The average G-dwarf binary tends to be farther separated than an M-dwarf binary. When fitted with a uni-modal log-normal function, the peaks are $\sim45$ AU and $\sim5$ AU respectively. The distribution for G-dwarfs is also broader, with a significant number of binaries lying between $10^3$ and $10^6$ AU \citep[e.g.][]{dk13}. Therefore, to probe a fixed range of separations for all masses may introduce a systematic skew in the space of multiplicity versus primary mass. 

Our search for companions to KM-dwarfs is effectively confined to under $\sim 300$ AU. The binaries in our sample have a comparable separation distribution to the field, with most of the density confined within 300 AU (see Figure \ref{fig:sepqdistr}). Therefore, if the true distribution is field-like, we expect our low-mass multiplicity measurement to be representative of the total multiplicity.   

The SACY study detects companions to FGK-dwarf members between 3 and 1000 AU, which we supplement with known SBs to enhance their completeness at shorter periods. Using kernel density estimation, SACY measures a separation distribution for their sample that is significantly narrower than the field, extending negligibly beyond 1000 AU. If YMG G-dwarf binaries really have this narrower range of separations, then SACY would be essentially complete as well. 

Nevertheless, \citet{ell15} cautions that bias against wider companion discovery may exist in the algorithm they use to identify bound systems in the survey, which could render their derived separation distribution too narrow. If we instead assume a field-like separation distribution for sun-like stars in YMGs, then SACY has missed $\sim20\%$ of all multiple systems from the wide end. Therefore, a factor of up to 1.25 may need to be applied to correct for the total multiplicity fraction for the higher-mass bins.   

%----------------------------------------------------------------
\subsubsection{Membership Bias}\label{subsubsec:membershipbias}
The methods by which memberships are determined for these targets are statistical and includes any RV information available for the target, attributed to its galactic motion. Therefore, SB1s have a higher chance of being assigned to the wrong YMG or the field. 

In our sample selection and YMG assignment, we have largely used the versions of the \citep{mal13} and \citep{kra14} catalogues that disregarded RV information. Therefore, we expect this bias to play a minimal role in our overall multiplicity values. It is possible that mis-assigment of stars to incorrect YMG could affect our evaluation of the environmental/age dependence of multiplicity, but so long as the SB1 bias is not systematically undermining a specific YMG, there should be no net effect.  

%----------------------------------------------------------------
\subsubsection{Chance Alignment Bias}\label{subsubsec:chancealignbias}
   
A visual binary could appear to be single if its components are chance aligned along our line of sight. Unlike semi-major axis, projected separation is not an intrinsic property of a binary system but a function of orientation and time of observation. Depending on the orbital elements of the system, binary components whose semi-major axes are theoretically resolvable may become momentarily chance aligned, i.e. their projected separation may drop below the resolution limit, in an arbitrary snapshot. On the surface, this phenomenon constitutes a potential source of bias that results in systematically missing binaries.      

Upon closer inspection, the issue is more subtle. Such a bias is only problematic if we were interested in retrieving the {\emph{true}} distribution of semi-major axes in our sample, which is difficult in practice. Semi-major axis is only statistically related to the projected separation measured at any give moment (see, e.g. \citealt{dm91}: Section 5.3; \citealt{tor99}: Fig 3; \citealt{bra06}: Appendix). While the average projected separation is within a factor of order unity to the true semi-major axis, a large dispersion exists.   

Therefore, despite semi-major axis being a more meaningful quantity to characterize a binary system, we do not attempt to perform a statistical retrieval. Nevertheless, we do investigate this issue for a fixed range of semi-major axes to which observations are most sensitive, namely 3 to 300 AU. Due to the range of possible orbital eccentricities, overall we expect comparable quantities of visual binaries to be scattered in as out by this bias, for a negligible net effect. 

% ----------------------------------------- 
\subsubsection{Stellar Model Uncertainties}\label{subsubsec:model-uncertainties}

Current state-of-the-art stellar evolution models tend to over-predict the radii (hence luminosity) of young, low-mass stars by 10-25\% compared to empirical measurements \citep[e.g.][]{dav16}. This is a source of systematic error in any derivations of stellar mass from luminosity. As the error acts differentially on stars of different masses, it could also affect mass ratio calculations of binary components, and therefore our physical contrast curves and estimates of incompleteness.

To examine the magnitude of such a bias, we consider the case of an M-dwarf binary for which one star is radiative ($0.5 M_\sun$) and the other fully convective ($0.2 M_\sun$). We expect any differential effect on mass ratio to be most pronounced for such a pairing, since the mass-luminosity scaling relation changes across the fully convective boundary. For a given over-estimation factor in stellar luminosity of 30\% for both components, the offset in mass ratio is only $\sim5\%$. Since 30\% is already a conservative upper bound in the model luminosity error, we conclude that its impact on our physical sensitivity curves would be minuscule.

% -----------------------------------------        
\subsubsection{Net Effect of Biases}\label{subsubsec:netbiases}

\begin{figure}[htb]
\center 
\hspace{-1.5cm}
\includegraphics[scale=0.55]{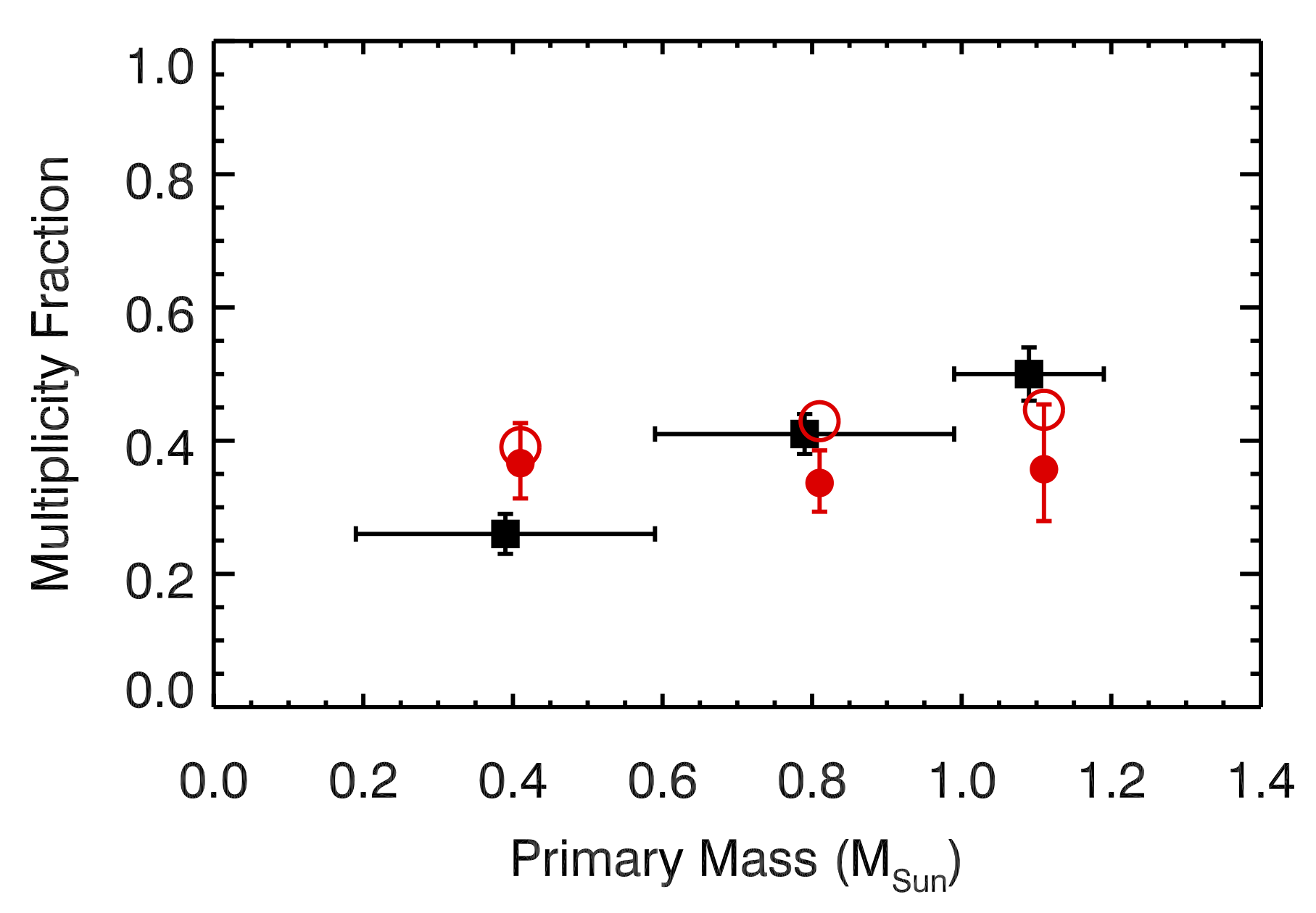}
\caption{A comparison of multiplicity fractions as a function of primary mass between the field and that measured for YMGs in this study. Black squares denote field points \citep{dk13}. Red solid circles represent raw measurements from our combined sample with SACY, and red empty circles are the post-bias correction values to our measurements. The bias-corrected values in the higher-mass bins agree well with the field. However, the low-mass bin is in mild tension.  }
\label{fig:MF_field}
\end{figure}

Figure \ref{fig:MF_field} compares the multiplicity fraction in the field versus our YMG measurements, as a function of primary mass. We divide the mass range spanned by the combined sample into 3 bins: $0.2 - 0.6$, $0.6 - 1.0$, and $>1.0 M_\sun$. The field values for the two higher mass bins are supplied by \citet{rag10}, whereas the lowest bin is an average from literature computed by \citet{dk13}. The lowest mass bin consists predominantly of MagAO stars, whereas the highest mass bin comprises objects almost exclusively from SACY. SACY targets also account for just over 60\% of the intermediate mass bin. In addition to the raw measured value, this plot shows a crude correction representing the combined effect of biases discussed. 

% --- Figure showing Branch Bias CDF ---
% --------------------------------------

In the low-mass bin, we begin by applying the detectability corrections (Section \ref{subsec:mf1}), which constitutes a $6\%$ additive boost. Then, we multiply the resulting fraction by the Branch bias factor of 0.8 (see Section \ref{subsubsec:branch}). Finally, we add an upward correction of 4\% for selection bias. The overall increase is $\sim2\%$. 

In the intermediate-mass bin, we compute detectability corrections to the Clio targets and assume perfect sensitivity in this mass range from the SACY survey. The Branch bias is negligible in this mass regime. We add only 2\% for selection bias because Clio targets contribute only $\sim 40\%$ to this mass bin. Lastly, we multiply the multiplicity fraction by factor 1.1 assuming a $\sim10\%$ contribution from binaries wider than our detection range.    

In the highest-mass bin, which is composed of SACY targets exclusively, we apply a correction factor of 1.25 to account for the contribution from wide binaries, as discussed under Section \ref{subsubsec:sepbias}).

After correction, multiplicity rates in the higher mass bins agree well with the field. However, the YMG M-dwarf binary rate has been increased by the bias corrections to greater ($>1\sigma$) than the field. Nevertheless, since the corrections are relatively crude and the discrepancy of limited significance, the YMG findings may be reconcilable with the field measurements.  

%--------------------------------------------------

\bigskip
%==============================================
\section{Discussion \& Conclusions}\label{sec:conclusion}

In this paper we have summarized our work to identify unresolved binaries in recently proposed low-mass members of local YMGs using NIR AO imaging. The campaign resulted in images of 27 visual binaries, 15 of which are previously unknown, between $0\farcs05$ and $4\farcs0$ (roughly 3 to 300 AU) in a sample of 104 KM-dwarf stars associated with 7 YMGs with ages from 10 to 200 Myr. For 25 systems we have measured astrometric and photometric properties, deriving stellar masses and projected physical separations. When combined with the literature, 18 visual doubles now have multi-epoch and multi-waveband data. Two additional suspected visual binaries (and J0102-6235 and J0903-6348) are rejected on astrometric grounds, and the rest verified. In conjunction with spectroscopic binarity information from literature, we also find that two of systems are in reality triples (J0030-6236 and J2247-6920), and two of the binaries potentially form a hierarchical quadruple (J0241-5259A and B).  

We have measured standard multiplicity statistics for this population in concert with SACY's GK-dwarfs \citep{tor08,ell14,ell15,ell16}. In so doing, we generate a broader picture of stellar multiplicity for YMG stars in the mass range of 0.1 and 1.2 $M_\sun$, roughly doubling the mass range and number of targets from previous efforts. 
The 58 visual multiples and 12 spectroscopic binaries known from literature with no visual counterparts in the joint sample of 203 stars gives a combined raw multiplicity rate of at least $35\pm3\%$ with separations below 1000 AU. The raw visual binarity rate (1 -- 1000 AU) is at least $26\pm3\%$, and the companion fraction is $38\pm3\%$. These are comparable to field FGK stars for the same separation range \citep{rag10,dk13}. 7 of the targets are triples, giving a triple rate of $3\%$ for the joint YMG GKM population. 

Our M-star sub-sample has an overall raw observed binarity rate of $30^{+5}_{-4}\%$ under 300 AU, and the raw rate of occurrence for visual binaries resolved by the Clio camera between 1 and 300 AU is $21^{+5}_{-4}\%$. The rates for stars with $< 0.6 M_\sun$ are $36\pm6\%$ and $27^{+6}_{-5}\%$, respectively. While the visually-resolved raw binarity appears to be roughly consistent with the most recent field measurement for M-dwarfs over comparable separation ranges (CF $\sim23\pm3\%$ for $3 - 10,000$ AU, \citealt{wd15}), the overall multiplicity fraction is discrepant (MF $\sim26\pm3\%$ overall, \citealt{dk13}).

Our combined analysis with SACY finds no evidence for the overall multiplicity fraction to be dependent on primary mass from $0.2$ to $1.2 M_\sun$ in the raw measurements. This would corroborate the findings of SACY \citep{ell15}, now with the statistics appreciably bolstered in the mass range of $0.2 - 0.7 M_\sun$. However, such a finding is in conflict with the well-known behaviour in the field, where more massive primaries are more likely to be multiple.     

As discussed in depth under Section \ref{subsec:bias}, the raw observed multiplicity rates can deviate from the true rates due to various biases. Detectability considerations (Section \ref{subsec:mf1}) and initial selection bias against known visual binaries (Section \ref{subsubsec:selectionbias}) could have rendered our measurement for M-star multiplicity an underestimate from the true rate by $\sim 10\%$. Meanwhile, Branch bias (Section \ref{subsubsec:branch}) could inflate our observed M-dwarf binarity by up to a factor of 1.25, and even more at the low-mass end. On the other hand, when separation bias (see Section \ref{subsubsec:sepbias}) is accounted for in the measurement of higher-massed SACY objects, the values for GK-dwarfs are boosted. The bias-adjusted multiplicities in the higher mass bins are consistent with those in the field, while in the low mass bin there is still tension at $\sim1.5\sigma$, subjected to various assumptions in corrections (see Figure \ref{fig:MF_field} in Section \ref{subsubsec:netbiases}).  

There is no significant indication of varying multiplicity fractions from YMGs despite their differences in age and possibly birth environments. Furthermore, no notable changes in binary separations are found as a function of age, though the statistics are based on small numbers. These results agree with the YMG solar-type star sample of SACY and suggests minimal dynamical evolution in the $\sim10 - 200$ Myr after their formation even for binaries systems of lower mass, and supports a similar formation pathway for all YMGs. 

Previous studies have used binary fraction as evidence for the contribution of a given population to the field. For example, \citet{pat02} compared the field binary separation distribution with that of open clusters and T-associations, concluding that dense and loose star-forming environments contribute 70\% and 30\% of the field stars, respectively. \citet{goo10} points out that, when deducing the origin of the field from comparing binary statistics, one must consider the effect of dynamical processing on young populations, which could modify their binary rates before they enter the field. 
If the low-mass stars in YMGs do exhibit greater multiplicity than field stars and are expected to retain their multiplicity properties due to lack of dynamical evolution before entering the field, then YMGs may not be a significant contributor to the field. This agrees with the assessment by \citet{ll03} that most field stars originate in OB associations.

%==============================================

%==============================================
%\begin{table*}[htb]
\clearpage
\begin{sidewaystable}
\vspace{-10cm}
%\centering
\caption{System Properties of Visual Binaries Observed by Clio}
\label{table:vbs}
\begin{tabular}{lccclcrccrrccccl}
\hline\hline
Target & SpT & Dist. & Epoch & Band & Ang. Sep. & PA\textsuperscript{a} & $\Delta$mag & Co- & $P_{\rm{unbound}}$ & Proj. Phys. & $M_{\rm prim}$\textsuperscript{b} & $M_{\rm sec}$\textsuperscript{b} & q & Known & Notes\\
Name &  & (pc) &  &  &  ($''$) & ($^\circ$) &  & move? & (\%) & Sep. (AU) & ($M_\sun$) & ($M_\sun$) & & Binary? &  \\
\hline\hline
 & & & & & & & {\bf{TucHor}}& & & & & & & & \\
\hline
      J0030-6236&  M2.2&    44$~\pm$    2&  15-11-26&    H&  0.121$~\pm$  0.001&   268.6$~\pm$    0.36&    0.14$~\pm$    0.04& CO&   0.0000&     5.3$~\pm$     0.2&   0.54&   0.49&   0.9& yes &  1, 2 \\
      J0102-6235&  M2.9&    46$~\pm$    2&  15-11-26&    H&  0.678$~\pm$  0.005&   325.7$~\pm$    0.35&    0.36$~\pm$    0.04&  B&   0.0000&    31.0$~\pm$     1.4&   \nodata &   \nodata &   \nodata & yes &  2 \\
      J0229-5541&  M4.8&    46$~\pm$    2&  14-11-30&    H&  0.159$~\pm$  0.001&   101.4$~\pm$    0.35&    0.21$~\pm$    0.04&  -&   0.0000&     7.3$~\pm$     0.3&   0.11&   0.09&   0.8& no &  - \\
     J0241-5259B&  M2.5&    44$~\pm$    3&  14-12-01&    H&  0.075$~\pm$  0.001&   174.5$~\pm$    0.36&    0.05$~\pm$    0.01&  -&   0.0000&     3.3$~\pm$     0.2&   0.56&   0.54&   1.0& no &  3 \\
      J0257-6341&  M3.6&    63$~\pm$    3&  14-12-01&    H&  0.423$~\pm$  0.003&   325.4$~\pm$    0.35&    0.07$~\pm$    0.01&  -&   0.0000&    26.7$~\pm$     1.3&   0.32&   0.30&   0.9& no  &  - \\
      J0331-4359&  K6.0&    45$~\pm$    1&  14-11-30&    H&  0.393$~\pm$  0.003&    92.3$~\pm$    0.34&    2.77$~\pm$    0.01& CO&   0.0001&    17.8$~\pm$     0.4&   0.73&   0.17&   0.2& yes &  2, 4, 5 \\
      J0405-4014&  M4.2&    48$~\pm$   3&  15-11-27&    H&  0.609$~\pm$  0.004&   102.6$~\pm$    0.34&    0.58$~\pm$    0.04&  -&   0.0001&    29.2$~\pm$     0.6&   0.30&   0.20&   0.7& no &  - \\
      J0447-5035&  M4.0&    55$~\pm$    3&  14-12-01&    H&  0.544$~\pm$  0.004&   121.3$~\pm$    0.34&    0.03$~\pm$    0.03& CO&   0.0000&    29.9$~\pm$     1.6&   0.28&   0.28&   1.0& yes & 2 \\
      J1708-6936&  M3.5&    49$~\pm$    3&  15-05-11&    H&  0.435$~\pm$  0.003&     9.1$~\pm$    0.34&    0.72$~\pm$    0.02&  C&   0.0001&    21.3$~\pm$     1.3&   0.54&   0.35&   0.7& yes & 2  \\
      J2108-4244&  M4.4&    44$~\pm$    2&  14-11-11&   Ks&  0.139$~\pm$  0.001&   302.3$~\pm$    0.35&    0.11$~\pm$    0.04& CO&   0.0000&     6.1$~\pm$     0.3&   0.20&   0.19&   0.9& no &  - \\
      J2244-5413&  M4.0&    49$~\pm$    4&  14-12-01&    H&  0.535$~\pm$  0.004&   298.8$~\pm$    0.35&    0.32$~\pm$    0.01& CO&   0.0000&    26.2$~\pm$     2.1&   0.42&   0.34&   0.8& yes & 2, 6 \\
      J2247-6920&  K6.0&    52$~\pm$    1&  15-05-11&    H&  0.246$~\pm$  0.002&   175.2$~\pm$    0.35&    3.50$~\pm$    0.11& CO&   0.0001&    12.8$~\pm$     0.3&   0.68&   0.08&   0.1& no &  7, 8 \\
\hline
 & & & & & & & {\bf{Col}}& & & & & & & & \\
\hline
      J0142-5126&  M6.5&    66$~\pm$   6&  15-11-26&    H&  2.014$~\pm$  0.014&   288.6$~\pm$    0.37&    0.89$~\pm$    0.23&  -&   0.0022&   132.9$~\pm$     2.2&   0.21&   0.12&   0.6& no &  - \\
      J0324-5901&  K7.0&    90$~\pm$    5&  14-11-30&    H&  0.466$~\pm$  0.003&   280.2$~\pm$    0.34&    2.33$~\pm$    0.01& CO&   0.0002&    42.0$~\pm$     2.3&   0.78&   0.26&   0.3& yes &  2, 4 \\
      J0332-5139&  M2.0&    88$~\pm$    5&  14-11-30&    H&  3.198$~\pm$  0.024&   115.5$~\pm$    0.34&    1.50$~\pm$    0.03&  I&   0.0064&   281.4$~\pm$    16.1&   0.62&   0.26&   0.4 & yes &  2, 4 \\
      J0717-6311&  M2.0&    58$~\pm$    4&  14-04-21&    H&  0.077$~\pm$  0.001&   203.2$~\pm$    0.41&    0.07$~\pm$    0.01&  -&   0.0000&     4.5$~\pm$     0.3&   0.39&   0.37&   1.0& no &  4 \\
\hline
 & & & & & & & {\bf{bPic}}& & & & & & & & \\
\hline
      J0533-4257&  M4.5&    16$~\pm$    4&  14-12-01&    H&  0.066$~\pm$  0.000&   222.2$~\pm$    1.09&    0.55$~\pm$    0.02&  -&   0.0000&     1.1$~\pm$     0.3&   0.17&   0.12&   0.7& no &  3 \\
      J1657-5343&  M3.0&    51$~\pm$    3&  15-05-10&    H&  0.051$~\pm$  0.000&   219.7$~\pm$    0.47&    0.20$~\pm$    0.01&  -&   0.0000&     2.6$~\pm$     0.2&   0.53&   0.47&   0.9& no &  4 \\
      J1729-5014&  M3.0&    64$~\pm$    5&  15-05-11&    H&  0.699$~\pm$  0.005&    16.7$~\pm$    0.34&    0.21$~\pm$    0.05& CO&   0.0003&    44.7$~\pm$     3.5&   0.64&   0.58&   0.9&yes & 3, 5, 9, 13 \\
      J2121-6655&  K7.0&    32$~\pm$    1&  15-05-11&    H&  0.063$~\pm$  0.001&   301.6$~\pm$    4.48&    0.90$~\pm$    0.15&  -&   0.0000&     2.0$~\pm$     0.1&   0.55&   0.30&   0.5& no &  4 \\
\hline
 & & & & & & & {\bf{Arg}}& & & & & & & & \\
\hline
      J1516-5855&  K7.0&    77$~\pm$    3&  15-05-10&    H&  2.337$~\pm$  0.015&   207.1$~\pm$    0.34&    0.12$~\pm$    0.06&  C&   0.0116&   179.9$~\pm$     7.1&   0.67&   0.65&   1.0& yes &  4, 13 \\
\hline
 & & & & & & & {\bf{Car}}& & & & & & & & \\
\hline
      J0754-6320&  M3.0&    80$~\pm$    5&  14-04-21&    H&  0.857$~\pm$  0.005&   156.9$~\pm$    0.34&    0.10$~\pm$    0.01&  C&   0.0001&    68.6$~\pm$     4.3&   0.45&   0.43&   0.9& yes &  2, 3 \\
      J0809-5652&  K0.0&   104$~\pm$    3&  14-12-01&    H&  0.782$~\pm$  0.005&    29.9$~\pm$    0.34&    2.73$~\pm$    0.01&  C&   0.0051&    81.2$~\pm$     2.4&   0.90&   0.27&   0.3& no &  - \\
      J0903-6348&  M0.5&    66$~\pm$    3&  14-04-21&    H&  1.132$~\pm$  0.007&    64.1$~\pm$    0.34&    5.02$~\pm$    0.03&  B&   0.0750&    74.7$~\pm$     3.4&   0.67&   0.04&   0.1& yes & 2, 4, 10 \\
      J0931-5314&  K5.0&   120$~\pm$   19&  15-05-12&    H&  0.264$~\pm$  0.002&     8.2$~\pm$    0.34&    3.48$~\pm$    0.10& CO&   0.0143&    31.6$~\pm$     5.0&   0.76&   0.14&   0.2& no &  - \\
\hline
 & & & & & & & {\bf{ABDor}}& & & & & & & & \\
\hline
      J0524-4223&  M0.5&    52$~\pm$    9&  14-12-01&    H&  0.248$~\pm$  0.002&    61.0$~\pm$    0.37&    0.88$~\pm$    0.01& CO&   0.0000&    12.9$~\pm$     2.2&   0.37&   0.22&   0.6& yes &  2, 4 \\
\hline
 & & & & & & & {\bf{TWA}}& & & & & & & & \\
\hline
      J1234-4538&  M1.5&    78$~\pm$    3&  15-05-12&    H&  0.591$~\pm$  0.004&   309.2$~\pm$    0.34&    0.03$~\pm$    0.03& CO&   0.0000&    46.2$~\pm$     1.8&   0.49&   0.48&   1.0& yes & 11, 12 \\
\hline
\hline
\end{tabular}
\\
~\\
Notes: \\
 \textsuperscript{a} Position Angle (PA) is measured in degrees east of north. \\
 \textsuperscript{b} Typical uncertainties are $0.02 M_\sun$ (see Section \ref{subsec:physpars}). \\
1. $4\farcs4\sim 200$ AU triple companion also detected in Clio \\
2. VB also imaged by \citet{jan16}\\
3. Flagged as probable binary by BANYAN \citep{mal13} based on YMG membership analysis (see text) \\
4. Not flagged as probable binary by BANYAN \citep{mal13} based on YMG membership analysis \\
5. VBs also detected by \citet{ell15}  \\
6. $0\farcs63$ VB detected by \citet{cha10} \\ 
7 SB1 according to \citep{mal14a} \\
8. Previously suggested by \citet{mal13} to belong to the field when RV data is incorporated into the YMG membership analysis \\
9. $0\farcs8$ VB according to \citet{tor08} \\
10. $7\farcs9$ VB detected by \citet{mal14a} \\ 
11. $0\farcs7$ VB suspected by \citet{zuc01} \\
12. $0\farcs62$ VB resolved by \citet{jan12} \\ 
%\end{table*}
\end{sidewaystable}
\clearpage

%==============================================

%==============================================
\section*{Acknowledgement}\label{sec:acknowledge}

In the process of compiling this work, the authors have engaged in helpful discussions with numerous friendly and knowledgeable persons in various capacities. We would like to thank Sean Andrews, Gaspard Duch{\^e}ne, Jason Eastman, Dan Foreman-Mackey, Robin Gong, Vinay Kashyap, Rainer K{\"o}hler, Charles Lada, Lison Malo, Ilya Mandel, Evgenya Shkolnik, and Willie Torres, for their patience, insights, assistance and  encouragement. 

We thank Adam Kraus for detailed feedback on an early version of this paper.

YS is supported by a Doctoral Postgraduate Scholarships from the Natural Science and Engineering Research Council (NSERC) of Canada. Support for B.P.B. was provided by NASA through Hubble Fellowship grant HST-HF2-51369.001-A awarded by the Space Telescope Science Institute, which is operated by the Association of Universities for Research in Astronomy, Inc., for NASA, under contract NAS5-26555. Work by B.T.M., K.M.M., and J.R.M. were performed under contract with the Jet Propulsion Laboratory (JPL) funded by NASA through the Sagan Fellowship Program executed by the NASA Exoplanet Science Institute. L.A.C. was supported by CONICYT-FONDECYT grant number 1171246 and the Millennium Science Initiative (Chilean Ministry of Economy), through grant “Nucleus RC130007. H.C.acknowledges support from the Spanish Ministerio de Econom\'ia y Competitividad under grant AYA 2014-55840-P. KMM's and LMC's work is supported by the NASA Exoplanets Research Program (XRP) by cooperative agreement NNX16AD44G.

\bigskip
This publication makes use of data products from the Two Micron All Sky Survey, which is a joint project of the University of Massachusetts and the Infrared Processing and Analysis Center/California Institute of Technology, funded by the National Aeronautics and Space Administration and the National Science Foundation. It also makes use of data products from the Wide-field Infrared Survey Explorer, which is a joint project of the University of California, Los Angeles, and the Jet Propulsion Laboratory/California Institute of Technology, funded by the National Aeronautics and Space Administration. The Digitized Sky Survey was produced at the Space Telescope Science Institute under U.S. Government grant NAG W-2166. The images of these surveys are based on photographic data obtained using the Oschin Schmidt Telescope on Palomar Mountain and the UK Schmidt Telescope. The plates were processed into the present compressed digital form with the permission of these institutions.

%==============================================
\vspace{+10pt}
%\baselineskip=12pt

%\begin{appendices}
\appendix

\iffalse
%==============================================
\section{Contrast Curves}\label{app:cc}

In Table \ref{table:cc} we provide contrast curves computed for each target in terms of $\Delta$mag at characteristic angular separations. 

\input{CC_table.tex}

%==============================================
\fi

\section{Astrometric Plots for Multiepoch Data}\label{app:astrometry}

We present plots depicting our astrometric analysis for Clio visual binaries with multi-epoch data to determine whether they are co-moving or stationary background objects (Table \ref{table:vb-epochs}).

\clearpage

%\vspace{-1cm}

% --- Figure showing Sensitivity Maps ---
\begin{figure*}[htb] 
\begin{minipage}{0.5\textwidth}
\centering
\includegraphics[scale=0.38]{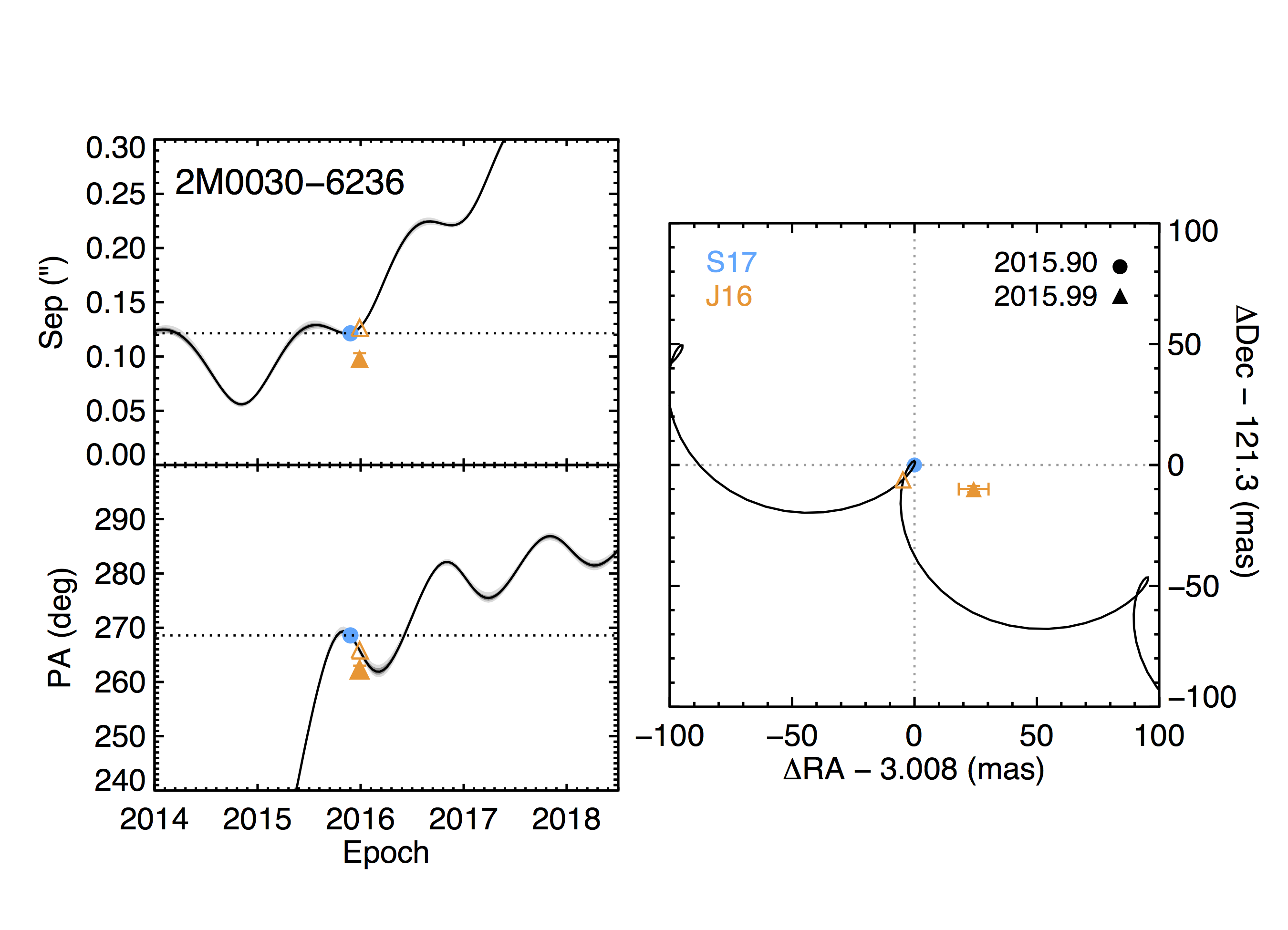}
\end{minipage}%
\begin{minipage}{0.5\textwidth}
\centering
%\vspace{-5cm}
\includegraphics[scale=0.38]{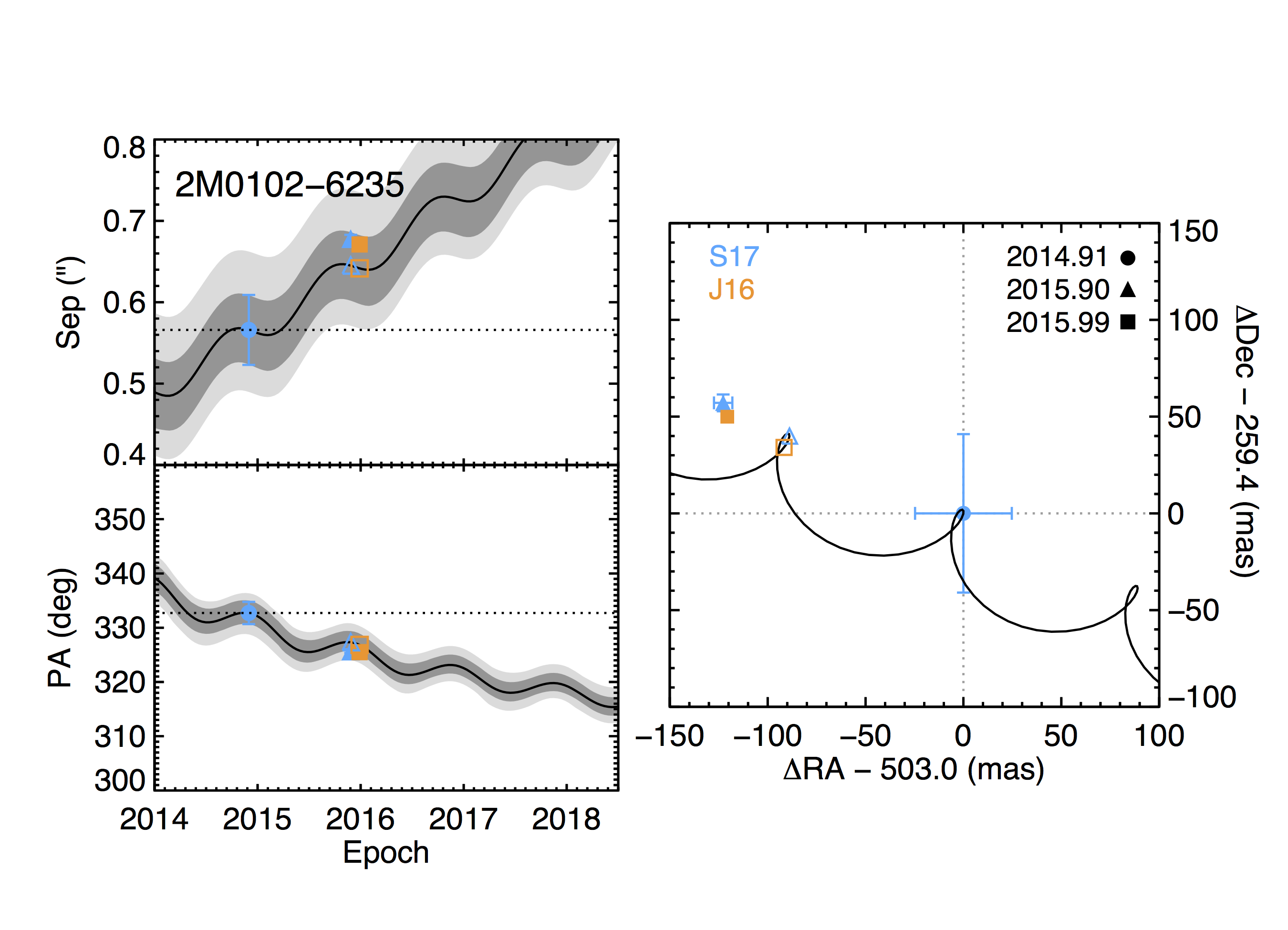}
\end{minipage}%
\begin{minipage}{0.5\textwidth}
\centering
\includegraphics[scale=0.38]{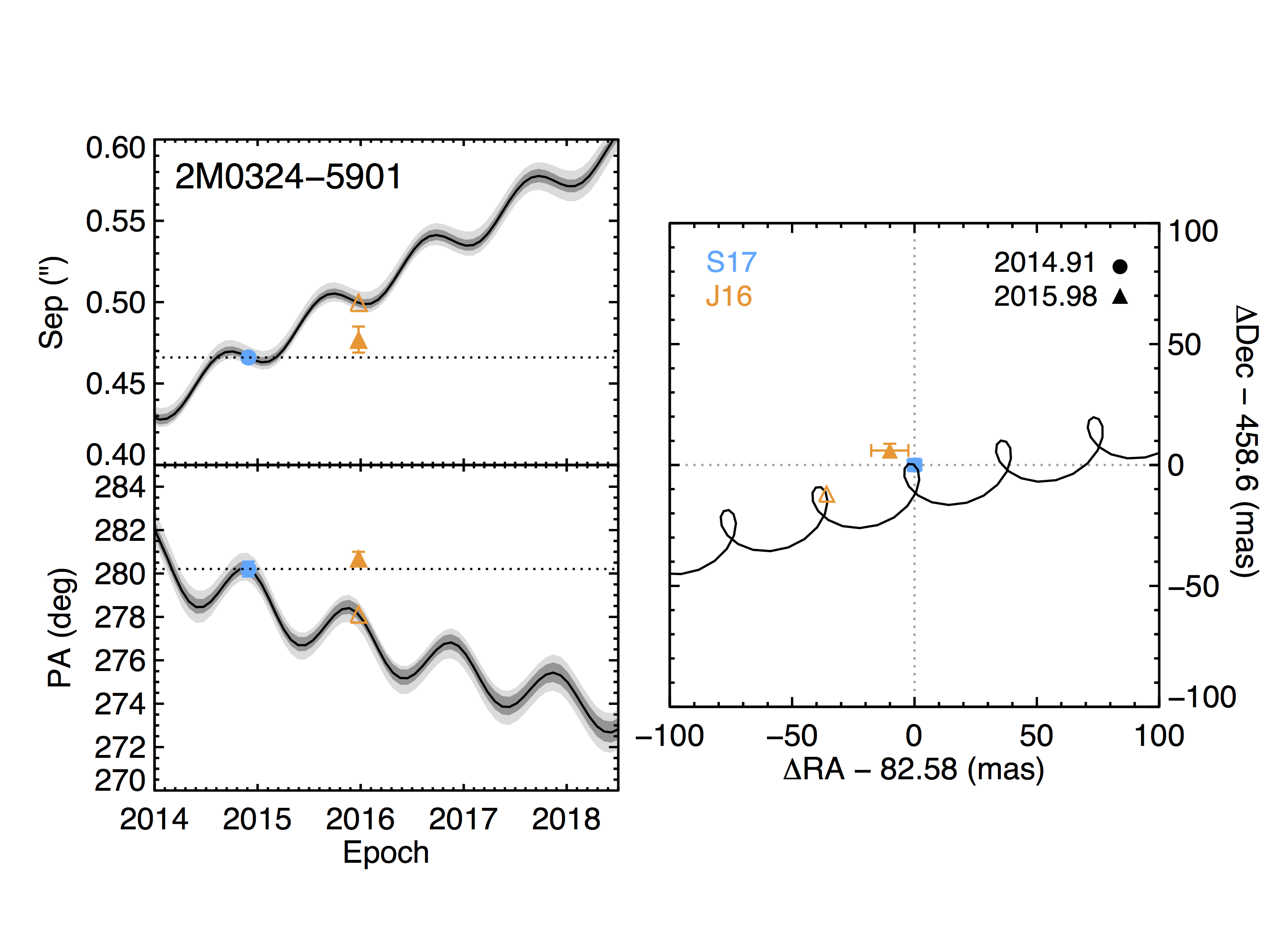}
\end{minipage}%
\begin{minipage}{0.5\textwidth}
\centering
\includegraphics[scale=0.38]{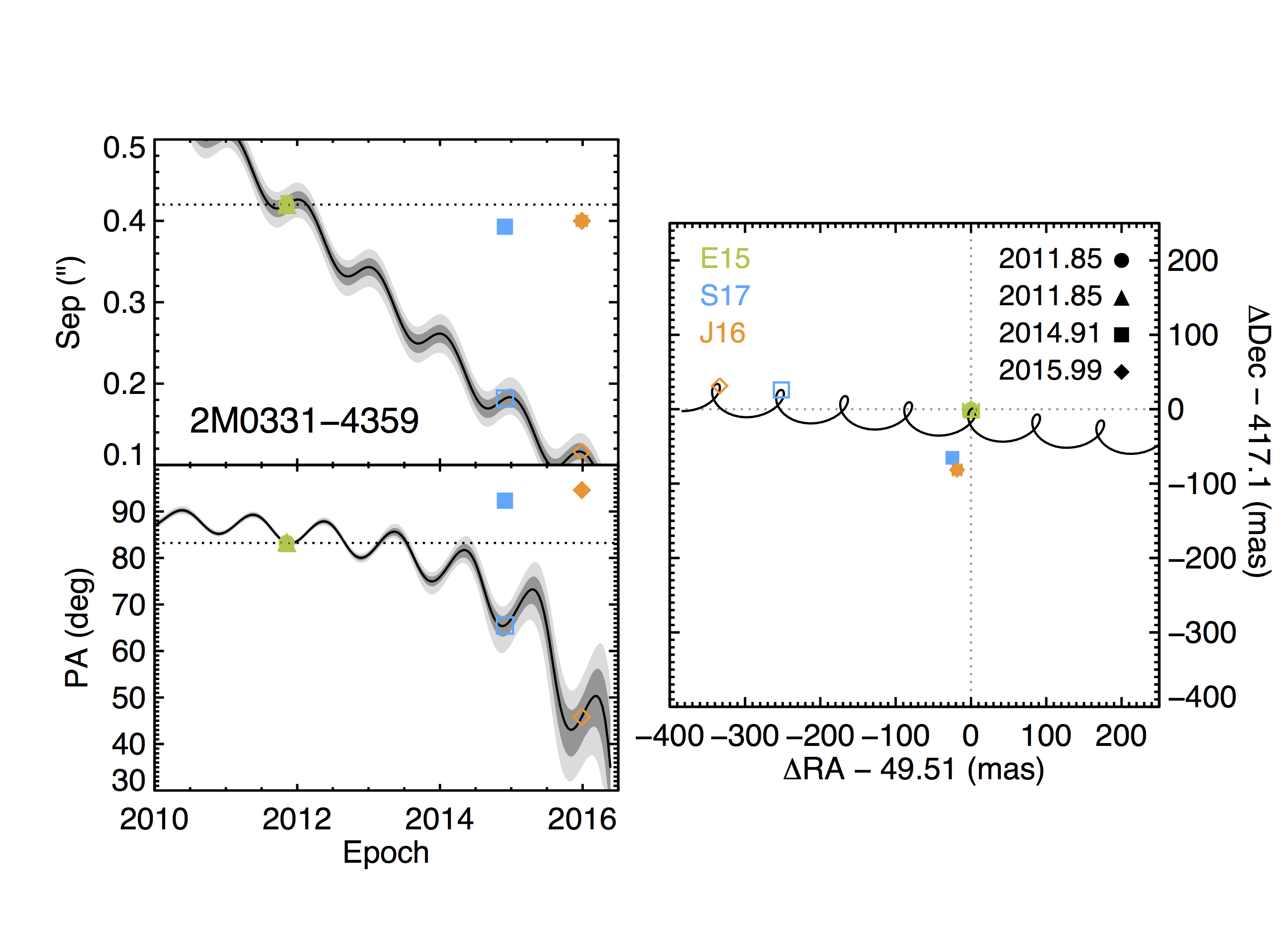}
\end{minipage}%
\begin{minipage}{0.5\textwidth}
\centering
\includegraphics[scale=0.38]{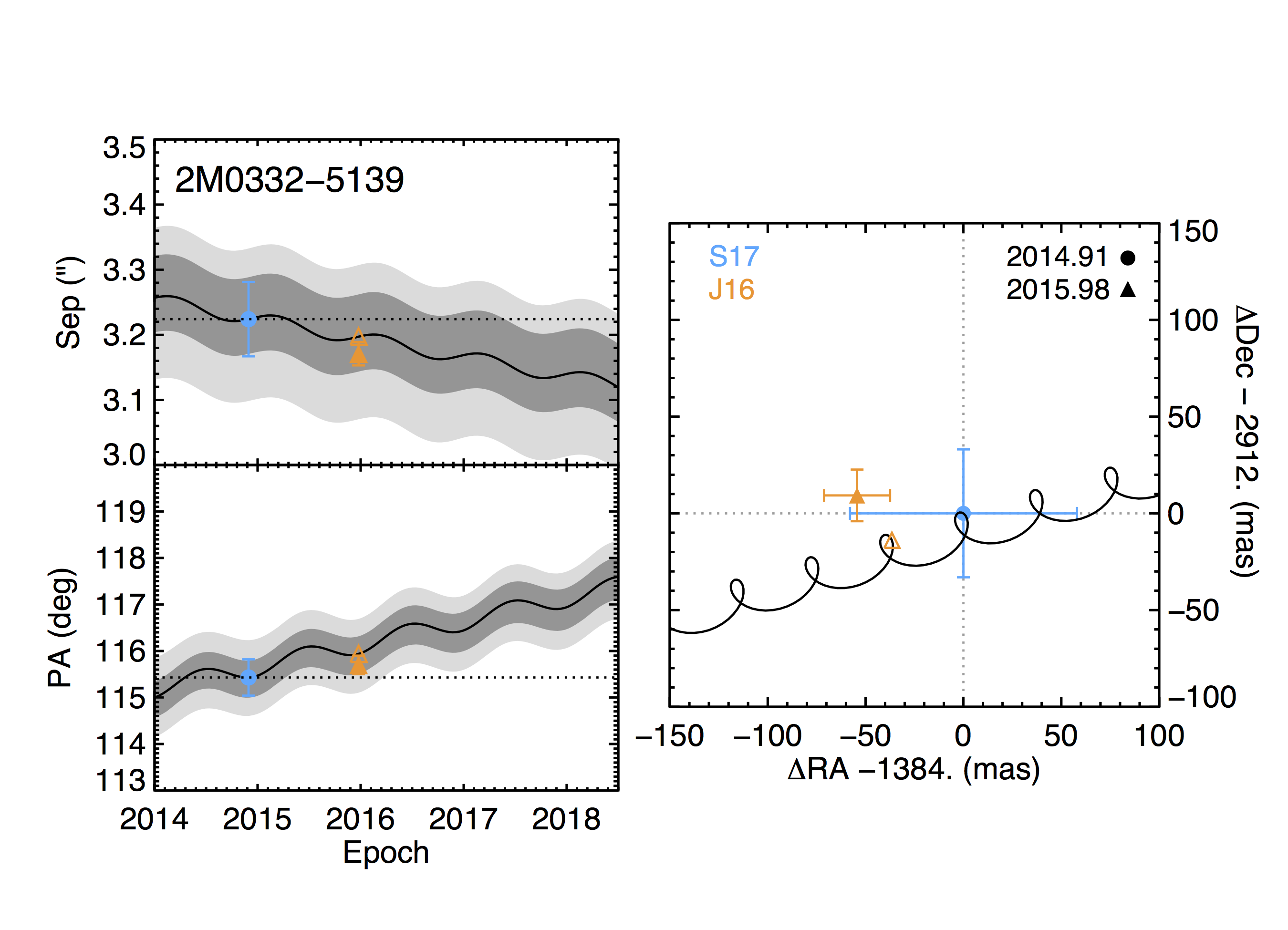}
\end{minipage}%
\begin{minipage}{0.5\textwidth}
\centering
\includegraphics[scale=0.38]{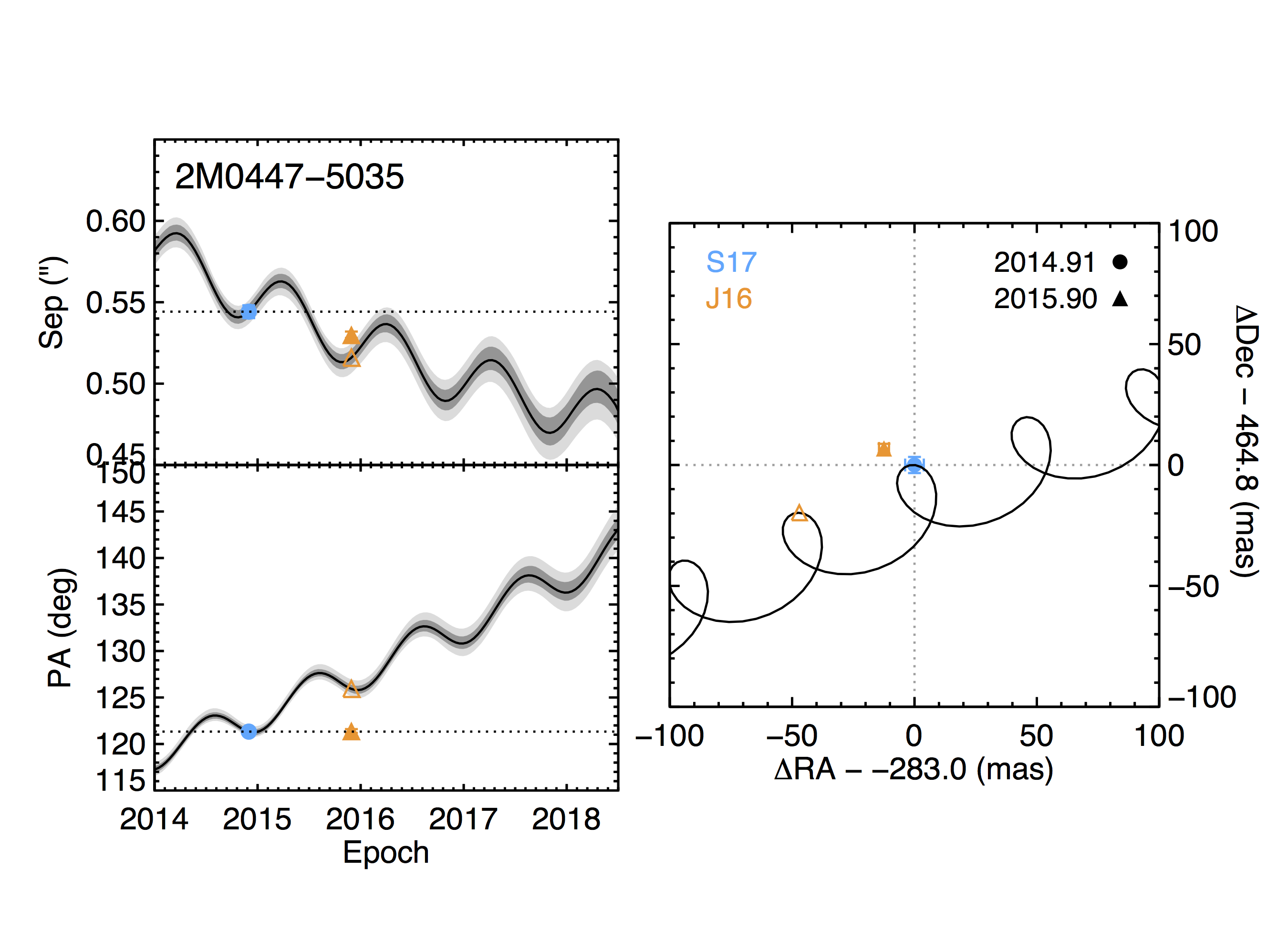}
\end{minipage}%
\caption{Co-moving analysis for Clio visual binaries with multi-epoch data. Each set of 3 plots display analysis for the target whose name appears in the upper-left corner. C10: \citet{cha10}, J12: \citet{jan12}, E15: \citet{ell15}, J16: \citet{jan16}, S17: this work. References and epochs of the data points plotted are listed in the right panel plot. Left: separation (top left) and position angle (bottom left) of the companion. The solid line shows the expected astrometric track of a distant background object as a result of proper and parallax motion of the primary, and open symbols indicate the expected measurement for such background object. The gray shaded regions represent $1\sigma$ and $2\sigma$ errors in the background tracks based on uncertainties in the proper motion, distance, and first epoch astrometry. The dotted lines show a perfectly comoving track. A real companion should not follow the background track and should move minimally relative to the primary (i.e. follow the dotted lines). Any movement between the epochs should be adequately explained by orbital motion. Right: $\Delta$RA and $\Delta$Dec as seen on the sky ($\Delta$ refers to primary - secondary position). }
\label{fig:ast1}
\end{figure*}

\clearpage
% --- Figure showing Sensitivity Maps ---
\begin{figure*}[htb] 
\begin{minipage}{0.5\textwidth}
\centering
\includegraphics[scale=0.4]{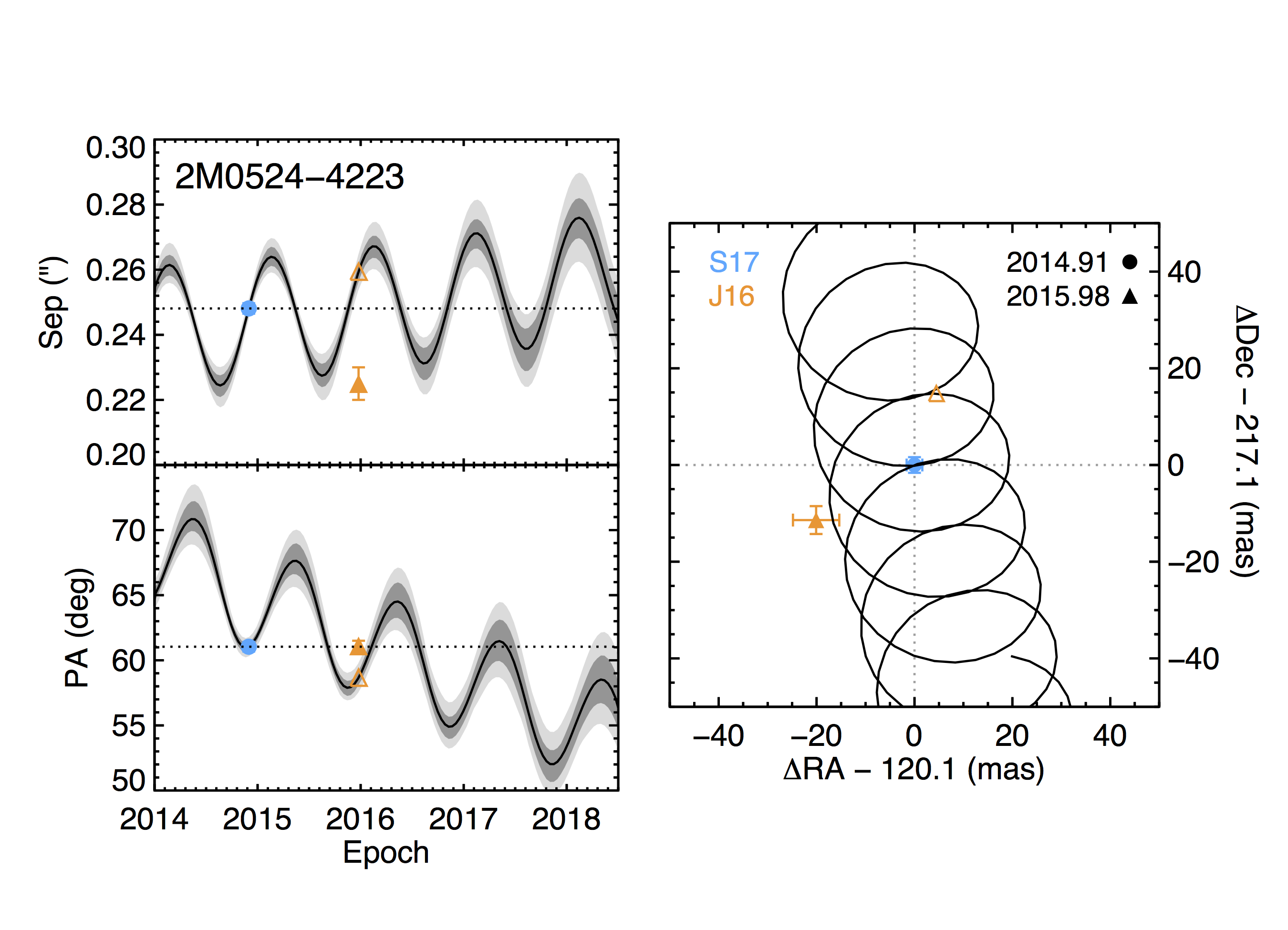}
\end{minipage}%
\begin{minipage}{0.5\textwidth}
\centering
%\vspace{-5cm}
\includegraphics[scale=0.4]{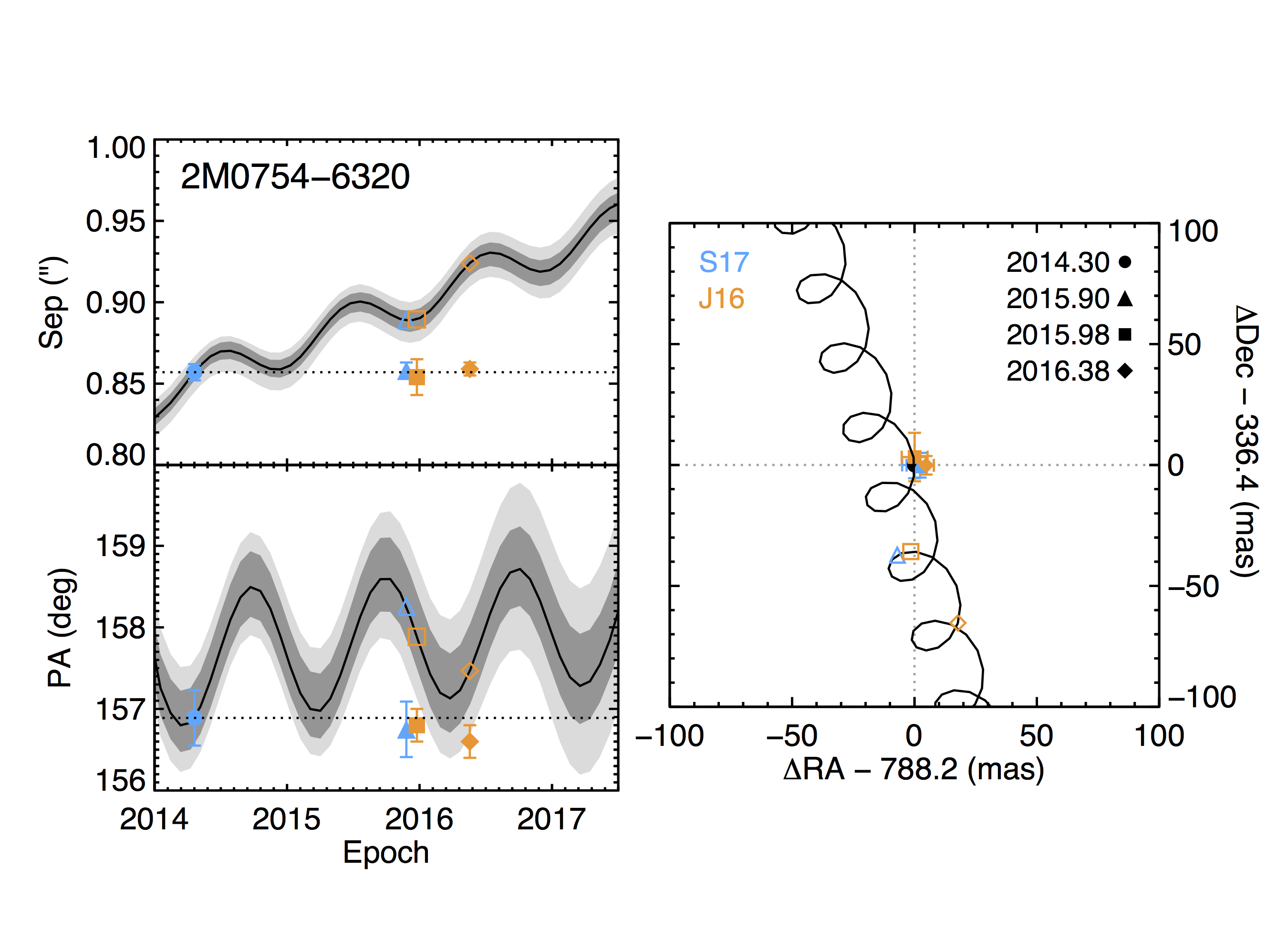}
\end{minipage}%
\begin{minipage}{0.5\textwidth}
\centering
\includegraphics[scale=0.4]{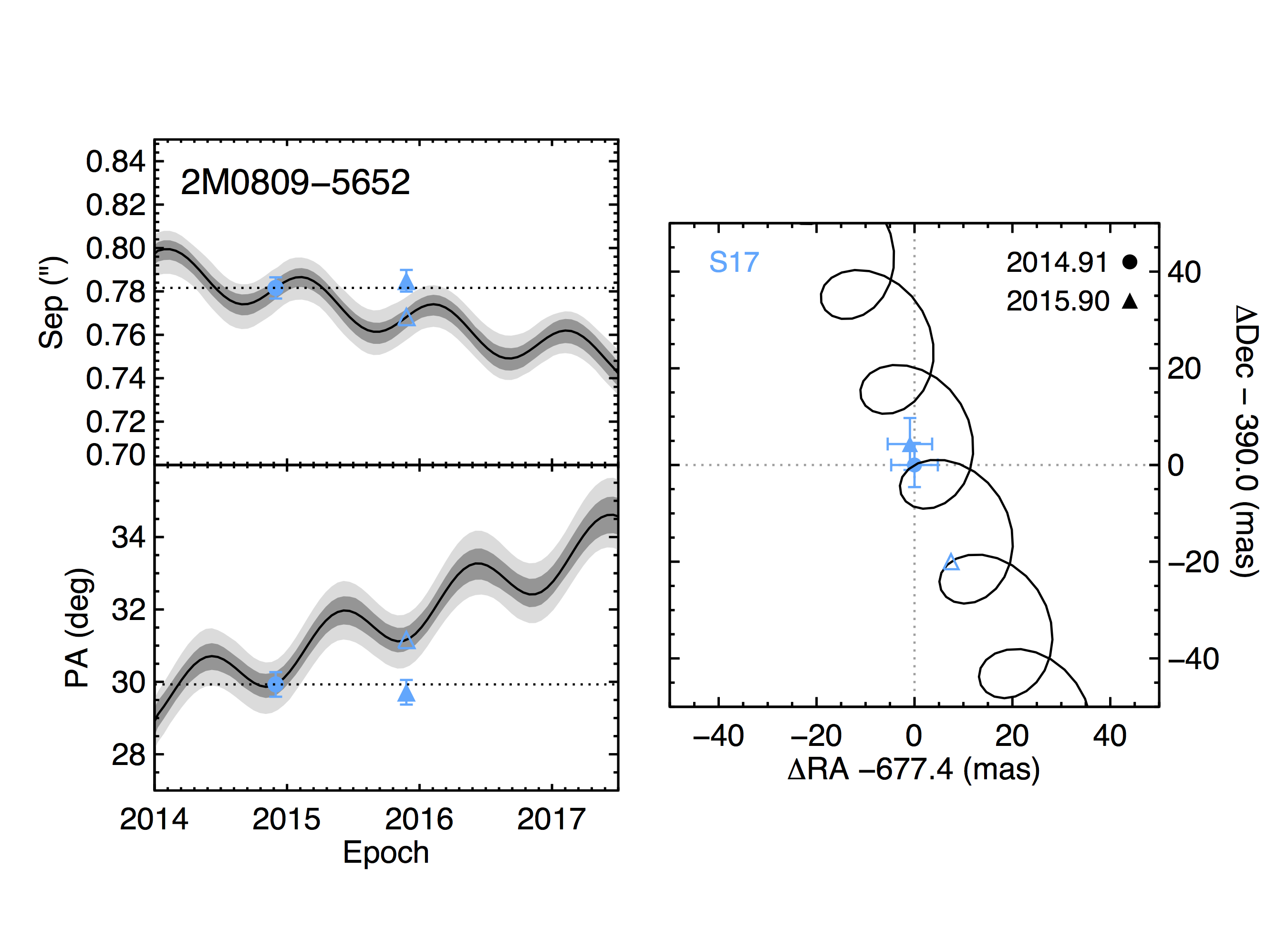}
\end{minipage}%
\begin{minipage}{0.5\textwidth}
\centering
\includegraphics[scale=0.4]{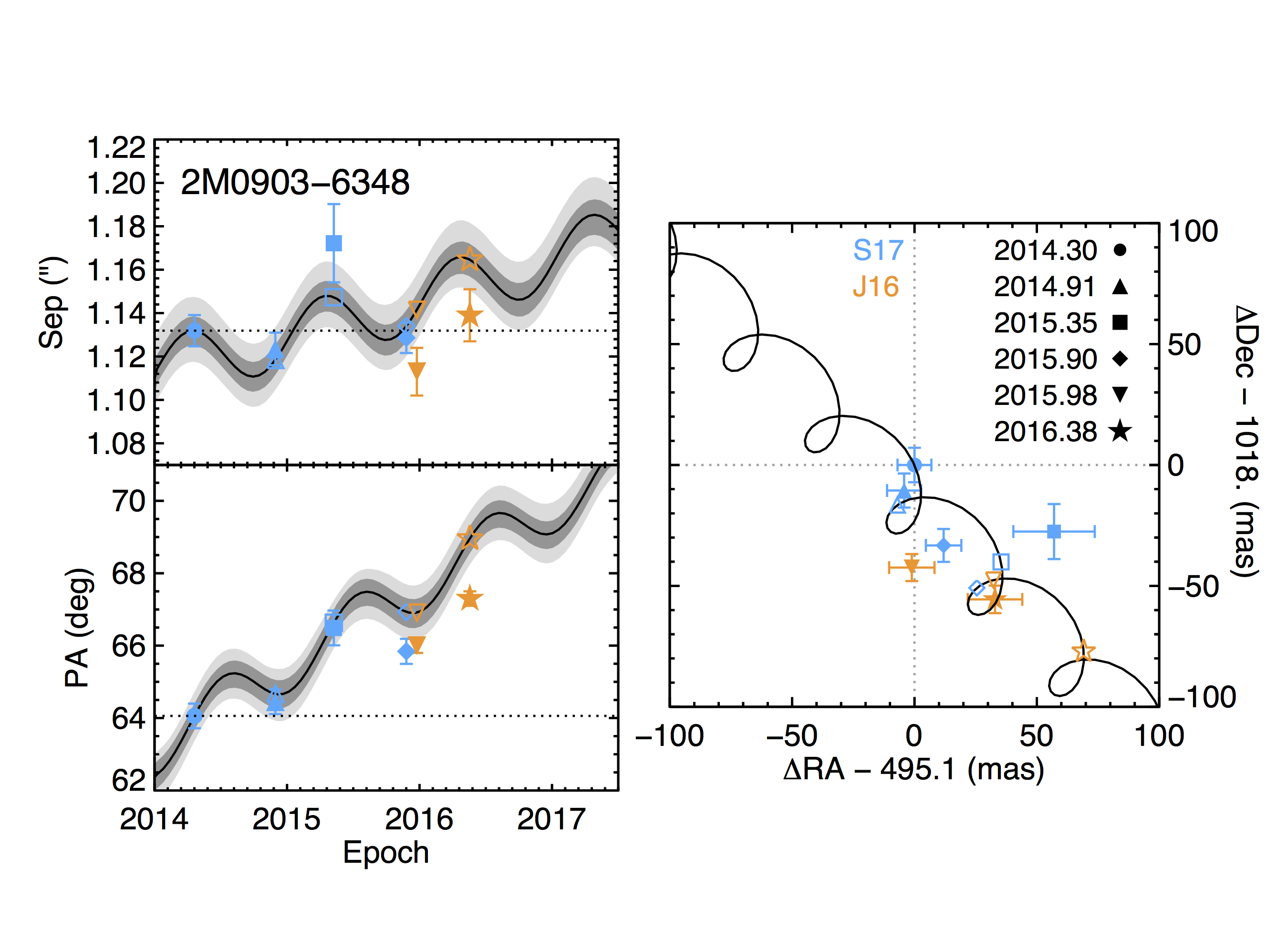}
\end{minipage}%
\begin{minipage}{0.5\textwidth}
\centering
\includegraphics[scale=0.4]{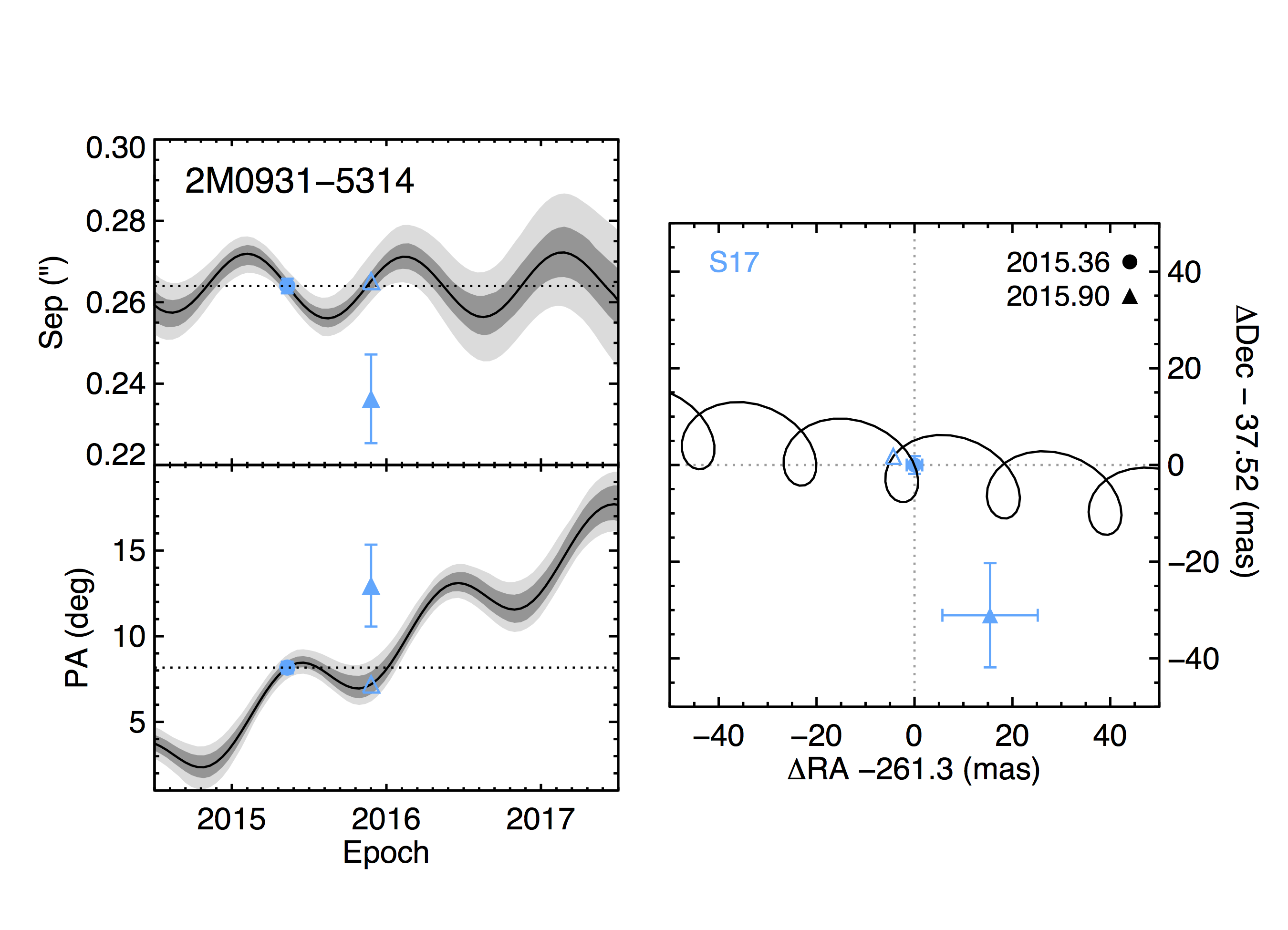}
\end{minipage}%
\begin{minipage}{0.5\textwidth}
\centering
\includegraphics[scale=0.4]{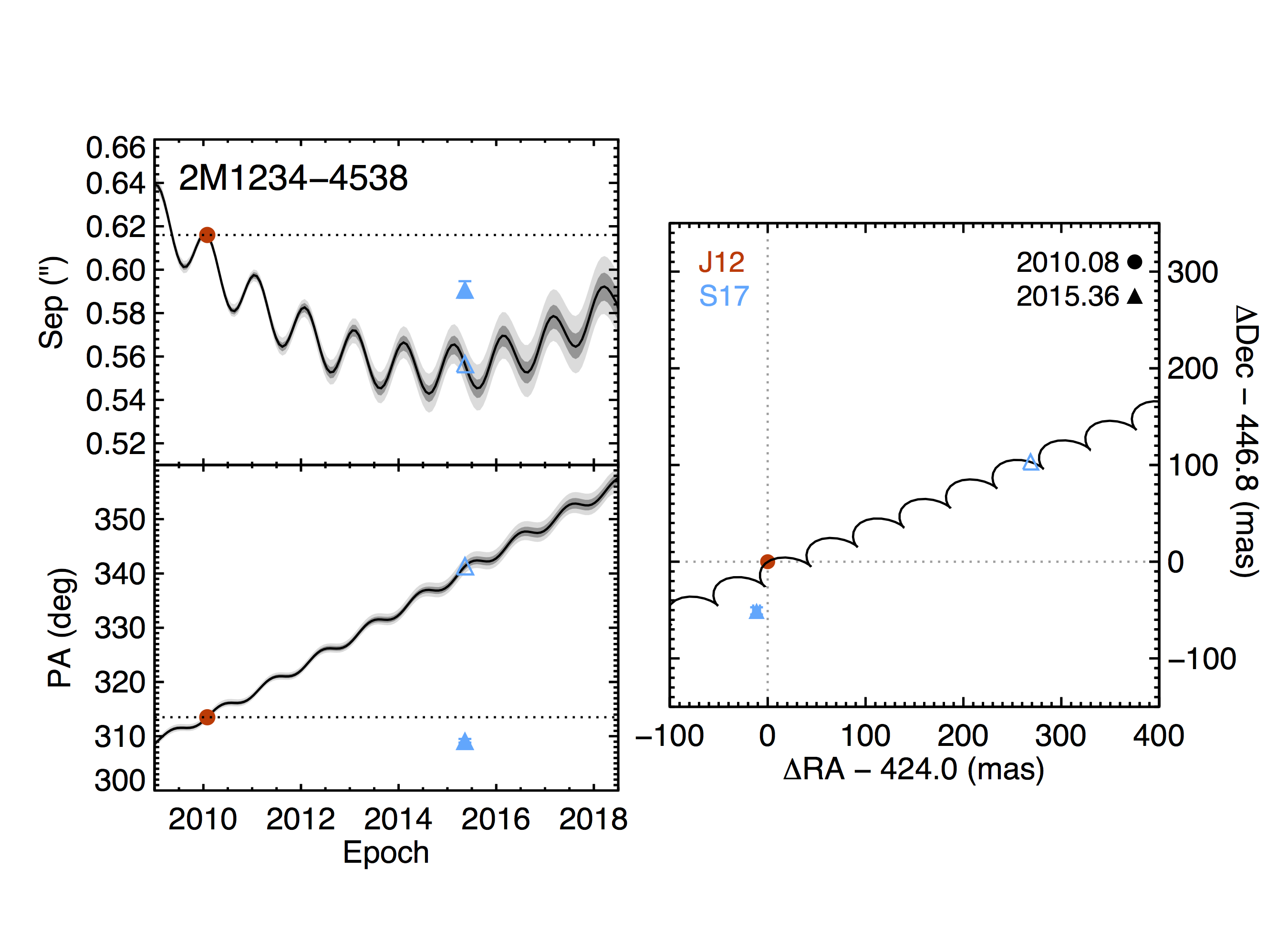}
\end{minipage}%
\caption{Same as in Figure \ref{fig:ast1}  }
\label{fig:ast2}
\end{figure*}

\clearpage
% --- Figure showing Sensitivity Maps ---
\begin{figure*}[htb] 
\begin{minipage}{0.5\textwidth}
\centering
\includegraphics[scale=0.4]{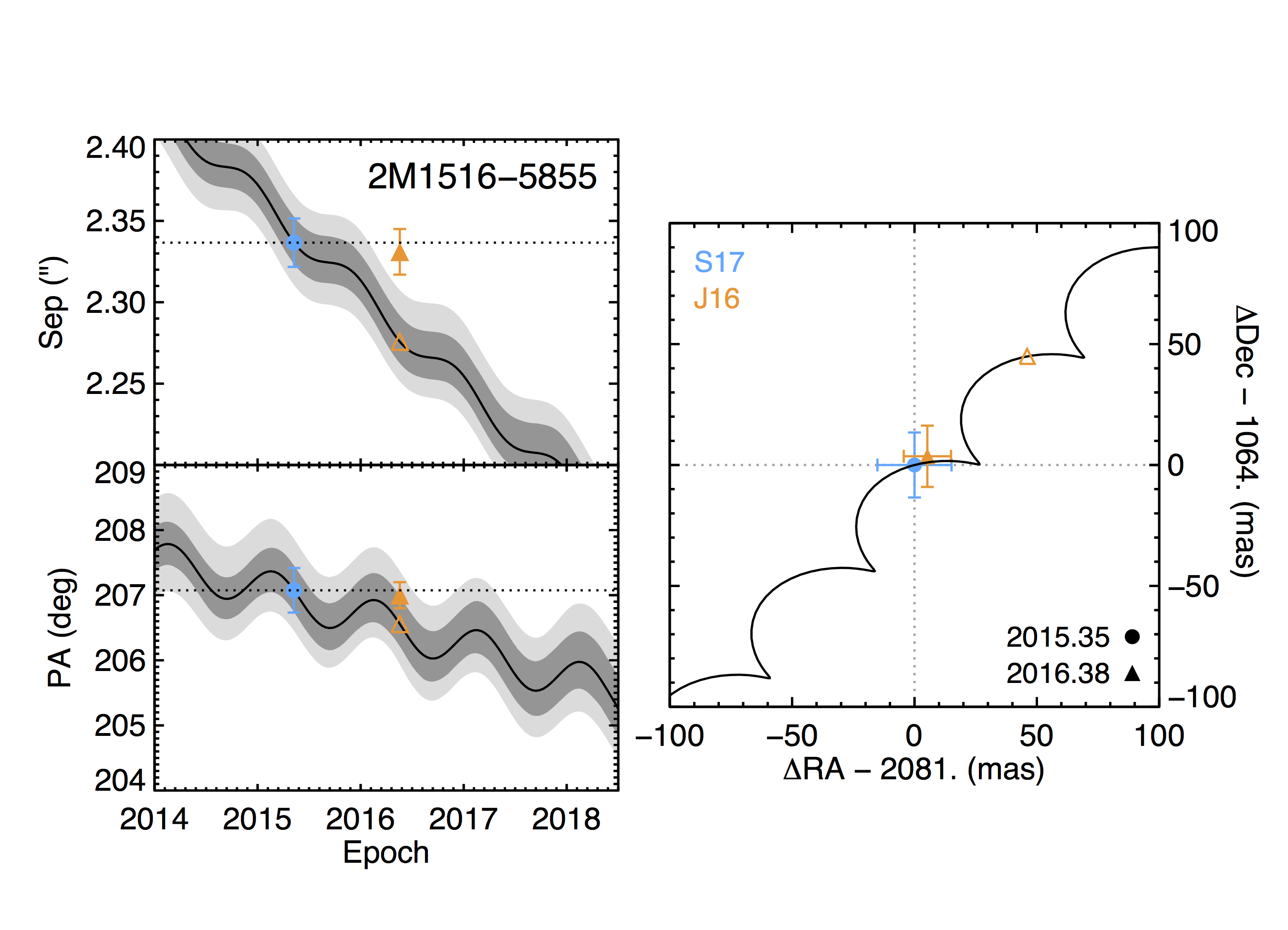}
\end{minipage}%
\begin{minipage}{0.5\textwidth}
\centering
%\vspace{-5cm}
\includegraphics[scale=0.4]{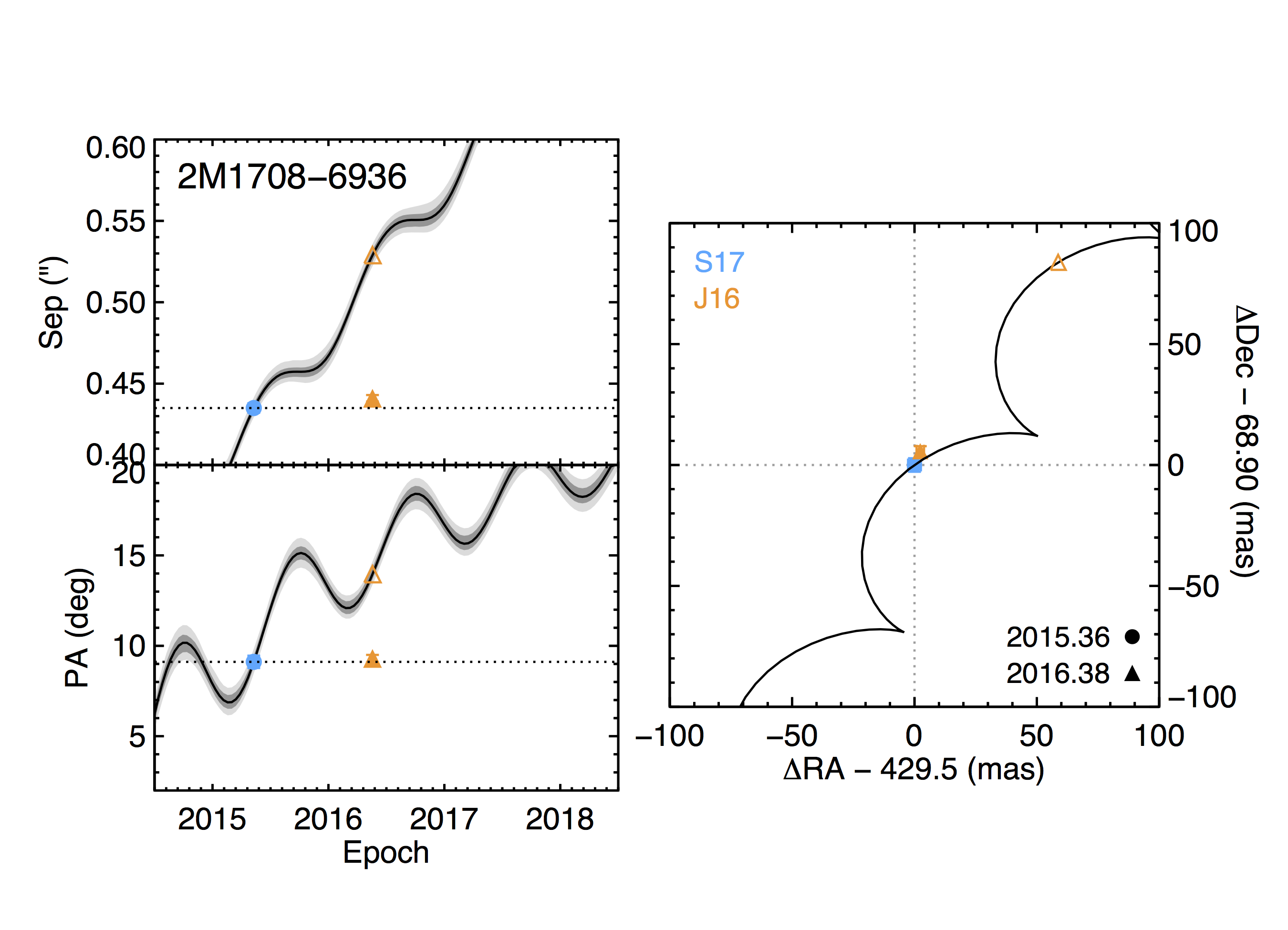}
\end{minipage}%
\begin{minipage}{0.5\textwidth}
\centering
\includegraphics[scale=0.4]{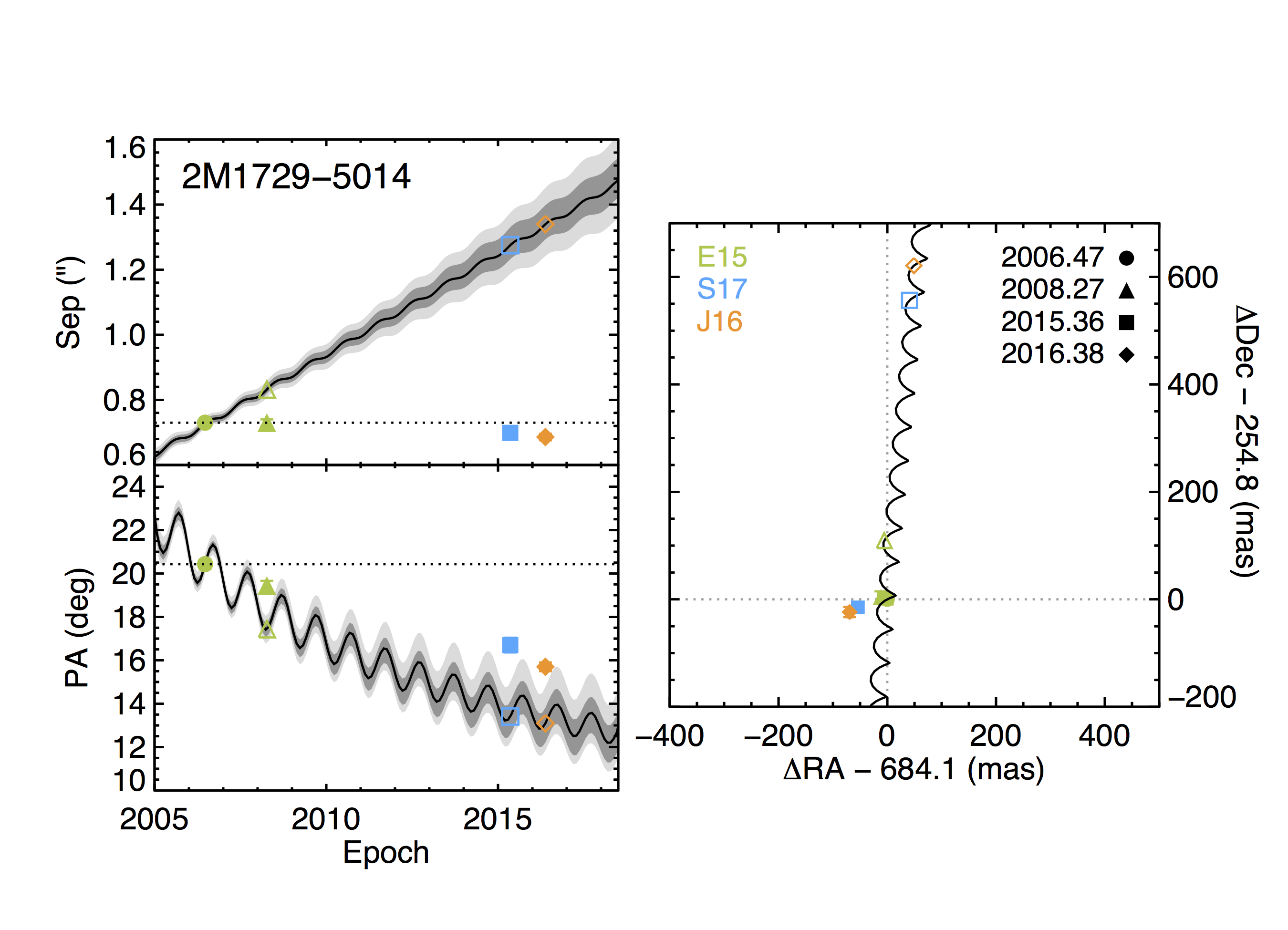}
\end{minipage}%
\begin{minipage}{0.5\textwidth}
\centering
\includegraphics[scale=0.4]{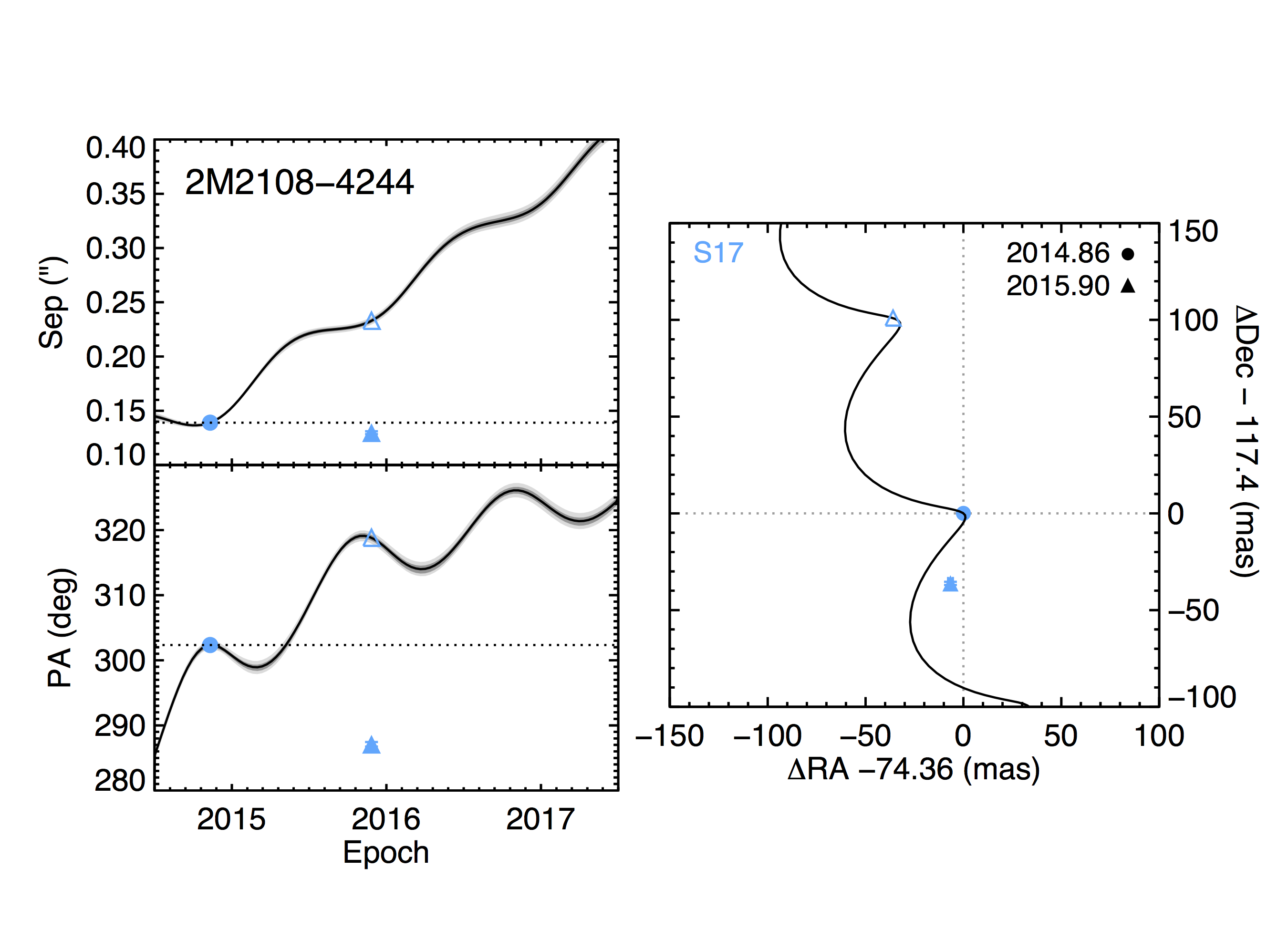}
\end{minipage}%
\begin{minipage}{0.5\textwidth}
\centering
\includegraphics[scale=0.4]{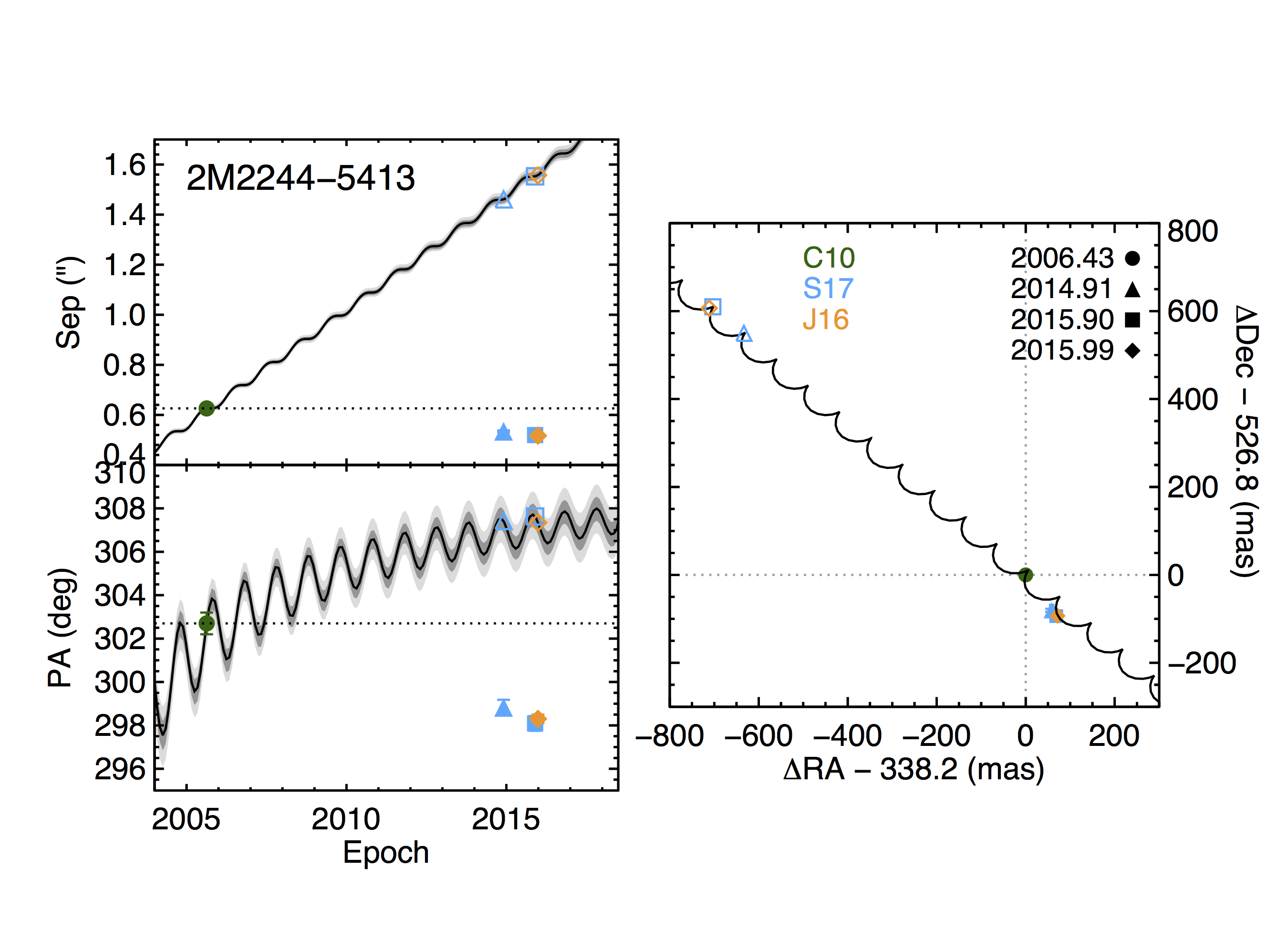}
\end{minipage}%
\begin{minipage}{0.5\textwidth}
\centering
\includegraphics[scale=0.4]{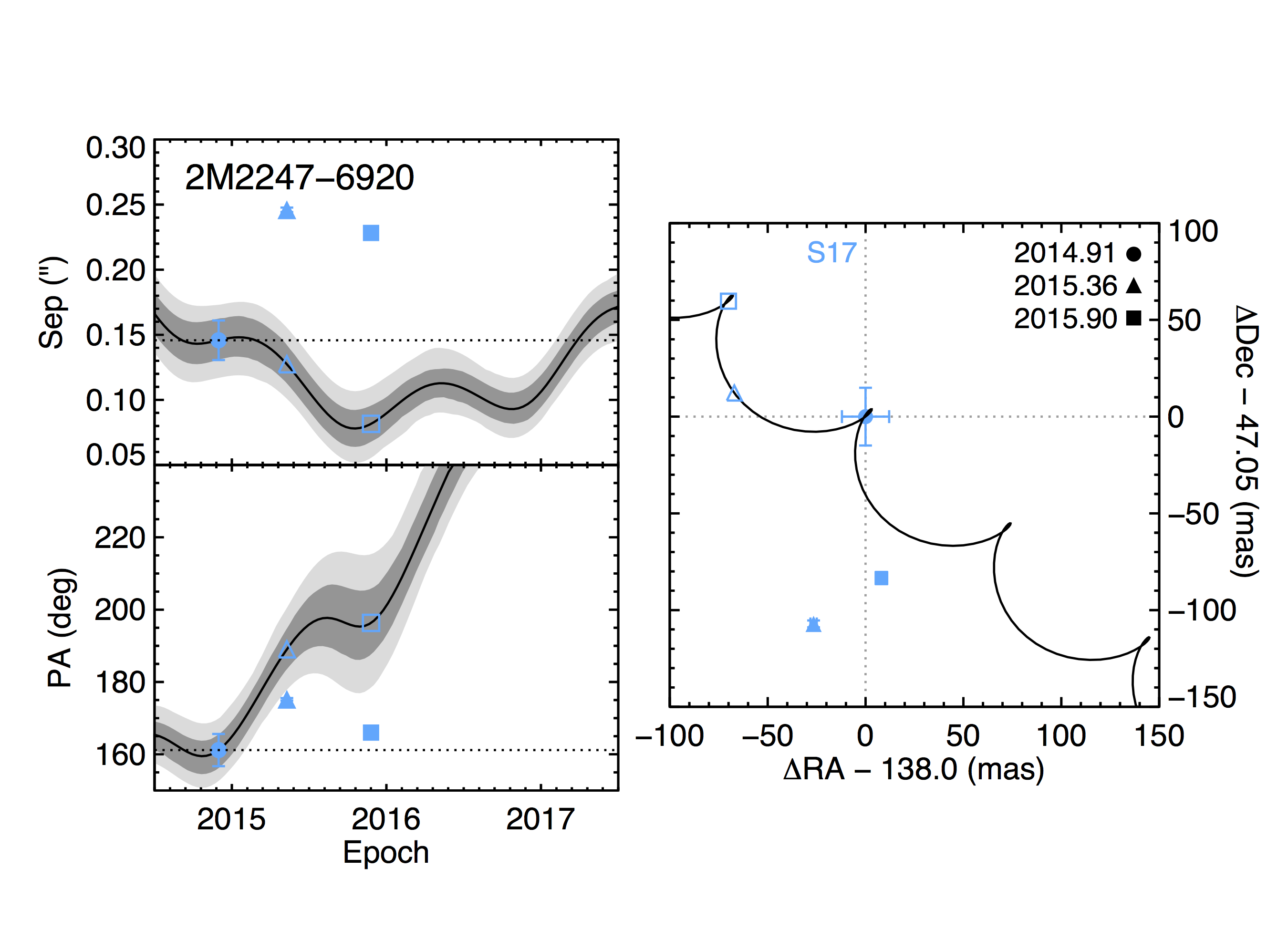}
\end{minipage}%
\caption{Same as in Figure \ref{fig:ast1}  }
\label{fig:ast3}
\end{figure*}

%==============================================
%\end{appendices}

\end{document}